\documentclass[
  journal=custom, 
  manuscript=research-paper, 
  year=2024,
  volume=00,
  dates=false,
]{scntfcjrnl-journal}

\usepackage{xargs}                      

\usepackage{tikz}
\usetikzlibrary{arrows}
\usepackage{colortbl}
\definecolor{BlueCustom}{HTML}{084594}
\definecolor{GreenCustom}{HTML}{00B980}
\definecolor{YellowCustom}{HTML}{FFAF00}
\definecolor{RedCustom}{HTML}{FF7E79}

\usepackage[utf8x]{inputenc}

\usepackage{hyperref}

\hypersetup{colorlinks,urlcolor=BlueCustom,citecolor=BlueCustom,linkcolor=gray}

\usepackage[switch, modulo]{lineno}

\usepackage[colorinlistoftodos,prependcaption,textsize=footnotesize]{todonotes}
\newcommandx{\improvement}[2][1=]{\todo[linecolor=red,backgroundcolor=red!25,bordercolor=red,#1]{#2}}
\newcommandx{\remark}[2][1=]{\todo[linecolor=GreenCustom,backgroundcolor=GreenCustom!25,bordercolor=GreenCustom,#1]{#2}}

\usepackage[bb=boondox]{mathalfa}
\usepackage{amsmath, amssymb}

\usepackage{xltabular}

\usepackage{microtype,siunitx,booktabs}
\usepackage{rotating}
\usepackage{pdflscape}
\usepackage{lscape}
\usepackage{graphicx}
\usepackage{booktabs}
\usepackage{multirow}
\usepackage[normalem]{ulem}
\useunder{\uline}{\ul}{}

\usepackage{titletoc}

\usepackage{cleveref}

\usepackage{caption}
\usepackage{float}
\usepackage{placeins}
\usepackage{makecell}  
\usepackage{array}     

\title[Within-vector viral dynamics challenges how to model the extrinsic incubation period for major arboviruses]{Within-vector viral dynamics challenges how to model the extrinsic incubation period for major arboviruses: dengue, Zika, and chikungunya}

\author{Léa Loisel}
\affiliation{Oniris, INRAE, BIOEPAR, Nantes, 44300, France}
\email[L. Loisel]{lea.loisel@inrae.fr}

\author{Vincent Raquin}
\affiliation{EPHE, Université PSL, INRAE, Université Claude Bernard Lyon 1, IVPC, UMR754, Lyon, F-69007, France}

\author{Maxime Ratinier}
\affiliation{EPHE, Université PSL, INRAE, Université Claude Bernard Lyon 1, IVPC, UMR754, Lyon, F-69007, France}

\author{Pauline Ezanno}
\affiliation{Oniris, INRAE, BIOEPAR, Nantes, 44300, France}

\author{Gaël Beaunée}
\affiliation{Oniris, INRAE, BIOEPAR, Nantes, 44300, France}


\begin{document}

\begin{abstract}
Arboviruses represent a significant threat to human, animal, and plant health worldwide. To elucidate transmission, anticipate their spread and efficiently control them, mechanistic modelling has proven its usefulness. However, most models rely on assumptions about how the extrinsic incubation period (EIP) is represented: the intra-vector viral dynamics (IVD), occurring during the EIP, is approximated by a single state. After an average duration, all exposed vectors become infectious. Behind this are hidden two strong hypotheses: (i) EIP is exponentially distributed in the vector population; (ii) viruses successfully cross the infection, dissemination, and transmission barriers in all exposed vectors. To assess these hypotheses, we developed a stochastic compartmental model which represents successive IVD stages, associated to the crossing or not of these three barriers. We calibrated the model using an ABC-SMC (Approximate Bayesian Computation - Sequential Monte Carlo) method with model selection. We systematically searched for literature data on experimental infections of \textit{Aedes} mosquitoes infected by either dengue, chikungunya, or Zika viruses. We demonstrated the discrepancy between the exponential hypothesis and observed EIP distributions for dengue and Zika viruses and identified more relevant EIP distributions  . We also quantified the fraction of infected mosquitoes eventually becoming infectious, highlighting that often only a small fraction crosses the three barriers. This work provides a generic modelling framework applicable to other arboviruses for which similar data are available. Our model can also be coupled to population-scale models to aid future arbovirus control.\\
\end{abstract}

\section{INTRODUCTION}
\label{sec:int}

Arboviruses, which are viruses transmitted to vertebrate hosts by the bite of arthropod vectors, caused almost a third of emerging infectious diseases over the last two decades \citep{jones_global_2008}. The prevalence of arbovirosis has increased considerably worldwide \citep{girard_arboviruses_2020}, especially due to intensified movements of people and goods as well as environmental changes \citep{mayer_emergence_2017,braack_mosquito-borne_2018,gould_emerging_2017}. This global expansion, combined with an increased resistance of mosquitoes to insecticides and the lack of vaccines \citep{moyes_contemporary_2017}, complicates the control of these diseases \citep{mayer_emergence_2017, braack_mosquito-borne_2018}. Among arbovirosis of major and growing importance for public health are those caused by the Zika (ZIKV), dengue (DENV), and chikungunya (CHIKV) viruses \citep{martinez_going_2019}. These viruses, mainly transmitted to humans by \textit{Aedes} mosquitoes, are currently distributed over both hemispheres of the globe, making them a worldwide threat \citep{bellone_role_2020}. A better understanding of the transmission dynamics of these viruses is therefore essential to anticipate and prevent future epidemics.

Arbovirus transmission is a complex process with multiple stages, from viral infection of the vector to virus spread within hosts populations, influenced by biotic and abiotic factors \citep{kramer_dissecting_2015}. Epidemiological models of transmission between vectors and hosts have been developed to better understand the stages of arbovirus transmission \citep{reiner_systematic_2013}. Accuracy and reliability of these models are needed to guide arbovirus surveillance and implement efficient arboviroses management \citep{chowell_basic_2013}. A key assumption of these models is how the extrinsic incubation period (EIP) is represented \citep{prudhomme_native_2019,christofferson_bridging_2016}. This EIP corresponds to the time required for a vector having acquired a virus during a blood meal on an infectious host to become infectious and able to transmit the virus to a susceptible host \citep{christofferson_estimating_2011}. Transmission thus varies according to the relationship between EIP and mosquito lifespan and biting rate. During the EIP, a virus must cross three barriers within the vector before reaching its saliva (Fig~\ref{IVD_diagram}). The intra-vector viral dynamic (hereafter referred to as IVD) corresponds to the dynamic of the virus crossing these three barriers: (i) the infection barrier, crossed when the virus enters the vector intestinal epithelium; (ii) the dissemination barrier, crossed when the virus exits from the vector intestine to reach the circulatory system, then spreads throughout the vector body until reaching and eventually crossing the (iii) transmission barrier, when the virus is excreted in the mosquito saliva \citep{black_flavivirus_2002}. At that stage, the virus can potentially be transmitted to a next host during a blood meal.

The IVD has been studied by experimental infection assays for various mosquitoes-arbovirus pairs \citep{aitken_vitro_1977, lambrechts_genetic_2009} but remains poorly considered in mechanistic models of vector-borne virus transmission. It has been shown recently, by combining experimental data and statistical modelling, that IVD variations for DENV impacts the epidemic size and probability of epidemic onset \citep{fontaine_epidemiological_2018}. Most transmission models do not explicitly represent the IVD but provides only an approximation with a single compartment whose exit rate is assumed constant and equal to the inverse of the average EIP \citep{chowell_basic_2013,christofferson_chikungunya_2014}. While parsimonious, this mathematical formalism relies on two strong assumptions: (i) an exponential distribution of the EIP within the mosquito population; and (ii) the successful crossing of all the three barriers by viruses in all exposed mosquitoes. It has been shown for measles that the incubation period in hosts is better represented by non-exponential distribution \citep{bailey_statistical_1954} or gamma distribution \citep{anderson_spread_1980}). In vector-borne diseases as well, the chosen EIP distribution can impact virus transmission between vector and hosts, e.g. inducing strong variations in the basic reproductive ratio for bluetongue \citep{brand_interaction_2016}  or shaping the height and timing of hosts epidemic peak \citep{chowell_basic_2013} and the early increase in infected hosts \citep{chowell_estimation_2007} for dengue. This strongly questions the use of an exponential distribution for the EIP duration in epidemiological models. In addition, vector competence data call into question the systematic virus crossing of all barriers in exposed mosquitoes \citep{obadia_zika_2022, kain_not_2022}. The use of a single compartment to approximate the EIP makes it impossible to distinguish the different stages of the IVD and thus assess the barriers impact on EIP.

By combining laboratory experiments and statistical models, \citet{lequime_modeling_2020} have studied the influence of biotic and abiotic factors on the IVD for DENV. However, to account for this new knowledge to better anticipate virus spread at a larger scale, a mechanistic model of the IVD is first needed. Such a intra vector mechanistic model would permit to explicitly relate the three barriers crossings to the (a)biotic factors, and thus to estimate the associated proportion of successful vector infection, dissemination and transmission. Three recent studies have proposed mechanistic models of within-mosquito viral dynamics, adapted to experimental data on DENV \citep{johnson_investigating_2024}  and ZIKV \citep{tuncer_determining_2021,lord_mechanistic_2023}.The authors studied the impact of the infectious dose on two of the three stages of the IVD (infection and dissemination, or infection and transmission), at the viral level for DENV, and at the viral and cellular levels for ZIKV. A comprehensive and more generic mechanistic model of the whole IVD is still missing, while it would help characterising IVD for various arboviruses of major importance to public health.

Our aim was to better characterise the IVD for DENV, ZIKV, and CHIKV, three arboviruses of major importance for public health. To assess the relevance of assuming a systematic virus crossing of all within-mosquito barriers and an exponentially distributed EIP, we proposed a stochastic mechanistic model of the IVD that explicitly represents the IVD stages and the crossing of the three within-mosquito barriers. To assess the variation of these processes according to biotic factors (virus species or strains, mosquito species, infectious doses), we calibrated this generic model with a comprehensive set of published experimental data using comparable protocols. We compared the results obtained using an exponential versus a beta distributed EIP and inferred the probability of crossing each of the IVD barriers. Our results clearly demonstrate the discrepancy between the exponential assumption and realistic EIP distributions. They also highlight that only a fraction of exposed mosquitoes eventually become infectious. This newly acquired knowledge may be employed to enhance the robustness of existing epidemiological models for mosquito-borne diseases.

\section{RESULTS}
\label{sec:res}

\subsection{Exponential distribution of EIP is rarely relevant for DENV and ZIKV}

\begin{figure*}[h!t]
\centering
\includegraphics[width=\textwidth]{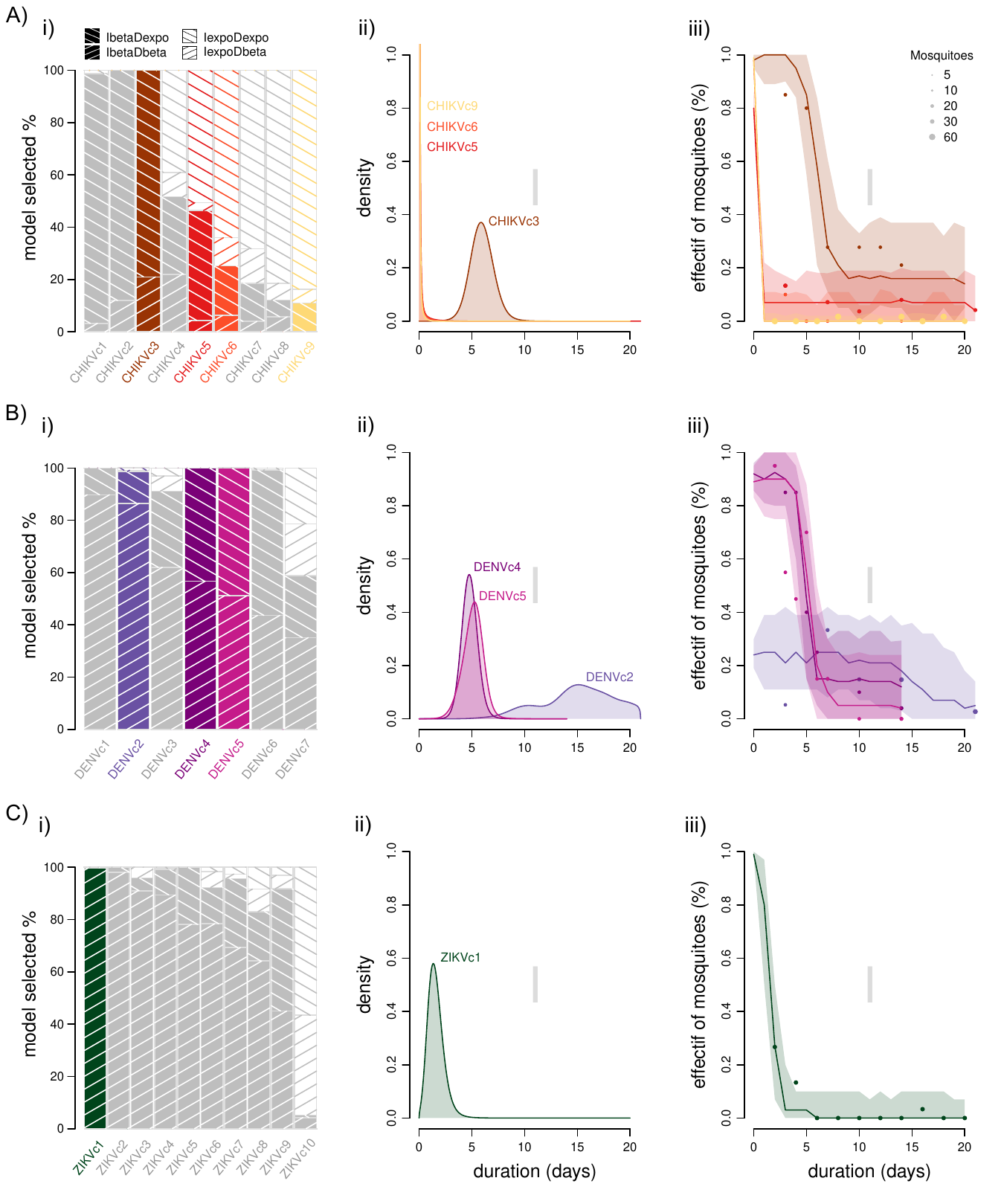}
\caption{Inference results for IVD stages distributions for all scenarios tested for CHIKV (A), DENV (B) and ZIKV (C) with SEIDT model: i) Selected proportion of each model (modBetaBeta, modBetaExpo,modExpoBeta, modExpoExpo) for each scenario with one colour per scenario. ii) Average of the selected distributions in the infected state for the main model selected (only selected scenarios* are represented). iii) Selected dynamics in the infected state for the main model selected (only selected scenarios* are represented). Dots represent observed data, line mean dynamics and  uncertainty ribbons (5\%-95\%) represent selected simulated dynamics for each scenario.
* \textit{scenarios with 5 or more observed Dpe}}
\label{result_inf_SEIDT}
\end{figure*}

Using vector competence data available in the literature for DENV, ZIKV and CHIKV, and selecting the best distributions for the duration in the infected and disseminated states while inferring model parameter values, we demonstrated that the beta distribution was much more frequently selected than the exponential one, especially for DENV and ZIKV  (Fig.~\ref{result_inf_SEIDT}). For the infected state, the main distribution selected was beta for 7/7 scenarios tested for DENV (Fig.~\ref{result_inf_SEIDT}Bi)) and for 9/10 scenarios tested for ZIKV (Fig.~\ref{result_inf_SEIDT}Ci)). For the disseminated state, the main distribution selected was as well beta for 6/7 scenarios tested for DENV (Fig.~\ref{result_inf_SEIDT}Bi)) and 8/10 scenarios tested for ZIKV (Fig.~\ref{result_inf_SEIDT}Ci)). Conversely, an exponential distribution was more often selected for chikungunya for both states (6/9 and 9/9 scenarios for the infected and the disseminated states, respectively  (Fig.~\ref{result_inf_SEIDT}Ai)). We obtained similar results when inferring parameters of the partial counterpart (SEID, see “~\ref{subsec:model_design} Model design” section) of the complete model (SEIDT, see “~\ref{subsec:model_design} Model design” section). SEID model was used when data on the last IVD stage was not available (Figs.~\ref{result_inf_SEID}).

\subsection{Intra-mosquito barriers are not systematically crossed}
\begin{figure*}[h!t]
\centering
\includegraphics[width=\textwidth]{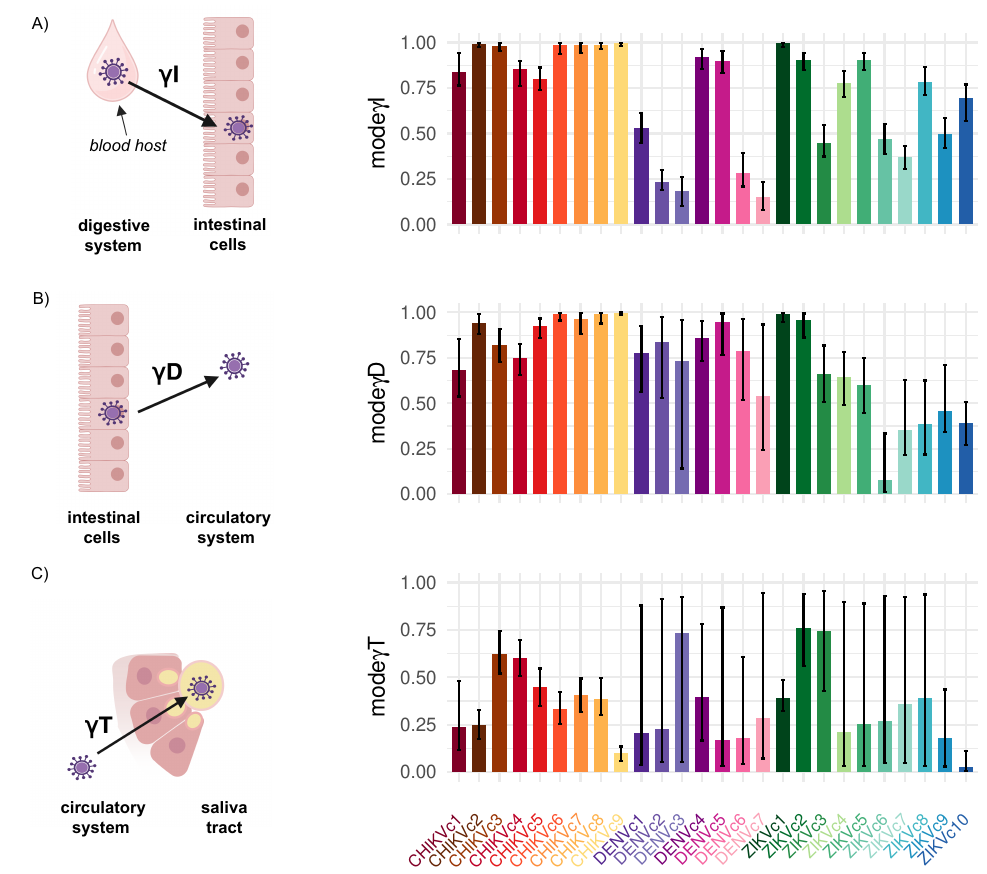}
\caption{Values of the crossing barrier parameters inferred for all scenarios tested for CHIKV, DENV and ZIKV with SEIDT model. A. infection barrier. B. Dissemination barrier. C. Transmission barrier. Each bar represents the parameter mode for a given scenario, with one colour per scenario and with the 90\% credibility interval represented by the error bar.}
\label{result_params_SEIDT}
\end{figure*}

To quantify the probability of crossing each of the three intra-mosquito barriers, we introduced specific parameters in the model, \(\gamma_I\), \(\gamma_D\), and \(\gamma_T\) being the proportion of mosquitoes for which the infection, dissemination, and transmission barriers were crossed, respectively. The values inferred for these three parameters indicated a non-systematic crossing of the three barriers. Indeed, the modes of \(\gamma_I\), \(\gamma_D\), and \(\gamma_T\) were different from 1, and their 90\%  credibility intervals (90\%  CI) did not include 1, for 19/26, 22/26, and 26/26 scenarios, respectively, using the complete SEIDT model (Fig.~\ref{result_params_SEIDT}; Table.~\ref{table_params_SEIDT}), and for 7/17 and 13/17 scenarios, respectively, using the partial SEID model (Fig.~\ref{result_params_SEID}; Table.~\ref{table_params_SEID}). However, the probabilities to cross the infection (\(\gamma_I\)) and dissemination (\(\gamma_D\)) barriers were sometimes very high  , being statistically (using a Wilcoxon test) above 0.9 for 8/26 and 9/26 scenarios, respectively, using the SEIDT model (Figs.~\ref{result_params_SEIDT}A-B; Tables~\ref{table_params_SEIDT} and \ref{table_wilcox_0.9_SEIDT}); and for 10/17 and 8/17 scenarios, respectively, using the SEID model (Figs.~\ref{result_params_SEID}A-B;Tables~\ref{table_params_SEID} and \ref{table_wilcox_0.9_SEID}). These results highlight a not systematic crossing of these two barriers when mosquitoes are exposed to an infectious blood meal. The probability to cross the transmission barrier \(\gamma_T\) was much lower, being statistically (using a Wilcoxon test) lower than 0.5 for 19/26 scenarios using the SEIDT model (Figs.~\ref{result_params_SEIDT}C; Tables~\ref{table_params_SEIDT} and \ref{table_wilcox_0.5_SEIDT})). This underlines the key role of this last transmission barrier during IVD \citep{merwaiss_chikungunya_2021}.

\subsection{IVD is influenced by biotic factors}

The type and shape of the main distributions selected for the durations in the infected and disseminated states and the mode values of \(\gamma_I\) , \(\gamma_D\) , and \(\gamma_T\) varied according to the scenarii studied (Figs.~\ref{result_inf_SEIDT}A--C~i and ii; ~\ref{result_inf_SEID}A--C~i and ii; Figs.~\ref{result_params_SEIDT} and \ref{result_params_SEID}).This suggests that factors such as the species and origin of the virus, the species and origin of the mosquito, or the infectious dose could have an impact on the characteristics of intra-vector viral dynamics. However, it was not possible with currently available published data, despite their richness, to characterise this impact more precisely, because of the disparity in the experimental conditions. 

\subsection{The inference quality depends on experimental data}

The quality of our model inference was satisfying based on an adequate level of visual fit between observed and simulated data for the models selected. (Figs.~\ref{result_inf_SEIDT}A-Ciii, ~\ref{result_inf_SEID}A-Ciii, ~\ref{fig_CHIKVc1}B-~\ref{fig_ZIKVp6}B). A total of 19/26 scenarios exhibited a satisfactory visual fit quality for the three IVD states with the complete SEIDT model (Table~\ref{table_visual_fit_SEIDT}), while 17/17 scenarios demonstrated a similar level of visual fit for the first two IVD states with the partial SEID model (Table~\ref{table_visual_fit_SEID}). This fit quality was also demonstrated for many scenarios by the low mean of root mean squared error (RSME) for each dynamic. The number of scenarios with a mean RMSE lower than 5 was 19/26, 19/26, and 21/26 scenarios for the infected, disseminated, and transmitter states, respectively, using the complete SEIDT model (Table~\ref{table_rmse_SEIDT}). All scenarios had a mean RSME lower than 5 for the infected and disseminated states using the partial SEID model (Table~\ref{table_rmse_SEID}). A decrease in the fit quality was observed (Figs.~\ref{fig_CHIKVc1}B-~\ref{fig_ZIKVp6}B)) when the dynamics of the infected, disseminated, and transmitter states deviated from a “classic” dynamics (i.e., increasing in the infected state, increasing and then decreasing in the disseminated state, and increasing in the transmitter state). For 6 scenarios (DENVc1, DENVc2, ZIKVc4, ZIKVc5, ZIKVc6, ZIKVc10), we observed an increasing then decreasing dynamics in the infected state (Table~\ref{table_dynamics_SEIDT}). In the same way, for 8 scenarios (CHIKVc1, CHIKVc2, CHIKVc5, CHIKVc6, CHIKVc7, CHIKVc8, DENVc7, ZIKVc1), we observed an increasing then decreasing dynamics in the transmitter state (Table~\ref{table_dynamics_SEIDT}). Our model, driven by up-to-date assumptions with regards to IVD, cannot represent such dynamics as it assumes no back-and-forth between states. This explains the limited visual adequacy obtained for these scenarios. In addition, we obtained 13/26 scenarios for which the infection, dissemination, and transmission rates, taken from literature, were within the 90\% credibility interval (90\% CI) obtained for the barrier crossing parameters \(\gamma_I\), \(\gamma_D\), and \(\gamma_T\) (Table~\ref{table_params_SEIDT}).

We obtained a low uncertainty in the estimated values for the majority of our parameters. For parameters involved in barrier crossing (i.e., \(\gamma_I\), \(\gamma_D\), and \(\gamma_T\) ), we observed a narrow 90\% CI for at least half of the scenarios studied (Fig.~\ref{result_params_SEIDT} and Fig.~\ref{result_params_SEID}). The 90\% CI were particularly narrow for \(\gamma_I\)(90\% CI < 0.15 for 16/26 scenarios with the SEIDT model, Table~\ref{table_params_SEIDT}); and 14/17 scenarios with the SEID model, Table~\ref{table_params_SEID}). The 90\% CI was slightly larger for \(\gamma_D\)(90\% CI < 0.15 for 8/26 scenarios with the SEIDT model, Table~\ref{table_params_SEIDT}); and 7/17 scenarios with the SEID model, Table~\ref{table_params_SEID}), while it was even larger for \(\gamma_T\) (90\% CI < 0.15 for only 2/26 scenarios with the SEIDT model, Table~\ref{table_params_SEIDT}). The uncertainty of inference results was influenced by the size of mosquito samples. It increased for scenarios where the mode of \(\gamma_I\) was low (except two, all  scenarios with mode of \(\gamma_I\)  < 0.5 had a 90\% CI for \(\gamma_D\) and \(\gamma_T\) > 0.4, Table~\ref{table_params_SEIDT}). For distributions of the IVD stage durations, we focused on the dispersion of the density distribution rather than of the distribution parameters themselves. For exponential distributions, we observed a low level of uncertainty across all scenarios, with the selected density distribution for each scenario demonstrating a high degree of superposition (Figs.~\ref{fig_CHIKVc1}A-~\ref{fig_ZIKVp6}A. For beta distributions, the visual appreciation was more challenging. Thus, we studied the closeness between all the selected distributions for each scenario using a Kolmogorov-Smirnov test and calculated the percentage of beta distributions that were statistically similar (p-value > 0.05). In the infected state, 12/19 scenarios with the SEIDT model and 9/15 scenarios with the SEID model (Table~\ref{table_ks_state_I_SEIDT}) exhibited more than 50\% of their beta distributions being similar. Conversely, the uncertainty appeared to be higher for the disseminated state, for which only 3/14 scenarios exhibited more than 50\% of the beta distributions being similar (Table~\ref{table_ks_state_D_SEIDT}). 

\section{DISCUSSION}
\label{sec:dis}
Our study – focused on DENV, CHIKV and ZIKV, three major human arboviruses - clearly demonstrates that two key assumptions commonly made in epidemiological models of vector-borne diseases transmission do not hold in most situations with regards to the virus species and strains, the mosquito species, and the infectious dose. Therefore, the representation of the intra-mosquito viral dynamics (IVD) should be chosen with caution when epidemiological models are used to assess surveillance and control strategies.
First, assuming an exponential distribution for the extrinsic incubation period (EIP) is often not realistic, especially for DENV and ZIKV. This assumption has already been questioned \citep{christofferson_chikungunya_2014, armstrong_successive_2020} and it have been shown for dengue that a log-normal distribution was preferentially selected for the EIP in a meta-analysis of vector competence experiments \citep{chan_incubation_2012}. Our generic and mechanistic modelling framework enables selecting the most relevant EIP distribution according to each of the experimental conditions and associated data tested, without the need to aggregate information across experiments. Used here for three major arboviruses, it can be easily used on others as long as similar experimental data are available.
Second, the three intra-mosquito barriers (infection, dissemination, and transmission) are not systematically crossed by viruses during the EIP of exposed mosquitoes. Although the existence of these barriers is already known \citep{franz_tissue_2015} the majority of epidemiological models of vector-borne diseases transmission do not take them into account and assume that all exposed mosquitoes eventually become infectious. Our modelling approach allows us to represent and quantify the probabilities of viruses crossing each of these barriers as a function of experimental conditions, thereby highlighting that the transmission barrier was the most challenging of the three to cross. 

A critical contribution is that our model allows relaxing these two assumptions if used as a building block in larger population-scale epidemiological models, thus proposing a new way to represent exposed vectors in such models. This will enable further studying the consequences these two assumptions could have on vector-borne disease transmission dynamics \citep{chowell_estimation_2007,chowell_basic_2013,brand_interaction_2016}. Besides influencing the EIP duration \citep{kramer_dissecting_2015,christofferson_bridging_2016}, biotic factors (virus and mosquito species and origin; infectious dose) seems also have an impact on the most relevant distributions of residence time in IVD stages, as well as probabilities of crossing intra-mosquito barriers. Therefore, we propose that modelling choices should be adapted to both biotic factors and exposed vector population in order to assess properly disease management measures.

Ideally, our approach would permit to improve the study of the possible functional link between IVD and the many biotic factors involved, in order to assess IVD outside tested experimental conditions. However, to do so, a considerable amount of additional data would be required. Such a functional link would be at core to account for an IVD varying with biotic factors, themselves varying in space and time, and could significantly impact recommendations made with epidemiological models to public health stakeholders. Hence, further laboratory work is needed to cover wider ranges of biotic factors while using comparable protocols, associated with an increased opening and sharing of experimental data. Indeed, despite experimental data are generated for addressing a given scientific question, they often could be reused, e.g., in meta-analyses or even to address new questions, as long as they are properly shared and standardized. Data used here were published data from studies on vector competence, each under specific experimental conditions. After selecting the most relevant ones from our scientific perspective (Fig.~\ref{prisma_diag}), we were able to re-use them for better understanding the intra-vector viral dynamics. However, a strong limitation for reusing published data lies in finding and properly using them. First, when we performed the systematic search and compared obtained results with review using a very close query \citep{chen_exploring_2023}, we obtained many different records (Fig.~\ref{prisma_diag}). This could be due to a variable terminology   \citep{chen_exploring_2023}. This highlights the difficulty of being comprehensive during a systematic review, due to the lack of standardization of study description and of a systematic use of an ontology \citep{chen_exploring_2023}. Furthermore, a number of the articles retrieved did not meet the necessary criteria for being used, either because they were not in open access or because they comprised solely figures, without the accompanying raw data. Sharing raw and standardized data following the FAIR (Findable-Accessible-Interoperable-Reusable) principle \citep{wilkinson_fair_2016} is crucial for increasing data value and usefulness \citep{wu_minimum_2022}. One potential solution to the issue of data sharing is the establishment of an online platform \citep{chen_exploring_2023}. An example of such a platform is the in-building COMET platform from the Verena program \citep{gallichotte_building_2024}.

One limitation of our study lies in the fact that data used to infer model parameter values comes from laboratory experiments, which sometimes are far from natural field conditions \citep{carrington_human_2014}. Indeed, parameter values governing IVD might be influenced by the type of blood feeding (artificial versus on an infected animal), as it is the case for mosquito infection rates \citep{azar_vector_2019}. Here, we used data from experiments performed with artificial blood feeding, due to a lack of suitable data from experiments performed with blood feeding on an infected vertebrate host. Such experiments are scarce and challenging to conduct due to evident ethical and logistical constraints \citep{achee_considerations_2015,morrison_acceptability_2019}. However, for the rare cases where this type of data is available, our modelling approach can be used without any change to improve knowledge on IVD while increasing model accuracy.

Experimental studies and modelling studies have reciprocal benefits. While models are fed on knowledge from experimental (and field) observations, experimental designs could also benefit from a modelling framework such as the one proposed here. First, our framework could help defining the number of time steps to be observed, often governed by cost and time limitations but which is important to provide a good understanding of vector competence \citep{kain_not_2022}. Second, our framework also enables the identification of the most relevant time points to observe for gaining as much information as possible. For CHIKV, we showed that vector EIP is often exponentially distributed and that the associated within-vector dynamics is generally very fast. As a result, access to several early time points would be more informative for this disease than later time points. In contrast, for ZIKV and DENV, characterised by a much slower IVD and a vector EIP barely exponentially distributed, access to later time points would be important. Our framework could then also be used to select the most appropriate duration for vector competence experiments. For example, the shapes of the exponential and beta distributions inferred for the infected and disseminated states indicate that most mosquitoes already went through these states long before the end of the experiments. This validates the model assumption on these maximum durations, and corroborates the commonly used time windows for vector competence experiments. A second added-value concerns mosquito best sample size determination. Indeed, this directly affects the uncertainty of the inference results, which can be qualified and quantified within our framework. According to \citet{tjaden_extrinsic_2013}, 20-30 mosquitoes is a typical sample size in vector competence experiments. In most cases, this sample size was sufficient to obtain a low uncertainty of our inference results. However, when the barriers to infection or dissemination were high, the number of mosquitoes in the final stage was insufficient to achieve inference results with low uncertainty. Our approach could be used to determine in advance how many mosquitoes should be sampled at each time point to ensure gathering sufficiently robust information, assuming we have a preliminary idea of the concerned IVD dynamics. 

Our model represents a foundational brick that may undergo future refinement to more accurately reflect the underlying processes involved in IVD. While the quality and uncertainty of the inference were satisfactory for a wide range of scenarios, there were instances where simulations and observations did not align. This discrepancy is potentially due to certain mechanisms not yet incorporated into the model. 

One possible avenue for further investigation would be to consider an additional state between exposed and infected vectors, to represent infected but not yet detectable individuals. Indeed, in few experiments, the number of mosquitoes observed in the infected state increased between the first and second day post exposition (Dpe). Our model does not allow such dynamics, as it assumes the infected barrier is crossed in the first few hours after the infected meal, in agreement with the digestion duration \citep{franz_tissue_2015,perrone_time_1988}. Consequently, mosquito numbers in the infected state can only decrease. A potential explanation of such an observed increase could lie in the latency period between the infection of epithelial cells and the replication of viruses within these cells. This could induce a delay to reach the detection threshold for viruses in mosquito bodies \citep{johnson_investigating_2023}. Infection of the mosquito midgut starts with the entry of virions into a few epithelial cells, where they replicate \citep{franz_tissue_2015}. This replication subsequently spreads the infection to neighbouring cells, ultimately resulting in the entire midgut being infected (by 7-10 days for DENV-2 \citep{salazar_dengue_2007}). The low initial number of infected epithelial cells increases the stochasticity of this process \citep{lord_mechanistic_2023}, especially at particular infectious doses (e.g., 7 log 10 FFU/mL \citep{lord_mechanistic_2023,johnson_investigating_2024}. The time required for infecting the midgut and reaching the detection threshold may vary among mosquitoes, depending on the initial number of infected cells. This could explain the increasing dynamics observed in some scenarios, in which infectious doses were 7 log10 FFU/mL for DENV and 7 or 7.2 log10 FFU for ZIKV, and observation days started at 4 Dpe for DENV and 3 or 6 Dpe for ZIKV. These Dpe are earlier than the usual first Dpe \citep{gutierrez-bugallo_vector_2020} and before the usual time to reach a large midgut infection \citep{salazar_dengue_2007} and the maximum proportion of infected mosquitoes \citep{tesla_temperature_2018}. On the one hand, adding a state in the model could facilitate a more comprehensive understanding of the dynamics in the infected state at the start of the experiment. On the other hand, it appears however challenging to parameterise the entry and exit from this new compartment, given the difficulty to experimentally distinguish a mosquito whose gut is "in the process of infection" from a mosquito whose midgut is fully infected. 
A second modification to better reflect observed dynamics could be to allow exiting the transmitter state. This has not been included in our model given the prevailing consensus that, once infected, the virus is not eliminated from the mosquito body \citep{lee_mosquito_2019}. Nevertheless, recent studies on CHIKV have highlighted a reduction in the proportion of mosquitoes with the virus present in their saliva at late time points, questioning this common assumption \citep{prudhomme_native_2019, robison_comparison_2020}. As such dynamics were not possible with our model, there was a discrepancy between observed and simulated data in a few scenarios for CHIKV. However, modelling exits from the transmitter state requires to determine the subsequent state reached by mosquitoes, among the disseminated state or other compartments. This is a challenging task, given the current uncertainty and the lack of knowledge about this mechanism \citep{prudhomme_native_2019, viginier_chikungunya_2023}.
Finally, another modification would be to consider the effect of multiple blood meals on IVD, with the aim of developing a more accurate representation of the natural behaviour of mosquitoes \citep{scott_detection_1993, zahid_biting_2023}. Recently, \citet{armstrong_successive_2020} demonstrated that two successive blood meals increased the dissemination of CHIKV, DENV2, and ZIKV in \textit{Aedes aegypti}. This suggests that parameter values associated to the disseminated state in the IVD model might be influenced by a second blood meal. If so, the dynamics in this state could have been slightly underestimated while neglecting blood meal recurrence. Incorporating this process into our model would lead to a change in the model type, an individual-based model then being more adapted. Furthermore, inferring this process would require experimental data not currently available.

This work provides a generic modelling framework that can be applied to other arboviruses and coupled with population-scale model. This could  enable to study the impact on IVD characteristic variation on vector-host transmission dynamics. Thus incorporating IVD and its variations into vector-host transmission models could enhance their predictive ability. This is crucial for implementing adequate control measures to fight arboviruses, which pose a significant threat to public health in many places of the world, increasing due to global changes.

\section{MATERIALS AND METHODS}
\label{sec:met}

\subsection{Experimental data}
\label{subsec:exp_data}
To identify relevant experimental data to calibrate our model, we performed a systematic review of the literature (Fig.~\ref{prisma_diag}), using the PRISMA protocol \citep{page_prisma_2021}, leading to 15 articles \citep{prudhomme_native_2019, fontaine_epidemiological_2018, merwaiss_chikungunya_2021, robison_comparison_2020, viginier_chikungunya_2023,lequime_modeling_2020,seixas_potential_2018,amraoui_potential_2019,bohers_recently_2020,calvez_assessing_2020, ritchie_explosive_2013, fortuna_assessing_2024,hall-mendelin_assessment_2016, richard_vector_2016,richard_vector_2016-1}(Tables.~\ref{exp_data_SEID} and~\ref{exp_data_SEIDT}) encompassing a total of 43 scenarios. Each scenario consisted in a vector competence experiment on female mosquitoes of a specific genus, species and origin, infected with a virus of a specific species, strain and origin, with a given infectious dose (log10 FFU/mL) obtained during an artificial blood meal. Selected articles used fairly similar experimental protocols (Fig.~\ref{exp_protocol}) and methods for identifying the IVD stages to which mosquitoes belong to. The fully engorged females were kept under consistent conditions of temperature (°c), humidity (\%), and light. They were sacrificed per group of 20-30 individuals on specific days post-exposure (Dpe). A minimum of 4 Dpe was required for experiments to be selected in the present study (maximum number of observed time points = 10). To detect the virus, different titration methods were used (reverse transcription polymerase chain reaction (RT-PCR) assay, focus-forming assay (FFA), plaque forming assay (PFA)). The virus presence was searched in different parts of mosquito bodies (saliva, thorax, abdomen, legs, wings, head) to determine the state of each individual mosquito (i.e., infected, disseminated, or transmitter) at each Dpe.

\subsection{Model design}
\label{subsec:model_design}

To represent the three stages of the IVD and assess when and in which proportions viruses crossed the within-mosquito barriers, we built a stochastic mechanistic compartmental model in discrete time (time step of one day), representing the same processes as observed in the vector competence experiments (Fig.~\ref{model_diag}). The model encompassed one compartment per IVD stage: E for exposed mosquitoes (i.e., virus located in their digestive system in the host blood); I for infected mosquitoes (i.e., virus located in their intestinal cells); D for disseminated mosquitoes (i.e., virus located in their circulatory system); and T for transmitter, infectious mosquitoes (i.e., virus located in their saliva). To account for the non-systematic barrier crossing \citep{black_flavivirus_2002}, we introduced three parameters, \(\gamma_I\), \(\gamma_D\), and \(\gamma_T\), to represent the proportions of mosquitoes for which the infection, dissemination, and transmission barriers were crossed, respectively. We assumed that once a barrier was crossed, there was no way back, thus the virus would not be eliminated from the mosquito body \citep{lee_mosquito_2019}. For mosquitoes for which one or two barriers were not crossed, we have added two compartments I\_s and D\_s (“s” for stop), denoting for mosquitoes for which the infection barrier but not the dissemination barrier, and the dissemination barrier but not the transmission barrier, have been crossed respectively.
To infer the distribution of the IVD stage durations, we divided respectively the I and D compartments into n and m sub-compartments of 1\-day duration each (n and m thus being the maximum length of stay, in days, in I and D). The maximum duration in I was here set at the experiment duration. The maximum duration in D was set at the experiment duration minus one day. These durations were chosen because experiment durations (between 12 and 28 days) were close to mosquito longevity \citep{lambert_meta-analysis_2022}. This approach ensured that experiment duration encompassed the mosquito biting period. In addition, a duration longer than that of the experiment for the maximum duration of stay in I and D could not be calibrated with the experimental data. When crossing a barrier and entering a new IVD stage (either I or D), mosquitoes were distributed among the sub-compartments of this new stage to represent the duration distribution in the stage. This distribution can be either mediated by an exponential distribution (of parameter \(\lambda\)) or by a beta distribution (of parameters \(\alpha\) and \(\beta\)). The beta distribution allowed us to obtain a wide range of shapes depending on \(\alpha\) and \(\beta\) values, without the need for prior knowledge. This distribution being defined on the interval [0,1], it was well suited to represent proportions. These proportions were employed as probabilities to distribute the number of mosquitoes in the n and m sub-compartments, utilising a multinomial distribution. We bounded \(\alpha\) and \(\beta\) between 1 and 100 to avoid distributions tending towards infinity in 0 and 1. 
Due to mortality, the number of sampled mosquitoes fluctuated among observation dates. We accounted for this using parameter Ntot defining the mosquito population size. Ntot was adapted for each scenario based on experimental data. The model was implemented in R. Each model stochastic replicate with a parameter set (\(\gamma_I\),\(\gamma_D\),\(\gamma_T\),(\(\alpha_I\),\(\beta_I\) )  or \(\lambda_I\),(\(\alpha_D\),\(\beta_D\) )  or \(\lambda_D\) ) represented one vector competence experiment and provided outputs comparable to those obtained in the laboratory experiments, i.e., mosquito numbers in infected, disseminated, and transmitter states for each observed Dpe.

\subsection{Model calibration}
\label{subsec:model_calibration}

We inferred model parameters using an Approximate Bayesian Computation (ABC) approach, with a Sequential Monte Carlo (SMC) sampler \citep{del_moral_sequential_2006}. This iterative algorithm improves the basic ABC algorithm by incorporating two main steps: weighted resampling of simulated particles and a gradual reduction in tolerance. As in the ABC rejection approach, a prior distribution is defined, aiming to estimate a posterior distribution. However, in ABC-SMC, this estimation is achieved sequentially by constructing intermediate distributions in each iteration, converging towards the posterior distribution. Our specific implementation of the algorithm (R package \href{https://gaelbn.github.io/BRREWABC/}{BRREWABC}) improves upon Del Moral et al.’s (2006) original algorithm \citep{del_moral_sequential_2006} in three ways: (i) an adaptive threshold schedule selection based on quantiles of distances between simulated and observed data \citep{del_moral_adaptive_2012, drovandi_estimation_2011}; (ii) an adaptive perturbation kernel width during the sampling step, dependent on the previous intermediate posterior distribution \citep{beaumont_adaptive_2009, toni_approximate_2009}; and (iii) the capability to use multiple criteria simultaneously.  We added a step to the inference process to select the best model among four (modBetaBeta, modBetaExpo, modExpoBeta, modExpoExpo) differing in the distribution (Expo: exponential, or Beta) used for the infected (first part of the name) and disseminated (second part) states. Depending on the experimental data used, we inferred either a partial (when no information was available on the transmitter stage) or a complete model (when information was available all over the IVD stages), which led to 4 or 7 parameters inferred, respectively: three proportions (\(\gamma_I\),\(\gamma_D\),\(\gamma_T\)) and distribution law parameters for I ((\(\alpha_I\),\(\beta_I\) ) or \(\lambda_I\)) and D ((\(\alpha_D\),\(\beta_D\)) or \(\lambda_D\)) states. The ranges of variation of each of these parameters and their justifications are presented Table.~\ref{table_model_parameters}. The summary statistics used corresponded to the number of mosquitoes in each state at each Dpe. We calculated a distance for each of the four states, equal to the sum of the squared errors between the observed and simulated data over the different Dpe. We then used an acceptance criterion for each distance, allowing us to accept a particle, providing that all the four criteria were met.

\subsection{Statistical analysis}
\label{subsec:stat}
To analyse the values of \(\gamma_I\), \(\gamma_D\), and \(\gamma_T\), a Wilcoxon test was performed using R to statistically compare these parameters to 0.9 in order to assess the high values, and to 0.5 in order to assess the low values.
To assess the fit quality between observed and simulated data, a visual assessment was initially conducted. Then, the mean root mean squared error (RMSE) was calculated for each dynamic in the infected, disseminated, and transmitter states, for all simulation results and each scenario, using the simulated and observed mosquito numbers at each Dpe. The mean RMSE was expressed in mosquito numbers, with a mean RMSE lower than 5 indicating a good fit. 
To study the uncertainty of parameter values, three methods were employed, depending on the concerned parameter. For the crossing probabilities (\(\gamma_I\), \(\gamma_D\), \(\gamma_t\)), the degree of confidence in the 90\% credibility interval (90\% CI) was evaluated by comparing its width to 0.15, a proportion corresponding to less than 5 mosquitoes. To assess the inference uncertainty for distribution parameters ( \(\alpha\) , \(\beta\)) and (\(\lambda\)), it was more informative to evaluate the density distribution dispersion than the parameter dispersion. To assess the density distribution dispersion, a visual appreciation was first realised. For the exponential distributions, it was relatively straightforward to visualise dispersion. However, the visual appreciation of the dispersion of the beta distributions was more challenging. Consequently, for each distribution of each scenario, a comparison of each density distribution was conducted using a Kolmogorov-Smirnov test, which was then corrected using the FDR (false discovery rate) method \citep{benjamini_controlling_1995}. This was followed by the calculation of the percentage of statistically similar distributions (p-value >0.05) for each scenario.

\section*{Data Availability} 
\label{sec:data_availability}
Code and data used for inference have been deposited on \href{https://forgemia.inra.fr/dynamo/ivd-modelling-and-inference}{GitHub}.

\begin{acknowledgement}
This work was supported by INRAE Metaprogramme  DIGIT-BIO (Digital biology to explore and predict living organisms in their environment), through the MIDIIVEC project.
Funded by the European Union (WiLiMan-ID, grant agreement 101083833). Views and opinions expressed are however those of the author(s) only and do not necessarily reflect those of the European Union or REA. Neither the European Union nor the granting authority can be held responsible for them. We also thank Anne Lehebel and Nadine Brisseau (BIOEPAR, Nantes) for their advice on the statistical analysis of the results.

\end{acknowledgement}


\bibliography{biblio}

\section{Supplementary Material}
\label{sec:sup}

\startcontents[subsections]
\stopcontents[subsections]
\resumecontents[subsections]
\printcontents[subsections]{l}{1}[3]{}

\newpage
\onecolumn


\renewcommand\thefigure{S\arabic{figure}}    
\setcounter{figure}{0} 

\renewcommand\thetable{S\arabic{table}}    
\setcounter{table}{0} 








\clearpage

\clearpage
\subsection{IVD modelling}
\label{subsec:mod}
\begin{center}
    \vspace*{\fill} 
    {\Huge \bfseries IVD modelling} 
    \vspace*{\fill} 
\end{center}
\clearpage

\begin{figure}[H]
\centering
\includegraphics[width=\textwidth]{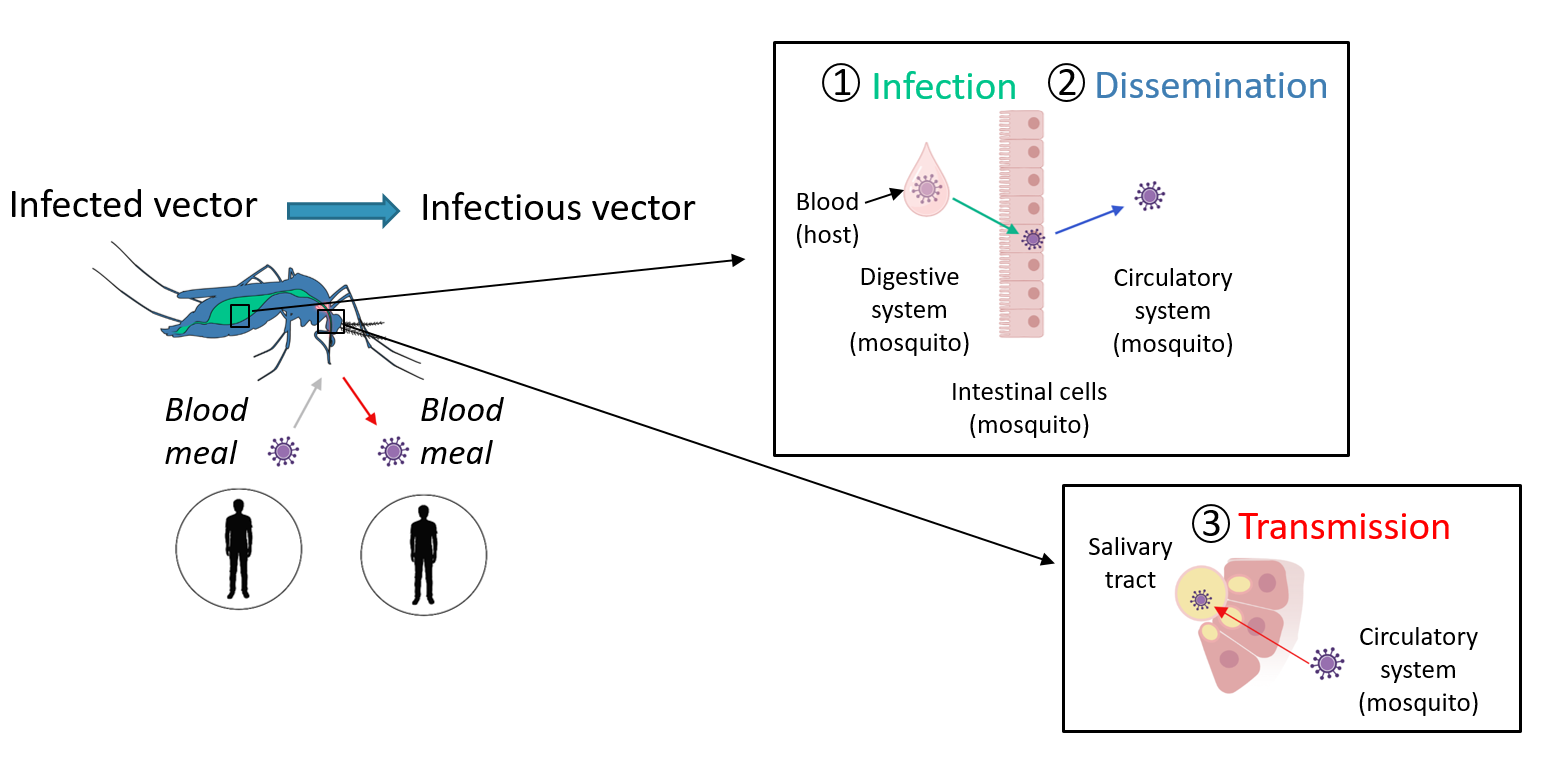}
\caption{Intra-vector viral dynamics conceptual diagram.}
\label{IVD_diagram}
\end{figure}
\FloatBarrier

\newpage
\begin{figure}[H]
\centering
\includegraphics[width=\textwidth]{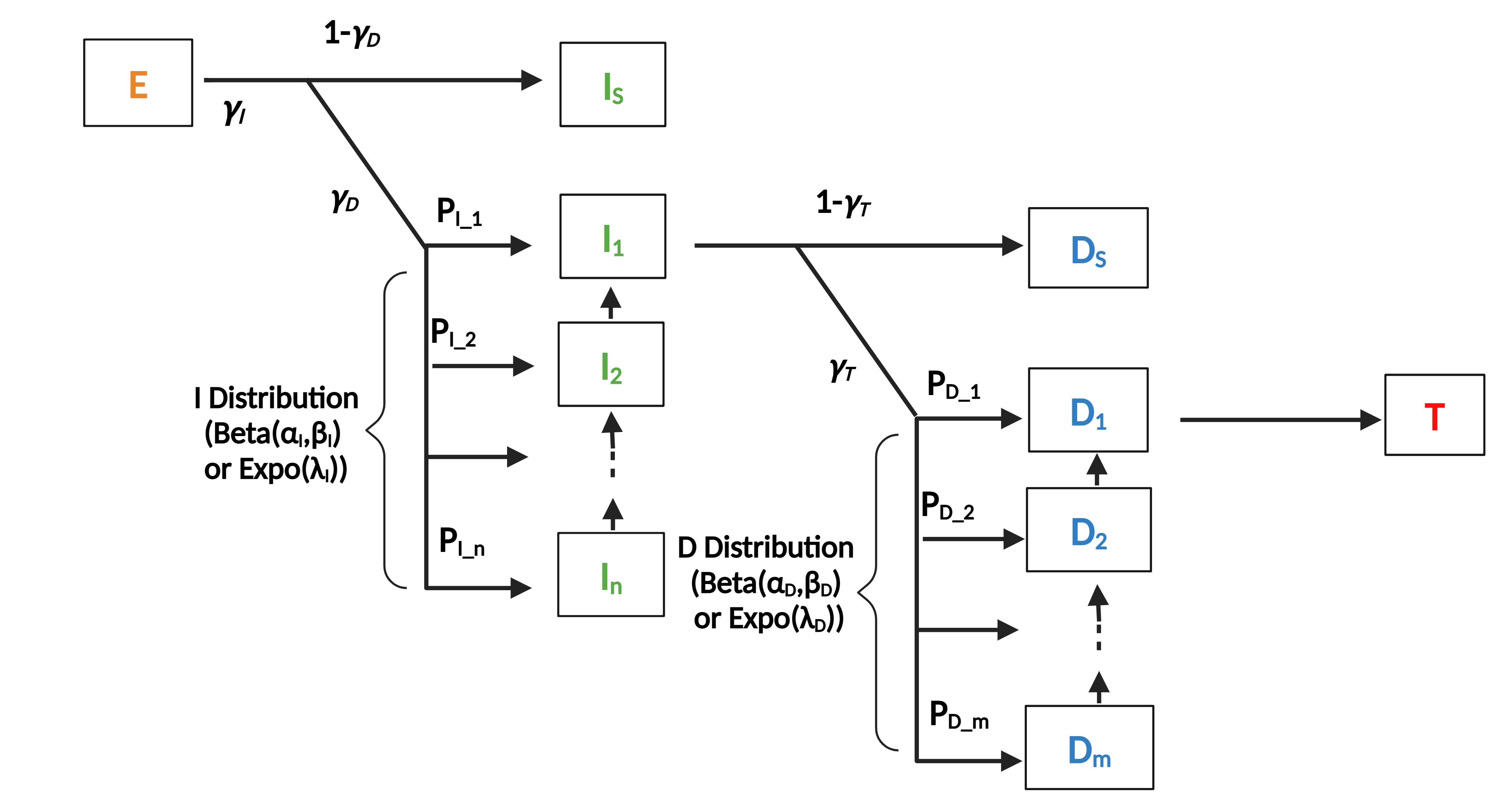}
\caption{Conceptual diagram of the intra-vector viral dynamic model. Each compartment represents a state of the vector (E: exposed vector, I: infected vector (with \(I_{\text{S}}\)): infected vector remaining in I, \(I_1\) to \(I_n\): infected vector remaining 1 to n days in I), D: disseminated vector (with D: infected vector remaining in D, \(D_1\) to \(D_m\): infected vector remaining 1 to m days in D); T: infectious vector = transmitter). The model parameters are: \(\gamma_I\), \(\gamma_D\), \(\gamma_T\) (proportion of mosquitoes for which respectively  the infection, dissemination and transmission barriers are crossed), \(\alpha\) and \(\beta\) (beta law parameters) or \(\lambda\) (exponential law parameters) and \(n\) and \(m\) (respectively maximum length of stay in I and D).}
\label{model_diag}
\end{figure}
\FloatBarrier

\begin{table}[H]
\centering
\caption{Model parameters}
\begin{tabular}{@{}p{0.07\textwidth}p{0.3\textwidth}p{0.13\textwidth}p{0.3\textwidth}@{}}
\toprule
Parameter & Description & Interval of variation & Justification \\ \midrule
\(\gamma_I\) & Proportion of mosquitoes for which the infection barrier will be passed  (=infection barrier) & {[}0,1{]} & Definition interval for proportional parameters \\
\(\gamma_D\) & Proportion of mosquitoes for which the dissemination barrier will be passed (= dissemination barrier) & {[}0,1{]} & Definition interval for proportional parameters \\
\(\gamma_T\) & Proportion of mosquitoes for which the transmission barrier will be passed (= transmission barrier) & {[}0,1{]} & Definition interval for proportional parameters \\
\(\alpha,\beta\) & Beta law parameters & \(\geq1\) \(\leq100\) & To avoid infinite distribution in 0 and to cover a large range of possible shapes \\
\(\lambda\) & Exponential law parameters & {[}0,100{]} & To cover a large range of possible shapes \\ \bottomrule
\end{tabular}%
\label{table_model_parameters}
\end{table}
\FloatBarrier

\clearpage
\subsection{Vectorial-competence experimental data} 
\begin{center}
    \vspace*{\fill} 
    {\Huge \bfseries Vectorial-competence experimental data} 
    \vspace*{\fill} 
\end{center}
\clearpage

\begin{figure}[H]
\centering
\includegraphics[width=\textwidth]
{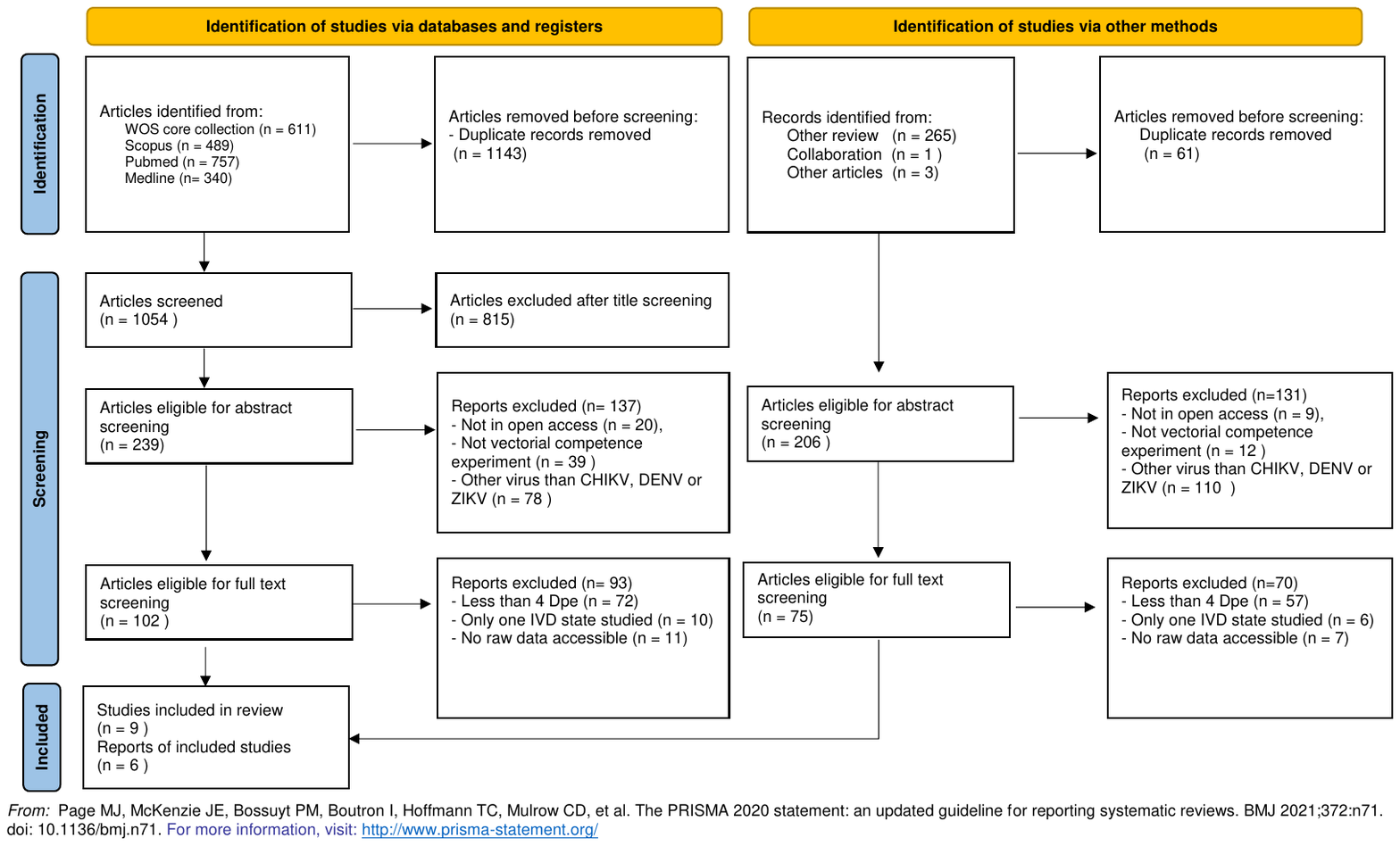}
\caption{PRISMA diagram of systematic review for literature data searching for inference of IVD model (diagram from \cite{page_prisma_2021}).}
\label{prisma_diag}
\end{figure}
\FloatBarrier

\clearpage
\begin{figure}[H]
\centering
\includegraphics[width=\textwidth]{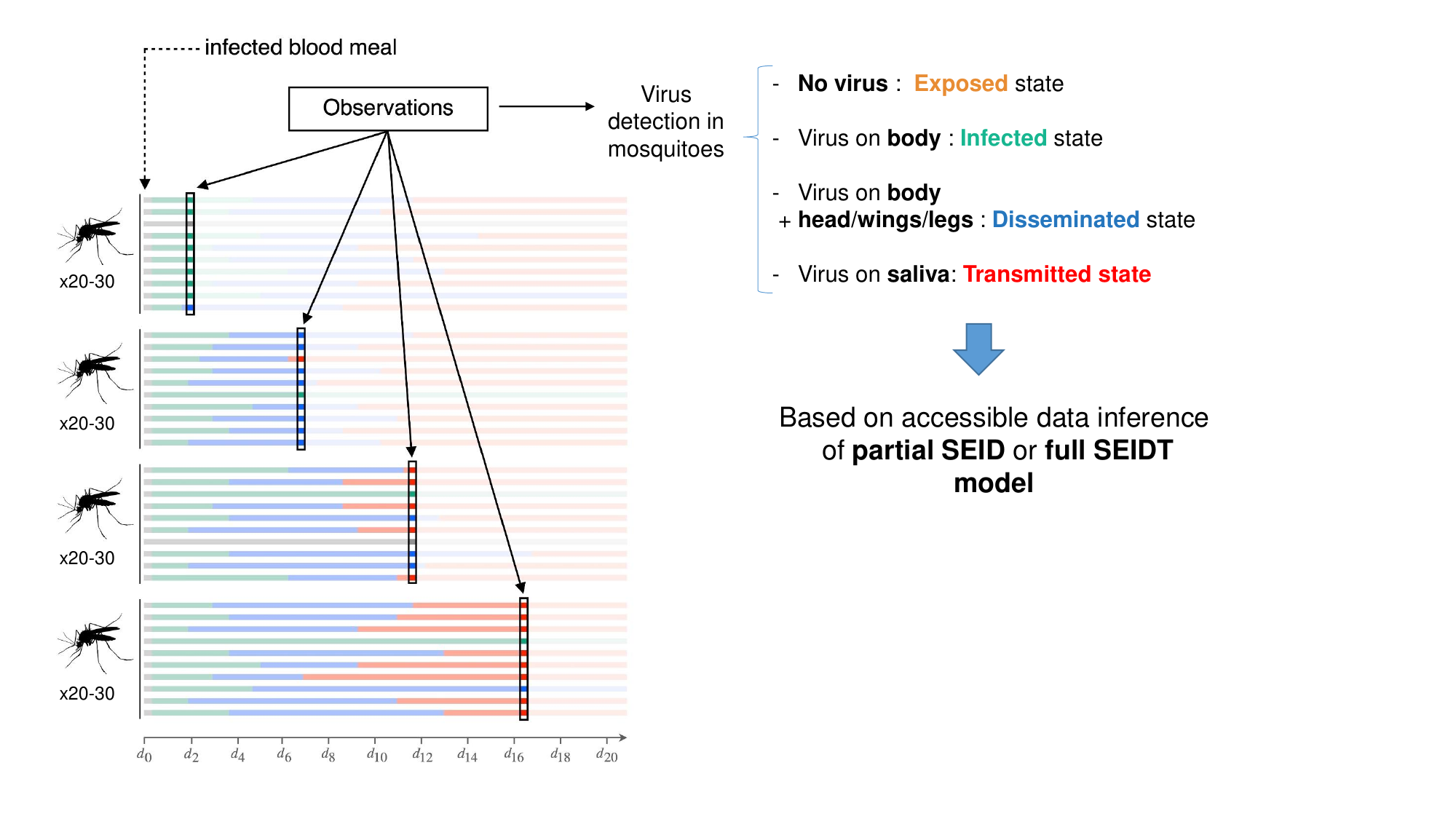}
\caption{Experimental protocol of vector competence experiment.}
\label{exp_protocol}
\end{figure}

\clearpage
\begin{sidewaystable}[!hp]
\centering
\caption{Summary of experimental data used to infer the partial SEID model. CHIKV:chikungunya virus, DENV-1:dengue virus serotype 1, ZIKV:Zika virus, ID:Infectious Dose, FFA : Fluorescent Focus Assay, RT-PCR:reverse transcription polymerase chain reaction, Dpe: Day post exposition, * Dpe no studied for ID=6.48Log10FFU/mL, ND:not defined }
\resizebox{\textwidth}{!}{%
\begin{tabular}{@{}lllllllllll@{}}
\toprule
\textbf{Virus species} & \textbf{Virus strain (origin)} & \textbf{Mosquito genus and species (origin)} & \textbf{ID} & \textbf{Ambiance conditions:} & \textbf{Dpe} & \textbf{Mosquito number (mean by Dpe)} & \textbf{Mosquitoes parts analyzed for} & \textbf{Scenario name} & \textbf{Short scenario name} & \textbf{Reference} \\ \midrule
 &  &  & (Log 10 FFU/mL) & - Temperature (°c) &  &  & - infection (I) &  &  &  \\
 &  &  &  & - Humidity (\%) &  &  & - dissemination(D) &  &  &  \\
 &  &  &  & - Light-dark cycle (hours) &  &  & (method) &  &  &  \\
\midrule
\multirow{3}{*}{\textbf{CHIKV}} & strain 06.21 (Indian Ocean) & \textit{Ae.albopictus} & 3.94 & - 26 °c & 2,6,9,14 & 43 & - I:body (FFA) & CHIKV\_IndOce\_albo\_Lyo\_3.94 & CHIKVp1 & \multirow{3}{*}{\cite{viginier_chikungunya_2023}} \\
 &  & (France, Lyon) & 6.07 & -70\% &  & 31 & - D : head (FFA) & CHIKV\_IndOce\_albo\_Lyo\_6.07 & CHIKVp2 &  \\
 &  &  & 8.63 & - ND &  & 19 &  & CHIKV\_IndOce\_albo\_Lyo\_8.63 & CHIKVp3 &  \\
\multirow{8}{*}{\textbf{DENV-1}} & Thailand 2010 a & \textit{Ae.aegypti} & 5.74 & - ND & 4,6,8,12, & 16 & - I:body (FFA) & DENV-1\_Tha2010a\_aeg\_Tha\_5.74 & DENVp1 & \multirow{8}{*}{\cite{fontaine_epidemiological_2018}} \\
 & Thailand 2010b & (Thailand, Kamphaeng Phet Province) & 5.70 & - ND & 18 & 21 & - D:head (FFA) & DENV-1\_Tha2010b\_aeg\_Tha\_5.70 & DENVp2 &  \\
 & Thailand 2013 &  & 5.79 & - ND &  & 17 &  & DENV-1\_Tha2013\_aeg\_Tha\_5.79 & DENVp3 &  \\
 & Laos 2012 &  & 5.84 &  &  & 25 &  & DENV-1\_Lao2012\_aeg\_Tha\_5.84 & DENVp4 &  \\
 & New Caledonia 2013 &  & 5.77 &  &  & 17 &  & DENV-1\_Nca2013\_aeg\_Tha\_5.77 & DENVp5 &  \\
 & Gabon 2012 &  & 5.82 &  &  & 15 &  & DENV-1\_Gab2012\_aeg\_Tha\_5.82 & DENVp6 &  \\
 & Haiti 2012 &  & 5.81 &  &  & 19 &  & DENV-1\_Hai2012\_aeg\_Tha\_5.81 & DENVp7 &  \\
 & Thailand 2012 &  & 5.80 &  &  & 19 &  & DENV-1\_Tha2012\_aeg\_Tha\_5.80 & DENVp8 &  \\
\multirow{6}{*}{\textbf{ZIKV}} & strain SL1602 & \textit{Ae.albopictus (France, Marseille)} & 6.48 & - 28°c & 5*,10,14,17,21 & 26 & - I:body(RT-PCR) & ZIKV\_Asi\_albo\_Mar\_6.48 & ZIKVp1 & \multirow{6}{*}{\cite{lequime_modeling_2020}} \\
 & (Asian lineage) &  & 6.87 & - 75 +/- 5\% &  & 19 & -D:head(RT-PCR) & ZIKV\_Asi\_albo\_Mar\_6.87 & ZIKVp2 &  \\
 &  & \textit{Ae.albopictus (France, Reunion Island)} & 5.90 & -16h:8h &  & 23 &  & ZIKV\_Asi\_albo\_LaR\_5.90 & ZIKVp3 &  \\
 &  &  & 6.87 &  &  & 23 &  & ZIKV\_Asi\_albo\_LaR\_6.87 & ZIKVp4 &  \\
 &  &  & 8.37 &  &  & 21 &  & ZIKV\_Asi\_albo\_LaR\_8.37 & ZIKVp5 &  \\
 &  &  & 6.48 &  &  & 13 &  & ZIKV\_Asi\_albo\_LaR\_6.48 & ZIKVp6 &  \\ 
\bottomrule \\
\end{tabular}%
}
\label{exp_data_SEID}
\end{sidewaystable}
\FloatBarrier

\clearpage
\begin{sidewaystable}[H]
\centering
\caption{Summary of experimental data used to infer the complete SEIDT model. CHIKV:chikungunya virus, DENV-1:dengue virus serotype 1, DENV-2:dengue virus serotype 2, DENV-3:dengue virus serotype 3, ZIKV:Zika virus, ID:Infectious Dose, FFA : Fluorescent Focus Assay, RT-PCR:reverse transcription polymerase chain reaction, Dpe: Day post exposition, ND:not defined }

\resizebox{\textwidth}{!}{%
\begin{tabular}{@{}lllllllllll@{}}
\toprule
Virus species & Virus strain (origin) & Mosquito genus and species (origin) & ID & Ambiance conditions: & Dpe & Mosquito number (mean by Dpe) & Mosquitoes parts analyzed for & ID\_scenario & ID\_scenario\_figures & Ref \\ 
 &  &  & (Log 10 FFU/mL) & Temperature (°c) &  &  & infection (I) &  &  &  \\
 &  &  &  & Humidity (\%) &  &  & dissemination(D) &  &  &  \\
 &  &  &  & Light-dark cycle (hours) &  &  & transmission (T) &  &  &  \\
 &  &  &  &  &  &  & (method) &  &  &  \\
 \midrule
CHIKV & 06.21 & \textit{Ae.albopictus} & 7 & 28 +/- 1°c & 3,7,14,21 & 30 & I:abdomen/thorax (FFA) & CHIKV\_LaR\_albo\_Rab\_7 & CHIKVc1 & \cite{amraoui_potential_2019} \\
 & (Reunion Island) & (Morocco, Rabat) &  & 80\% &  &  & D:head (FFA) &  &  &  \\
 &  &  &  & 16h:8h &  &  & T:saliva (FFA) &  &  &  \\
CHIKV & Carribean strain (French carribean Island) & \textit{Ae.aegypti} & 6 & 28°c & 3,6,9,12 & 48 & I:body(PFA) & CHIKV\_FrCarIs\_aeg\_Tha\_6 & CHIKVc2 & \cite{merwaiss_chikungunya_2021} \\
 &  & (Thailand, Kamphaeng Phet Province) &  & 70\% &  &  & D:head(PFA) &  &  &  \\
 &  &  &  & 12h:12h &  &  & saliva(PFA) &  &  &  \\
CHIKV & 06.21 & \textit{Ae.geniculatus} & 8 & 28 +/-1°c & 3,5,7,10,12,14,20 & 19 & I:body(FFA) & CHIKV\_Ind\_gen\_Tir\_8 & CHIKVc3 & \cite{prudhomme_native_2019} \\
 & (India) & (Albania, Tirana) &  & 80\% &  &  & D:head(FFA) &  &  &  \\
 &  &  &  & 16h:8h &  &  & T:saliva(FFA) &  &  &  \\
CHIKV & PF14/300914-109 & \textit{Ae.aegypti} & 7 & 27°c & 6,9,14,21 & 39 & I:thorax/abdomen (RT-PCR) & CHIKV\_Tah\_aeg\_Tah\_7 & CHIKVc4 & \cite{richard_vector_2016} \\
 & (Tahiti) & (Tahiti island, Toahotu) &  & 80\% &  &  & D:legs (RT-PCR) &  &  &  \\
 &  &  &  & 12h:12h &  &  & T:saliva (FFA) &  &  &  \\
CHIKV & 06.21 & \textit{Ae.albopictus} & 7 & 28 +/- 1°c & 3,7,10,14,21 & 27 & I:abdomen (FFA) & CHIKV\_LaR\_albo\_Tun\_7 & CHIKVc5 & \cite{bohers_recently_2020} \\
 & (Reunion Island) & (Tunisia,Carthage, Amilcar, La Marsa) &  & 80\% &  &  & D:thorax/head (FFA) &  &  &  \\
 &  &  &  & 16h:8h &  &  & T:saliva (FFA) &  &  &  \\
CHIKV & 06.21 & \textit{Ae.albopictus} & 8 & 28 +/-1°c & 3,5,7,10,12,14,20 & 18 & I:body(FFA) & CHIKV\_Ind\_albo\_Tir\_8 & CHIKVc6 & \cite{prudhomme_native_2019} \\
 & (India) & (Albania, Tirana) &  & 80\% &  &  & D:head(FFA) &  &  &  \\
 &  &  &  & 16h:8h &  &  & T:saliva(FFA) &  &  &  \\
CHIKV & NC/2011-568 (New Caledonia) & \textit{Ae.aegypti} & 7.3 & 28 +/- 1°c & 3,6,9,14 & 20 & I:thorax/abdomen (ND) & CHIKV\_Ncal\_aeg\_Mad\_Fun\_7.3 & CHIKVc7 & \cite{seixas_potential_2018} \\
 &  & (Madeira Island, urban Funchal) &  & 80\% &  &  & I:head(FFA) &  &  &  \\
 &  &  &  & ND &  &  & T:saliva(FFA) &  &  &  \\
CHIKV & NC/2011-568 (New Caledonia) & \textit{Ae.aegypti} & 7.3 & 28 +/- 1°c & 3,6,9,14 & 20 & I:thorax/abdomen (ND) & CHIKV\_Ncal\_aeg\_Mad\_Pa\_do\_Ma\_7.3 & CHIKVc8 & \cite{seixas_potential_2018} \\
 &  & (Madeira Island, rural Paul do Mar) &  & 80\% &  &  & I:head(FFA) &  &  &  \\
 &  &  &  & ND &  &  & T:saliva(FFA) &  &  &  \\
CHIKV & 99659 & \textit{Ae.aegypti} & 6.9 & 28°c & 2,4,6,8,10,12,14,16,18,20 & 60 & I:midgut(RT-PCR) & CHIKV\_BriVirIsl\_aeg\_PozRic\_6.9 & CHIKVc9 & \cite{robison_comparison_2020} \\
 & (British Virgin Islands) & (Mexico, Poza Rica) &  & 70,8 &  &  & D:legs/wings (RT-PCR) &  &  &  \\
 &  &  &  & 12h:12h &  &  & T:saliva (PFA) &  &  &  \\
 \midrule
DENV-2 & Prof Leon Rosen & \textit{Ae.albopictus} & 7 & 28 +/- 1°c & 3,7,14,21 & 28 & I:abdomen/thorax (FFA) & DENV-2\_Bang\_albo\_Rab\_7 & DENVc1 & \cite{amraoui_potential_2019}  \\
 & (Thailand, Bangkok) & (Morocco, Rabat) &  & 80\% &  &  & D:head (FFA) &  &  &  \\
 &  &  &  & 16h:8h &  &  & T:saliva (FFA) &  &  &  \\
DENV-2 & (Thailand, Bangkok) & \textit{Ae.albopictus} & 7 & 28 +/- 1°c & 3,7,10,14,21 & 28 & I:abdomen (FFA) & DENV-2\_Bang\_albo\_Tun\_7 & DENVc2 & \cite{bohers_recently_2020} \\
 &  & (Tunisia,Carthage, Amilcar, La Marsa) &  & 80\% &  &  & D:thorax/head (FFA) &  &  &  \\
 &  &  &  & 16h:8h &  &  & T:saliva (FFA) &  &  &  \\
DENV-1 & SG (EHI)D1/30889Y14 & \textit{Ae.albopictus} & 5 & ND & 7,14,21,28 & 17 & I:body(PFA) & DENV-1\_Sing\_albo\_Rey\_5 & DENVc3 & \cite{fortuna_assessing_2024} \\
 & (Singapore) & (Italia, Reynosa) &  & ND &  &  & D:legs/wings(PFA) &  &  &  \\
 &  &  &  & ND &  &  & T:saliva (PFA) &  &  &  \\
DENV-3 & 1998- GenBank JN406514 & \textit{Ae aegypti} & 10\textasciicircum{}4.9 CCID50 = & 28°c & 2,3,4,5,6,7,10,14 & 21 & I:body(ELISA) & DENV-3\_Cair\_1998\_aeg\_Cair\_4.7 & DENVc4 & \cite{ritchie_explosive_2013} \\
 &  & ( Australia,Cairns) & 4.7 & 75\% &  &  & D:legs/wings (ELISA) &  &  &  \\
 &  &  &  & 12h:12h &  &  & T:saliva(ELISA) &  &  &  \\
DENV-3 & 2008a - GenBank JN406515 & \textit{Ae aegypti} & 10\textasciicircum{}5.1 CCID50 = 4.9 & 28°c & 2,3,4,5,6,7,10,14 & 21 & I:body(ELISA) & DENV-3\_Cair\_2008\_aeg\_Cair\_4.9 & DENVc5 & \cite{ritchie_explosive_2013}\\
 & (Cairns) & ( Australia,Cairns) &  & 75\% &  &  & D:legs/wings (ELISA) &  &  &  \\
 &  &  &  & 12h:12h &  &  & T:saliva(ELISA) &  &  &  \\
DENV-1 & SG (EHI)D1/30889Y14 & \textit{Ae.albopictus} & 5 & ND & 7,14,21,28 & 18 & I:body(PFA) & DENV-1\_Sing\_albo\_Rom\_5 & DENVc6 & \cite{fortuna_assessing_2024}\\
 & (Singapore) & (Italia, Rome) &  & ND &  &  & D:legs/wings(PFA) &  &  &  \\
 &  &  &  & ND &  &  & T:saliva (PFA) &  &  &  \\
DENV-1 & SG (EHI)D1/30889Y14 & \textit{Ae.albopictus} & 5 & ND & 7,14,21,28 & 18 & I:body(PFA) & DENV-1\_Sing\_albo\_Mon\_5 & DENVc7 & \cite{fortuna_assessing_2024} \\
 & (Singapore) & (Italia, Montecchio) &  & ND &  &  & D:legs/wings(PFA) &  &  &  \\
 &  &  &  & ND &  &  & T:saliva (PFA) &  &  &  \\
 \midrule
ZIKV & PRVABC59 & \textit{Ae.aegypti} & 7.2 & 28°c & 2,4,6,8,10,12,14,16,18,20 & 30 & I:midgut(RT-PCR) & ZIKV\_PueRic\_aeg\_PozRic\_7.2 & ZIKVc1 & \cite{robison_comparison_2020}\\
 & (United States Puerto Rico) & (Mexico, Poza Rica) &  & 70,8 &  &  & D:legs/wings (RT-PCR) &  &  &  \\
 &  &  &  & 12h:12h &  &  & T:saliva (PFA) &  &  &  \\
ZIKV & ZIKV strain PF13/251013-18 & \textit{Ae.aegypti} & 7 (TCID50/mL)=6.8 & 27°c & 6,9,14,21 & 40 & -I : abdomen/thorax (RT-PCR) & ZIKV\_FrP\_aeg\_Tah\_6.8 & ZIKVc2 &  \\
 & (French Polynesia) & (Tahiti island, Toahotu) &  & 80\% &  &  & D: legs(RT-PCR) &  &  & \cite{richard_vector_2016-1} \\
 &  &  &  & 12h:12h &  &  & T : saliva(FFA) &  &  &  \\
ZIKV & MR 766 & \textit{Ae.aegypti} & 6.7 +/- 0.2 TCID50=6.5 & 28°c & 5,7,10,14 & 24 & I:body (RT-PCR) & ZIKV\_Ugan\_aeg\_Tow\_6.5 & ZIKVc3 & \cite{hall-mendelin_assessment_2016}\\
 & (Uganda) & (Australia,Townsville) &  & high relative Humidity &  &  & D:legs/wings (RT-PCR) &  &  &  \\
 &  &  &  & 12h:12h &  &  & T:saliva (RT-PCR) &  &  &  \\
ZIKV & NC-2014-5132 & \textit{Ae.aegypti} & 7 & 28°c & 6,9,14,21 & 32 & I:abdomen/thorax (PFA) & ZIKV\_Ncal\_aeg\_FrP\_7 & ZIKVc4 & \cite{calvez_zika_2018}  \\
 & (New Caledonia) & (French Polynesia) &  & 80\% &  &  & D:head PFA) &  &  &  \\
 &  &  &  & 12h:12h &  &  & T:saliva (PFA) &  &  &  \\
ZIKV & NC-2014-5132 & \textit{Ae.polynesiensis} & 7 & 28°c & 6,9,14,21 & 38 & I:abdomen/thorax (PFA) & ZIKV\_Ncal\_pol\_Wal\_7 & ZIKVc5 & \cite{calvez_zika_2018} \\
 & (New Caledonia) & (Wallis) &  & 80\% &  &  & D:head PFA) &  &  &  \\
 &  &  &  & 12h:12h &  &  & T:saliva (PFA) &  &  &  \\
ZIKV & NC-2014-5132 & \textit{Ae.albopictus} & 7.2 & 28 +/- 1°c & 3,7,14,21 & 28 & I: abdomen/thorax (PFA) & ZIKV\_Ncal\_albo\_Rab\_7.2 & ZIKVc6 & \cite{amraoui_potential_2019} \\
 & (New Caledonia) & (Morocco, rabat) &  & 80\% &  &  & D:head (PFA) &  &  &  \\
 &  &  &  & 16h:8h &  &  & T:saliva (PFA) &  &  &  \\
ZIKV & NC-2014-5132 & \textit{Ae.aegypti} (Samoa) & 7 & 28°c & 6,9,14,21 & 39 & I:abdomen/thorax (PFA) & ZIKV\_Ncal\_aeg\_Sam\_7 & ZIKVc7 & \cite{calvez_zika_2018} \\
 & (New Caledonia) &  &  & 80\% &  &  & D:head PFA) &  &  &  \\
 &  &  &  & 12h:12h &  &  & T:saliva (PFA) &  &  &  \\
ZIKV & NC-2014-5132 & \textit{Ae.aegypti} & 7 & 28°c & 6,9,14,21 & 26 & I:abdomen/thorax (PFA) & ZIKV\_Ncal\_aeg\_Ncal\_7 & ZIKVc8 & \cite{calvez_zika_2018}\\
 & (New Caledonia) & (New Caledonia) &  & 80\% &  &  & D:head PFA) &  &  &  \\
 &  &  &  & 12h:12h &  &  & T:saliva (PFA) &  &  &  \\
ZIKV & NC-2014-5132 & \textit{Ae.albopictus} & 7 & 28 +/- 1°c & 7,10,14,21 & 23 & I:abdomen (PFA) & ZIKV\_Ncal\_albo\_Tun\_7 & ZIKVc9 & \cite{bohers_recently_2020} \\
 & (New Caledonia) & (Tunisia,Carthage, Amilcar, La Marsa) &  & 80\% &  &  & D:thorax/head (PFA) &  &  &  \\
 &  &  &  & 16h:8h &  &  & T:saliva (PFA) &  &  &  \\
ZIKV & NC-2014-5132 & \textit{Ae.polynesiensis} & 7 & 28°c & 6,9,14,21 & 30 & I:abdomen/thorax (PFA) & ZIKV\_Ncal\_pol\_FrP\_7 & ZIKVc10 & \cite{calvez_zika_2018} \\
 & (New Caledonia) & (French Polynesia) &  & 80\% &  &  & D:head PFA) &  &  &  \\
 &  &  &  & 12h:12h &  &  & T:saliva (PFA) &  &  &  \\ \bottomrule
\end{tabular}%
}
\label{exp_data_SEIDT}
\end{sidewaystable}
\FloatBarrier

\clearpage
\subsection{Figures of additional results} 
\begin{center}
    \vspace*{\fill} 
    {\Huge \bfseries Figures of additional results} 
    \vspace*{\fill} 
\end{center}
\clearpage

\begin{figure}[H]
\centering
\includegraphics[scale=0.7]{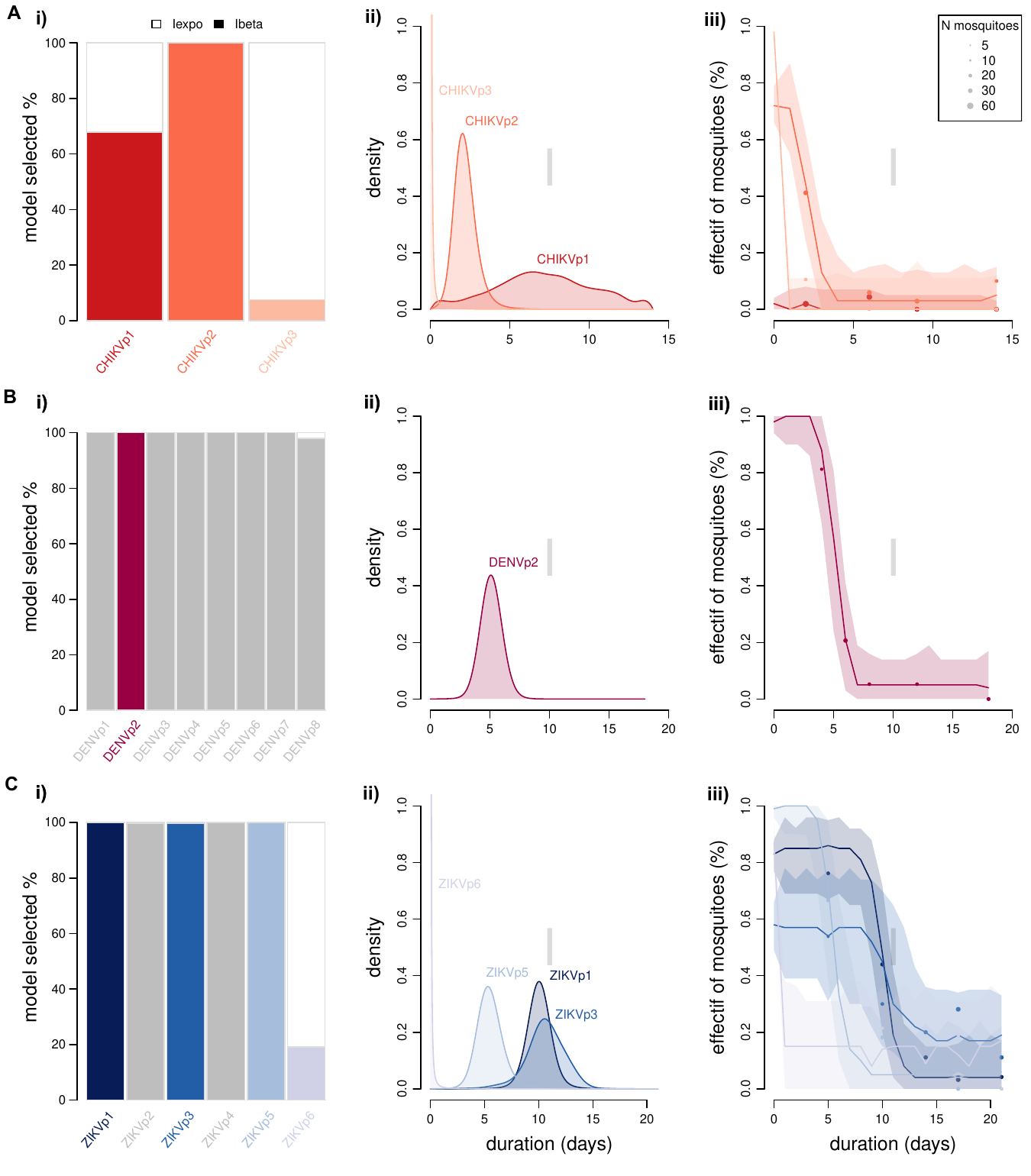}
\caption{Inference results for IVD stages distributions for all scenarios tested for CHIKV (A), DENV (B) and ZIKV (C) with SEID model: i) Selected proportion of each model (modBeta, modExpo,) for each scenario with one colour per scenario. ii) Average of the selected distributions in the infected state for the main model selected (only selected scenarios* are represented). iii) Selected dynamics in the infected state for the main model selected (only selected scenarios* are represented). Dots represent observed data, line mean dynamics and  uncertainty ribbons (5\%-95\%) represent selected simulated dynamics for each scenario.* \textit{scenarios with 5 or more observed Dpe}}
\label{result_inf_SEID}
\end{figure}

\begin{figure}[H]
\centering
\includegraphics[width=\textwidth]{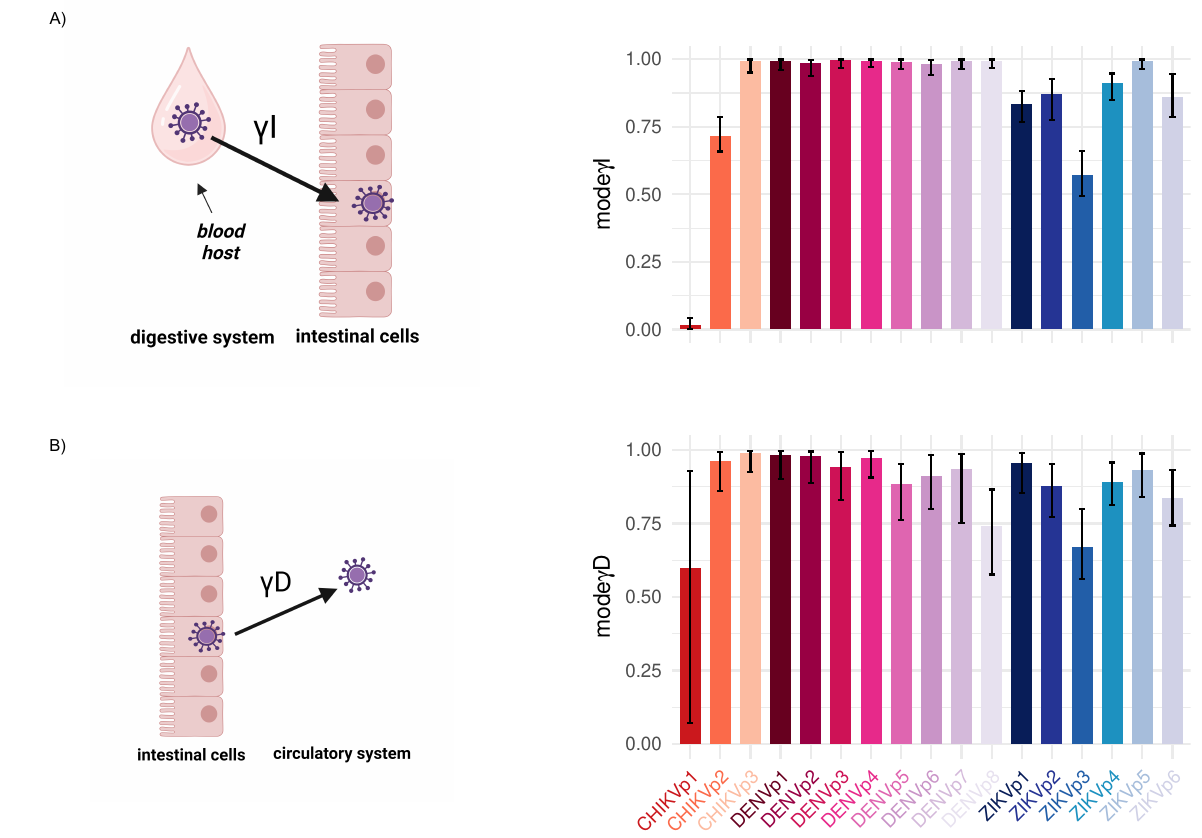}
\caption{Values of the crossing barrier parameters inferred for all scenarios tested for CHIKV, DENV and ZIKV. A. infection barrier. B. Dissemination barrier. Each bar represents the parameter mode for a given scenario, with one colour per scenario and with the 90\% credibility interval represented by the error bar.}
\label{result_params_SEID}
\end{figure}

\begin{figure}[H]
\centering
\includegraphics[width=\textwidth]{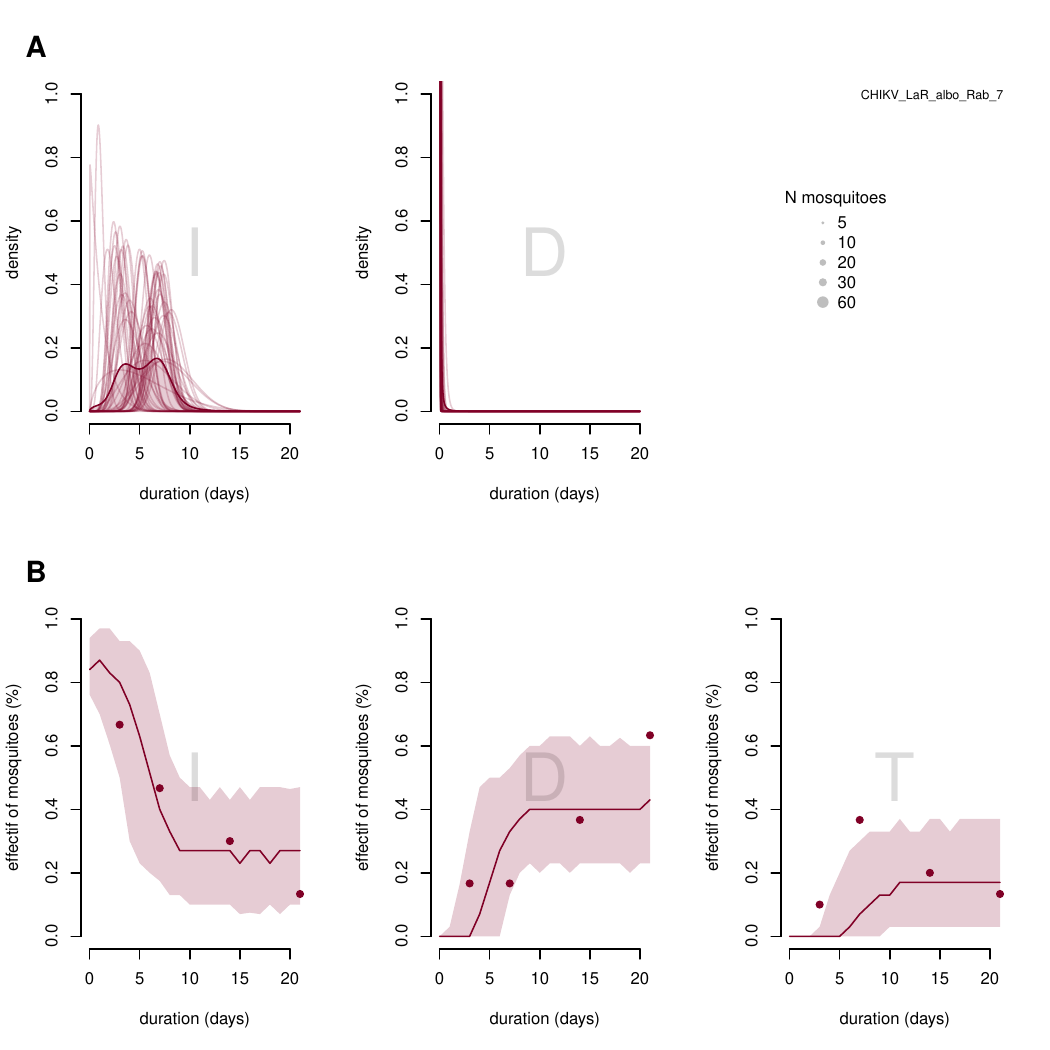}
\caption{Inference results for IVD stages distributions for scenario CHIKVc1(CHIKV\_LaR\_albo\_Rab\_7): \textit{Aedes.albopictus} from Rabat infected by chikungunya virus from Reunion Island with an infectious dose of 7 log10 FFU/mL : A) Selected distributions in the infected and disseminated states for the main model selected. The dark line represents the mean of distributions and light lines represent a random sample of 50 distribution among all selected distribution. B) Selected dynamics in the infected (I), disseminated (D), and transmitter (T) states for the main model selected. The dots represent the observed data, the line (mean dynamics), and the uncertainty ribbons (5\%-95\%) represent selected simulated dynamic.}
\label{fig_CHIKVc1}
\end{figure}

\begin{figure}[H]
\centering
\includegraphics[width=\textwidth]{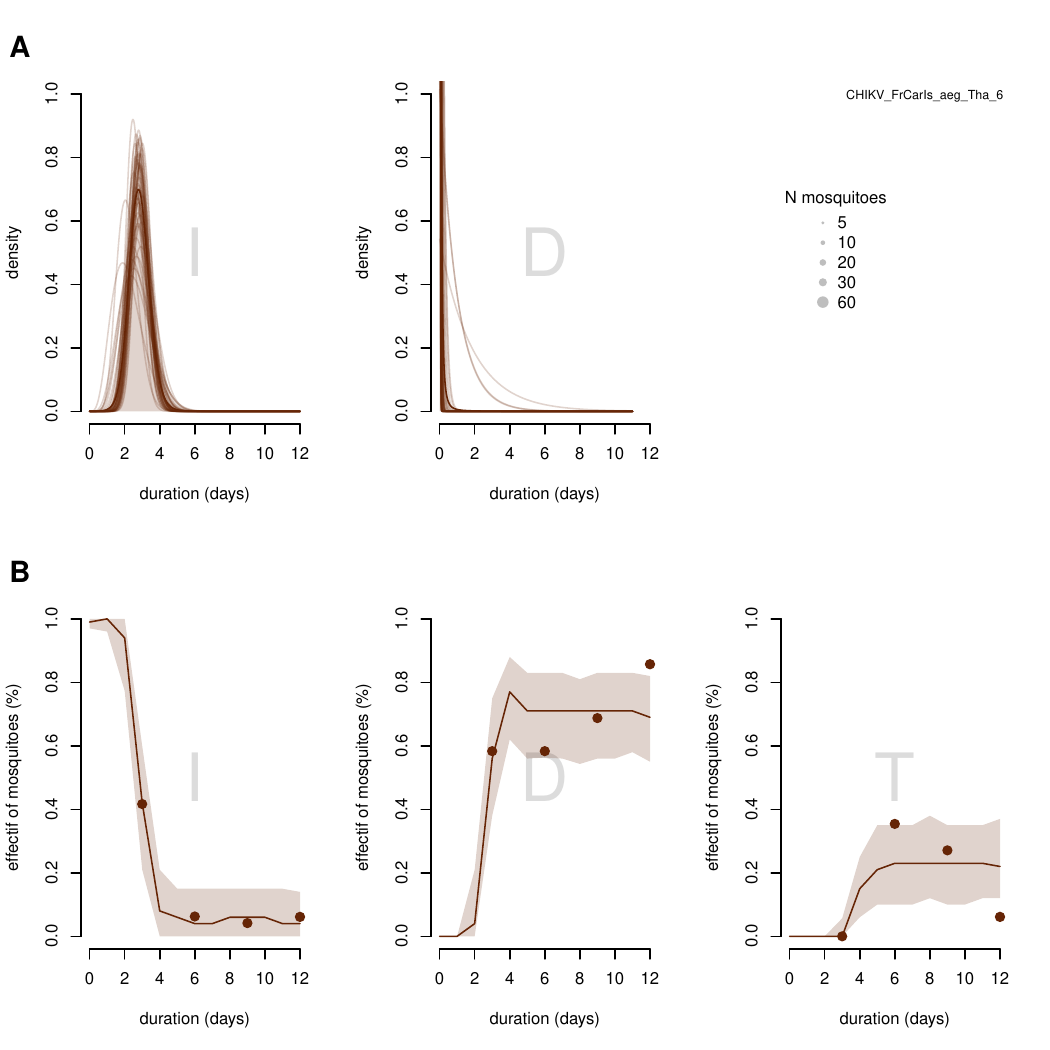}
\caption{Inference results for IVD stages distributions for scenario CHIKVc2(CHIKV\_FrCarIs\_aeg\_Tha\_6): \textit{Aedes.aegypti} from Thailand infected by chikungunya virus from French Caribbean Island with an infectious dose of 6 log10 FFU/mL : A) Selected distributions in the infected and disseminated states for the main model selected. The dark line represents the mean of distributions and light lines represent a random sample of 50 distribution among all selected distribution. B) Selected dynamics in the infected (I), disseminated (D), and transmitter (T) states for the main model selected. The dots represent the observed data, the line (mean dynamics), and the uncertainty ribbons (5\%-95\%) represent selected simulated dynamic.}
\label{fig_CHIKVc2}
\end{figure}

\begin{figure}[H]
\centering
\includegraphics[width=\textwidth]{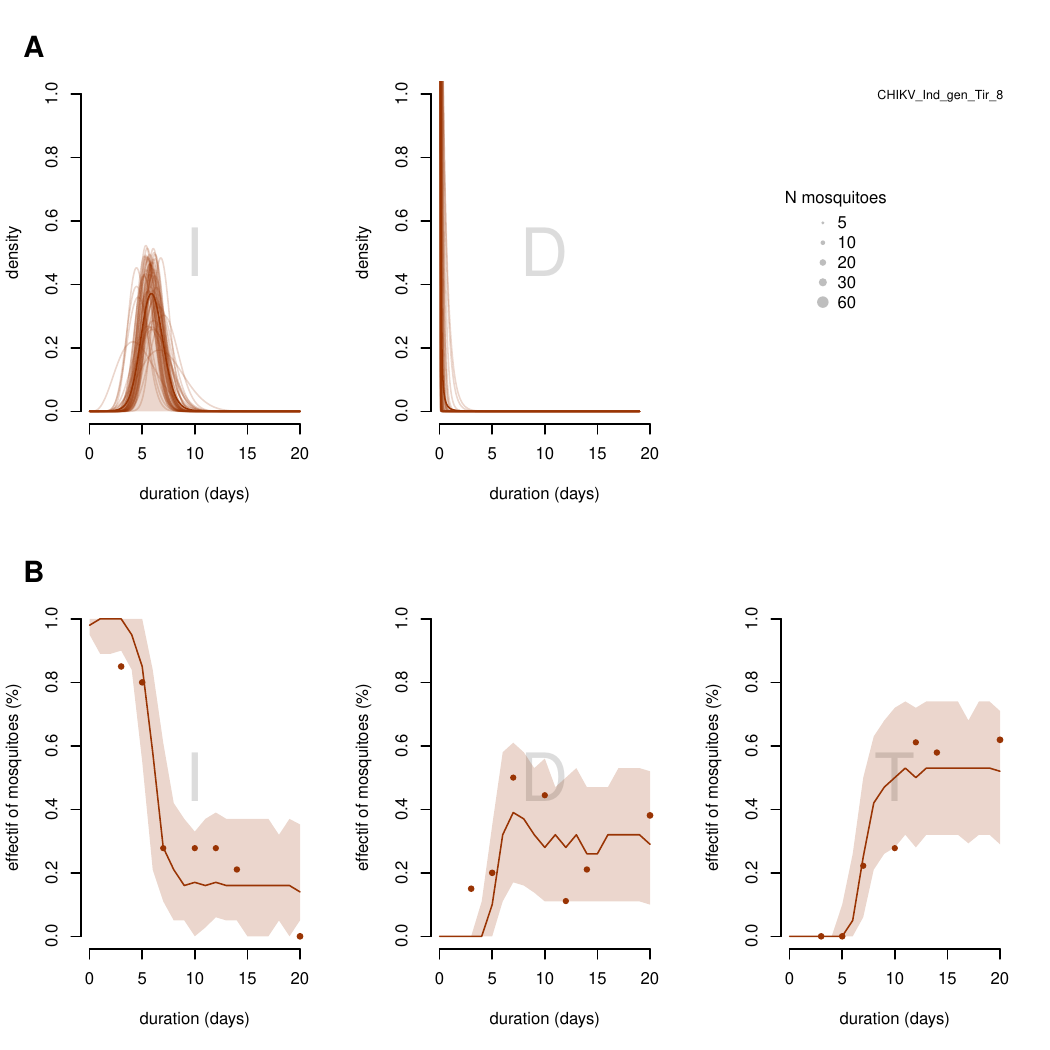}
\caption{Inference results for IVD stages distributions for scenario CHIKVc3(CHIKV\_Ind\_gen\_Tir\_8): \textit{Aedes.geniculatus} from Tirana infected by chikungunya virus from India with an infectious dose of 8 log10 FFU/mL : A) Selected distributions in the infected and disseminated states for the main model selected. The dark line represents the mean of distributions and light lines represent a random sample of 50 distribution among all selected distribution. B) Selected dynamics in the infected (I), disseminated (D), and transmitter (T) states for the main model selected. The dots represent the observed data, the line (mean dynamics), and the uncertainty ribbons (5\%-95\%) represent selected simulated dynamic.}
\label{fig_CHIKVc3}
\end{figure}

\begin{figure}[H]
\centering
\includegraphics[width=\textwidth]{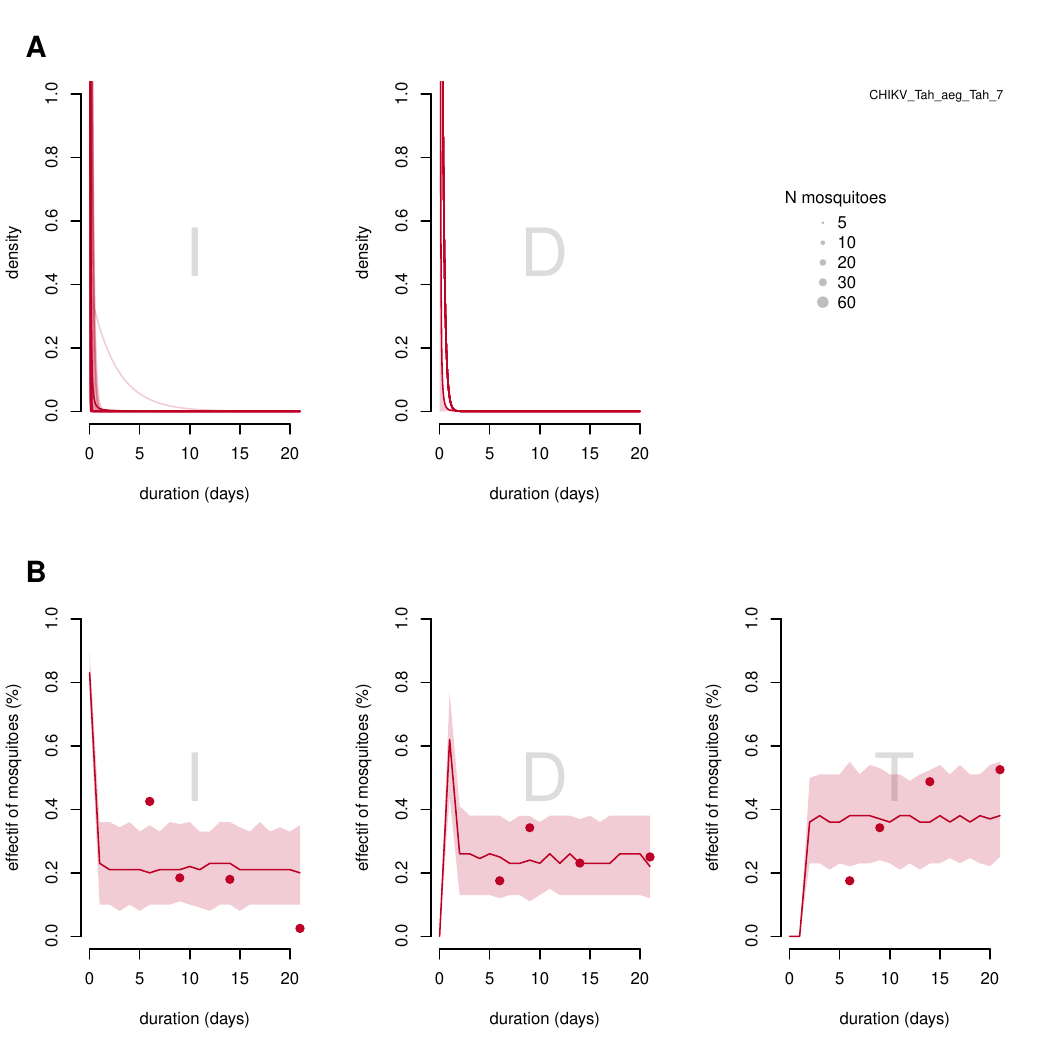}
\caption{Inference results for IVD stages distributions for scenario CHIKVc4(CHIKV\_Tah\_aeg\_Tah\_7): \textit{Aedes.aegypti} from Tahiti infected by chikungunya virus from Tahiti with an infectious dose of 7 log10 FFU/mL : A) Selected distributions in the infected and disseminated states for the main model selected. The dark line represents the mean of distributions and light lines represent a random sample of 50 distribution among all selected distribution. B) Selected dynamics in the infected (I), disseminated (D), and transmitter (T) states for the main model selected. The dots represent the observed data, the line (mean dynamics), and the uncertainty ribbons (5\%-95\%) represent selected simulated dynamic.}
\label{fig_CHIKVc4}
\end{figure}

\begin{figure}[H]
\centering
\includegraphics[width=\textwidth]{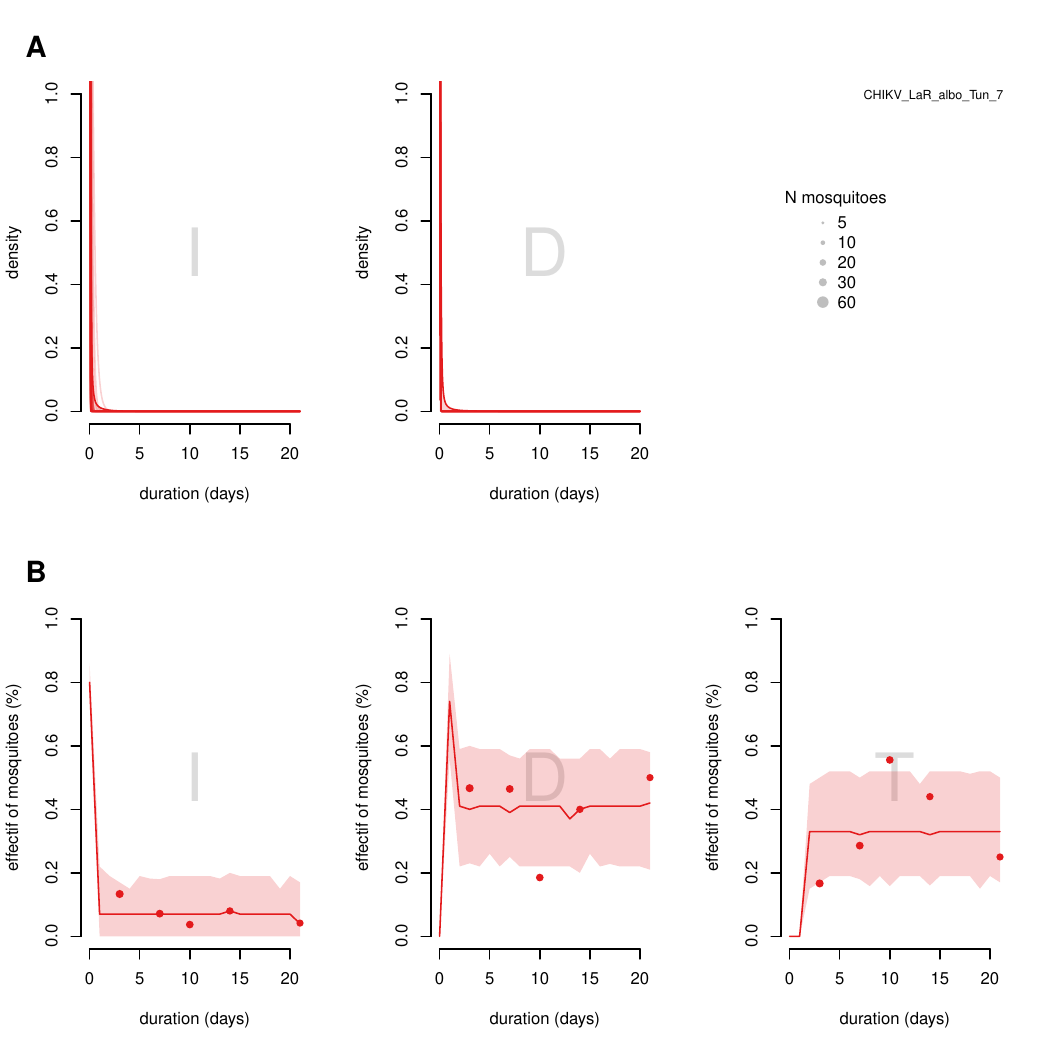}
\caption{Inference results for IVD stages distributions for scenario CHIKVc5(CHIKV\_LaR\_albo\_Tun\_7): \textit{Aedes.albopictus} from Tunisia infected by chikungunya virus from Reunion Island with an infectious dose of 7 log10 FFU/mL : A) Selected distributions in the infected and disseminated states for the main model selected. The dark line represents the mean of distributions and light lines represent a random sample of 50 distribution among all selected distribution. B) Selected dynamics in the infected (I), disseminated (D), and transmitter (T) states for the main model selected. The dots represent the observed data, the line (mean dynamics), and the uncertainty ribbons (5\%-95\%) represent selected simulated dynamic.}
\label{fig_CHIKVc5}
\end{figure}

\begin{figure}[H]
\centering
\includegraphics[width=\textwidth]{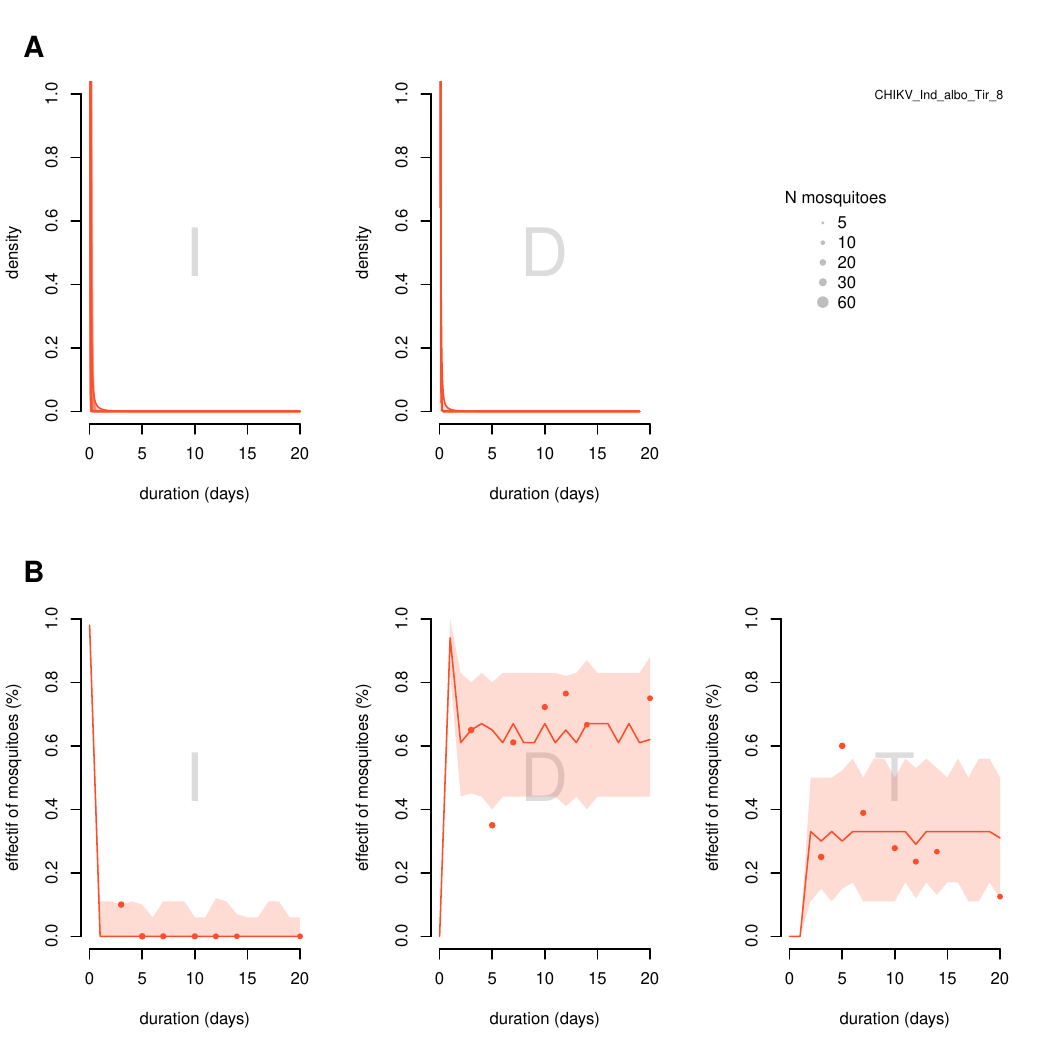}
\caption{Inference results for IVD stages distributions for scenario CHIKVc6(CHIKV\_Ind\_albo\_Tir\_8): \textit{Aedes.albopictus} from Tirana infected by chikungunya virus from India with an infectious dose of 8 log10 FFU/mL : A) Selected distributions in the infected and disseminated states for the main model selected. The dark line represents the mean of distributions and light lines represent a random sample of 50 distribution among all selected distribution. B) Selected dynamics in the infected (I), disseminated (D), and transmitter (T) states for the main model selected. The dots represent the observed data, the line (mean dynamics), and the uncertainty ribbons (5\%-95\%) represent selected simulated dynamic.}
\label{fig_CHIKVc6}
\end{figure}

\begin{figure}[H]
\centering
\includegraphics[width=\textwidth]{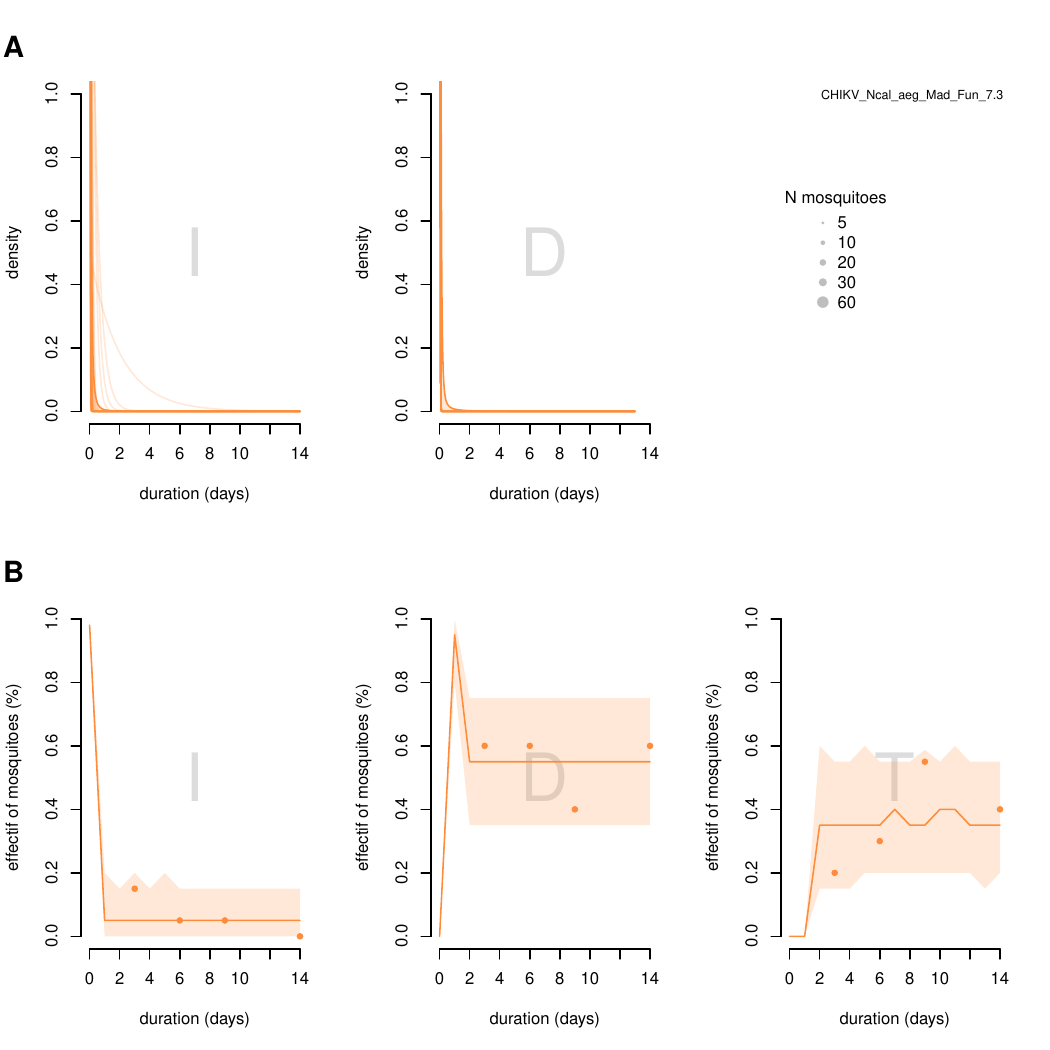}
\caption{Inference results for IVD stages distributions for scenario CHIKVc7(CHIKV\_Ncal\_aeg\_Mad\_Fun\_7.3): \textit{Aedes.aegypti} from Madeira Island, Funchal infected by chikungunya virus from New Caledonia with an infectious dose of 7.3 log10 FFU/mL : A) Selected distributions in the infected and disseminated states for the main model selected. The dark line represents the mean of distributions and light lines represent a random sample of 50 distribution among all selected distribution. B) Selected dynamics in the infected (I), disseminated (D), and transmitter (T) states for the main model selected. The dots represent the observed data, the line (mean dynamics), and the uncertainty ribbons (5\%-95\%) represent selected simulated dynamic.}
\label{fig_CHIKVc7}
\end{figure}

\begin{figure}[H]
\centering
\includegraphics[width=\textwidth]{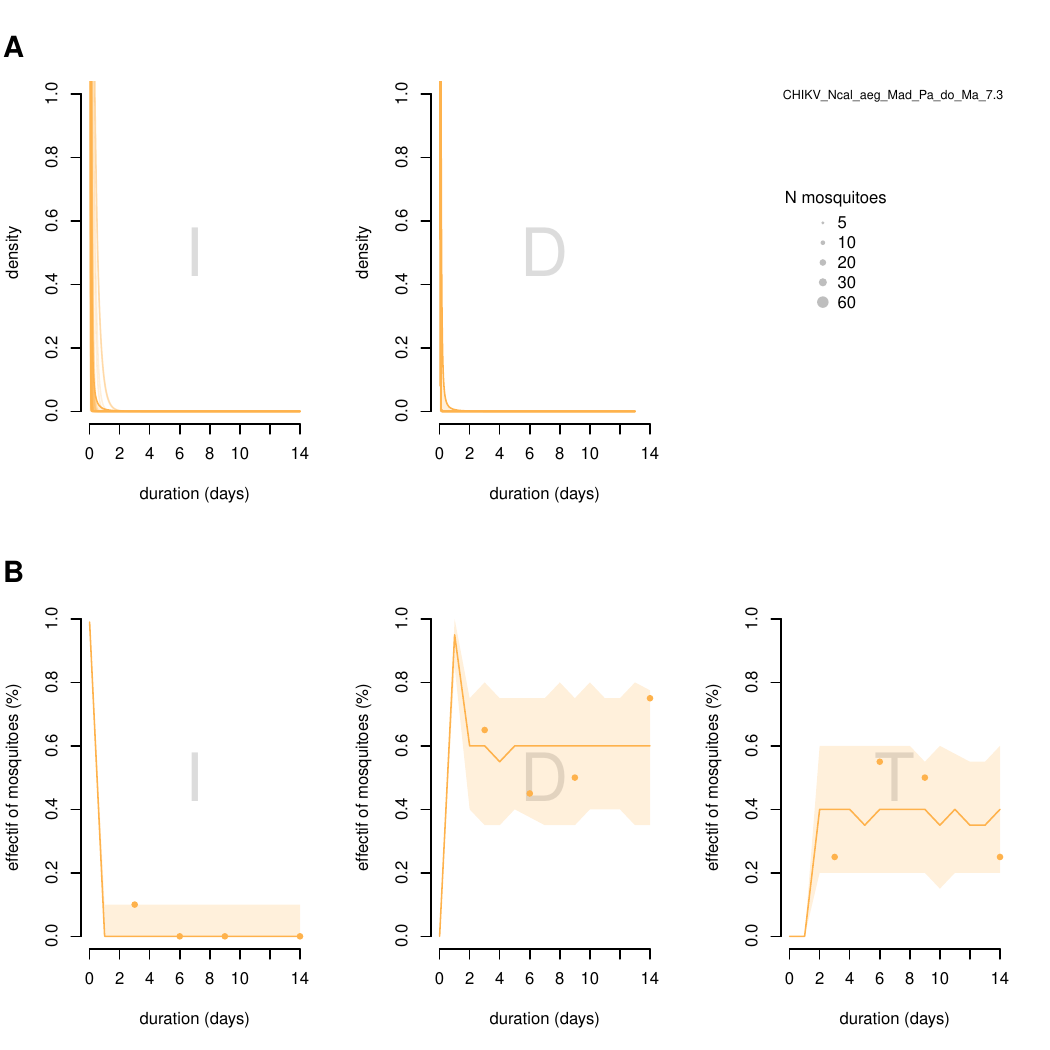}
\caption{Inference results for IVD stages distributions for scenario CHIKVc8(CHIKV\_Ncal\_aeg\_Mad\_Pa\_do\_Ma\_7.3): \textit{Aedes.aegypti} from Madeira Island, Paul do Mar infected by chikungunya virus from New Caledonia with an infectious dose of 7.3 log10 FFU/mL : A) Selected distributions in the infected and disseminated states for the main model selected. The dark line represents the mean of distributions and light lines represent a random sample of 50 distribution among all selected distribution. B) Selected dynamics in the infected (I), disseminated (D), and transmitter (T) states for the main model selected. The dots represent the observed data, the line (mean dynamics), and the uncertainty ribbons (5\%-95\%) represent selected simulated dynamic.}
\label{fig_CHIKVc8}
\end{figure}

\begin{figure}[H]
\centering
\includegraphics[width=\textwidth]{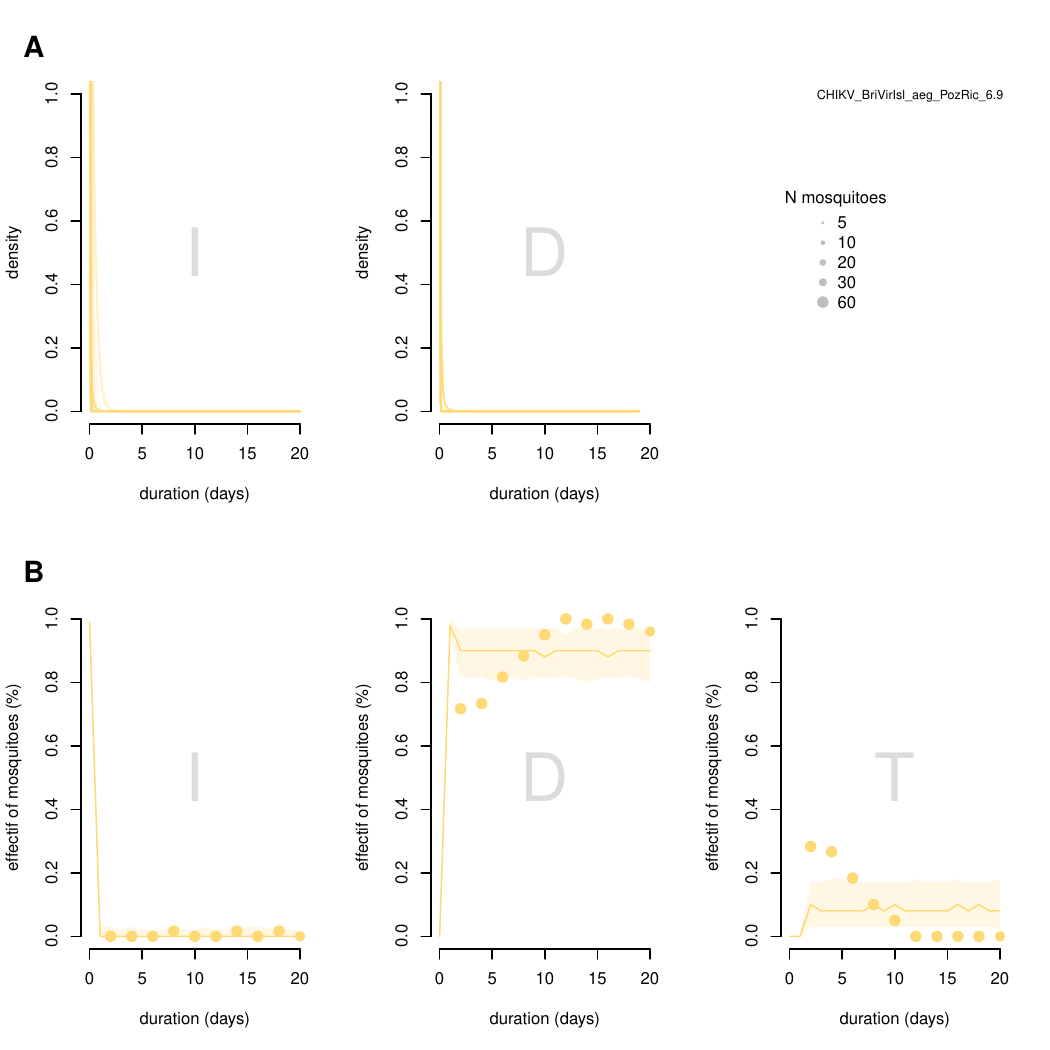}
\caption{Inference results for IVD stages distributions for scenario CHIKVc9(CHIKV\_BriVirIsl\_aeg\_PozRic\_6.9): \textit{Aedes.aegypti} from Mexico, Poza Rica, infected by chikungunya virus from British Virgin Island with an infectious dose of 6.9 log10 FFU/mL : A) Selected distributions in the infected and disseminated states for the main model selected. The dark line represents the mean of distributions and light lines represent a random sample of 50 distribution among all selected distribution. B) Selected dynamics in the infected (I), disseminated (D), and transmitter (T) states for the main model selected. The dots represent the observed data, the line (mean dynamics), and the uncertainty ribbons (5\%-95\%) represent selected simulated dynamic.}
\label{fig_CHIKVc9}
\end{figure}

\begin{figure}[H]
\centering
\includegraphics[width=\textwidth]{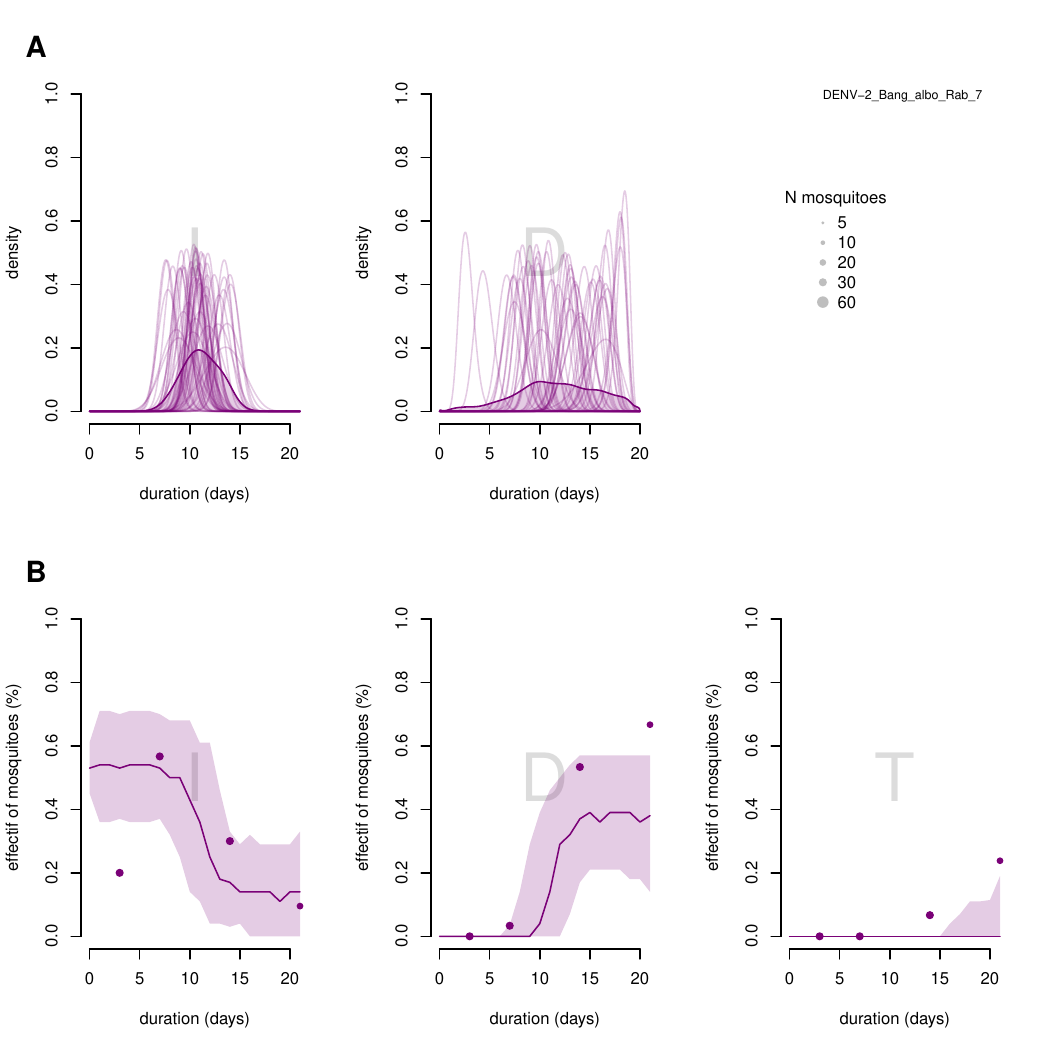}
\caption{Inference results for IVD stages distributions for scenario DENVc1(DENV\-2\_Bang\_albo\_Rab\_7): \textit{Aedes.albopictus} from Rabat infected by dengue virus from Bangkok with an infectious dose of 7 log10 FFU/mL : A) Selected distributions in the infected and disseminated states for the main model selected. The dark line represents the mean of distributions and light lines represent a random sample of 50 distribution among all selected distribution. B) Selected dynamics in the infected (I), disseminated (D), and transmitter (T) states for the main model selected. The dots represent the observed data, the line (mean dynamics), and the uncertainty ribbons (5\%-95\%) represent selected simulated dynamic.}
\label{fig_DENVc1}
\end{figure}

\begin{figure}[H]
\centering
\includegraphics[width=\textwidth]{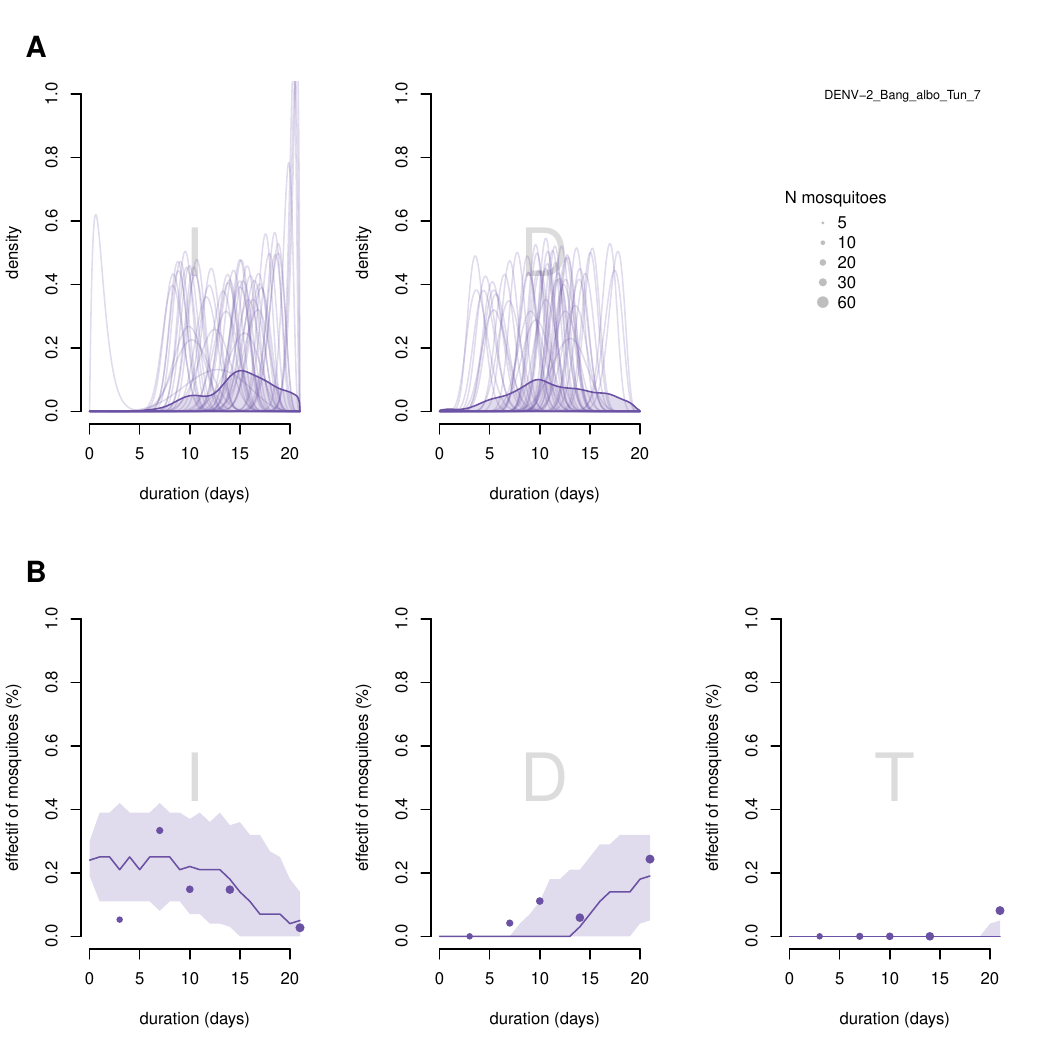}
\caption{Inference results for IVD stages distributions for scenario DENVc2(DENV\-2\_Bang\_albo\_Tun\_7): \textit{Aedes.albopictus} from Tunisia infected by dengue virus from Bangkok with an infectious dose of 7 log10 FFU/mL : A) Selected distributions in the infected and disseminated states for the main model selected. The dark line represents the mean of distributions and light lines represent a random sample of 50 distribution among all selected distribution. B) Selected dynamics in the infected (I), disseminated (D), and transmitter (T) states for the main model selected. The dots represent the observed data, the line (mean dynamics), and the uncertainty ribbons (5\%-95\%) represent selected simulated dynamic.}
\label{fig_DENVc2}
\end{figure}

\begin{figure}[H]
\centering
\includegraphics[width=\textwidth]{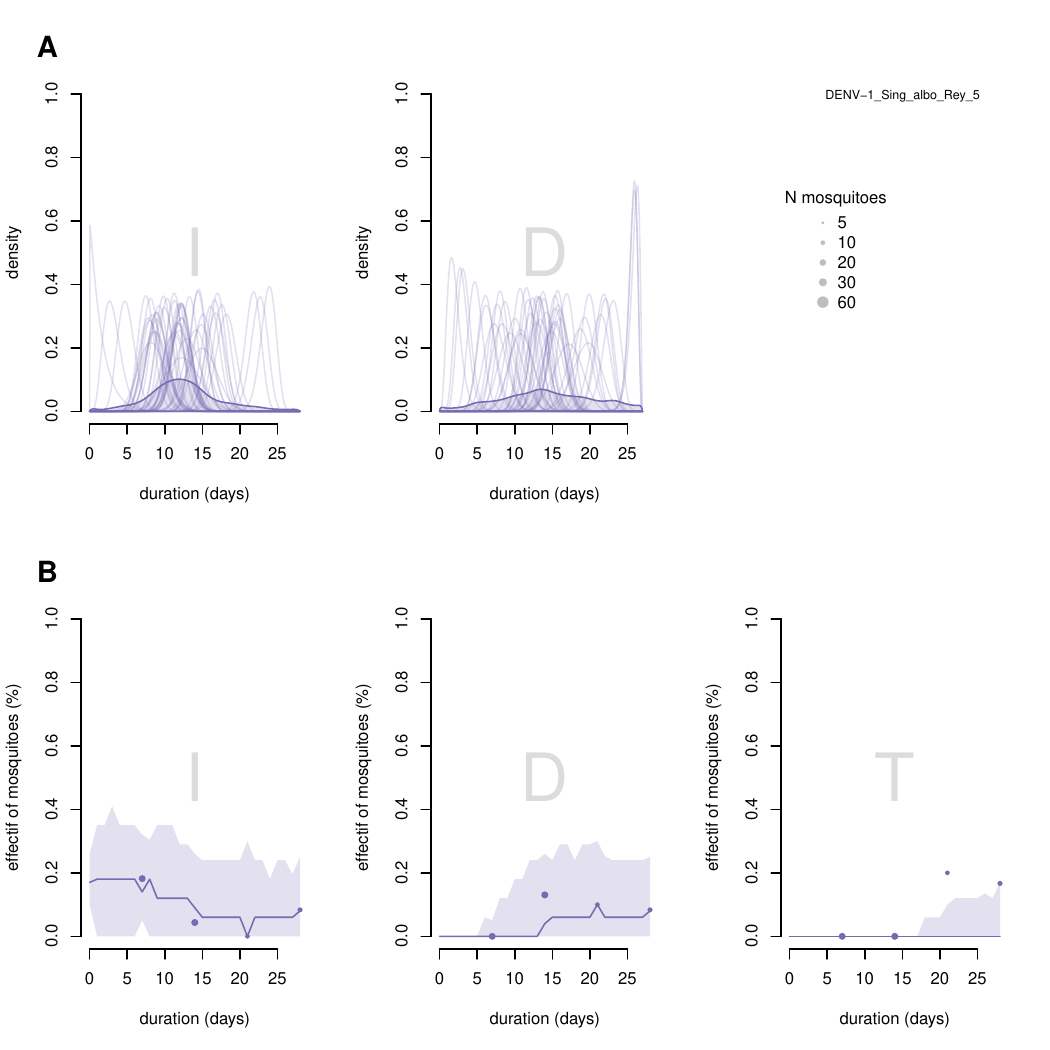}
\caption{Inference results for IVD stages distributions for scenario DENVc3(DENV\-1\_Sing\_albo\_Rey\_5): \textit{Aedes.albopictus} from Reynosa infected by dengue virus from Singapore with an infectious dose of 5 log10 FFU/mL : A) Selected distributions in the infected and disseminated states for the main model selected. The dark line represents the mean of distributions and light lines represent a random sample of 50 distribution among all selected distribution. B) Selected dynamics in the infected (I), disseminated (D), and transmitter (T) states for the main model selected. The dots represent the observed data, the line (mean dynamics), and the uncertainty ribbons (5\%-95\%) represent selected simulated dynamic.}
\label{fig_DENVc3}
\end{figure}

\begin{figure}[H]
\centering
\includegraphics[width=\textwidth]{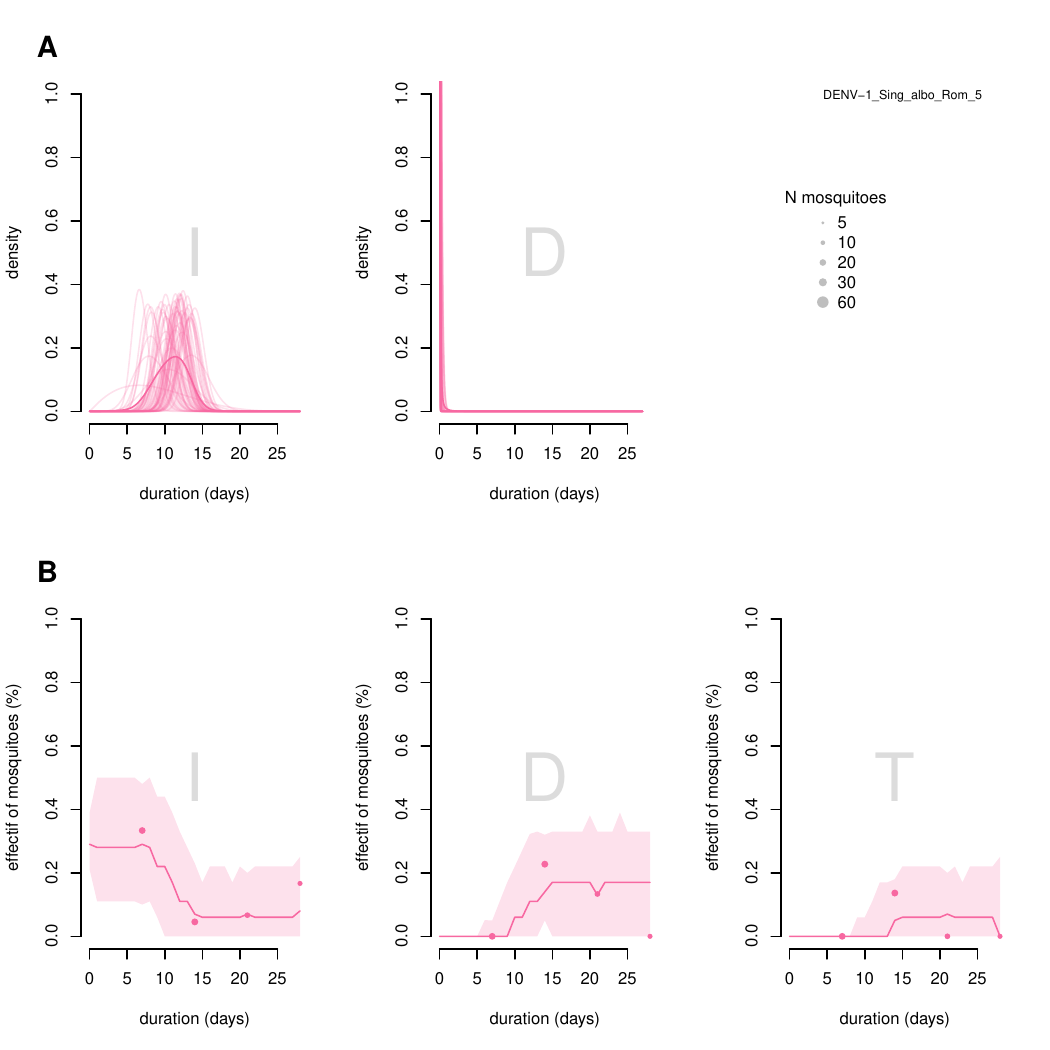}
\caption{Inference results for IVD stages distributions for scenario DENVc4(DENV\-1\_Sing\_albo\_Rom\_5): \textit{Aedes.albopictus} from Roma infected by dengue virus from Singapore with an infectious dose of 5 log10 FFU/mL : A) Selected distributions in the infected and disseminated states for the main model selected. The dark line represents the mean of distributions and light lines represent a random sample of 50 distribution among all selected distribution. B) Selected dynamics in the infected (I), disseminated (D), and transmitter (T) states for the main model selected. The dots represent the observed data, the line (mean dynamics), and the uncertainty ribbons (5\%-95\%) represent selected simulated dynamic.}
\label{fig_DENVc4}
\end{figure}

\begin{figure}[H]
\centering
\includegraphics[width=\textwidth]{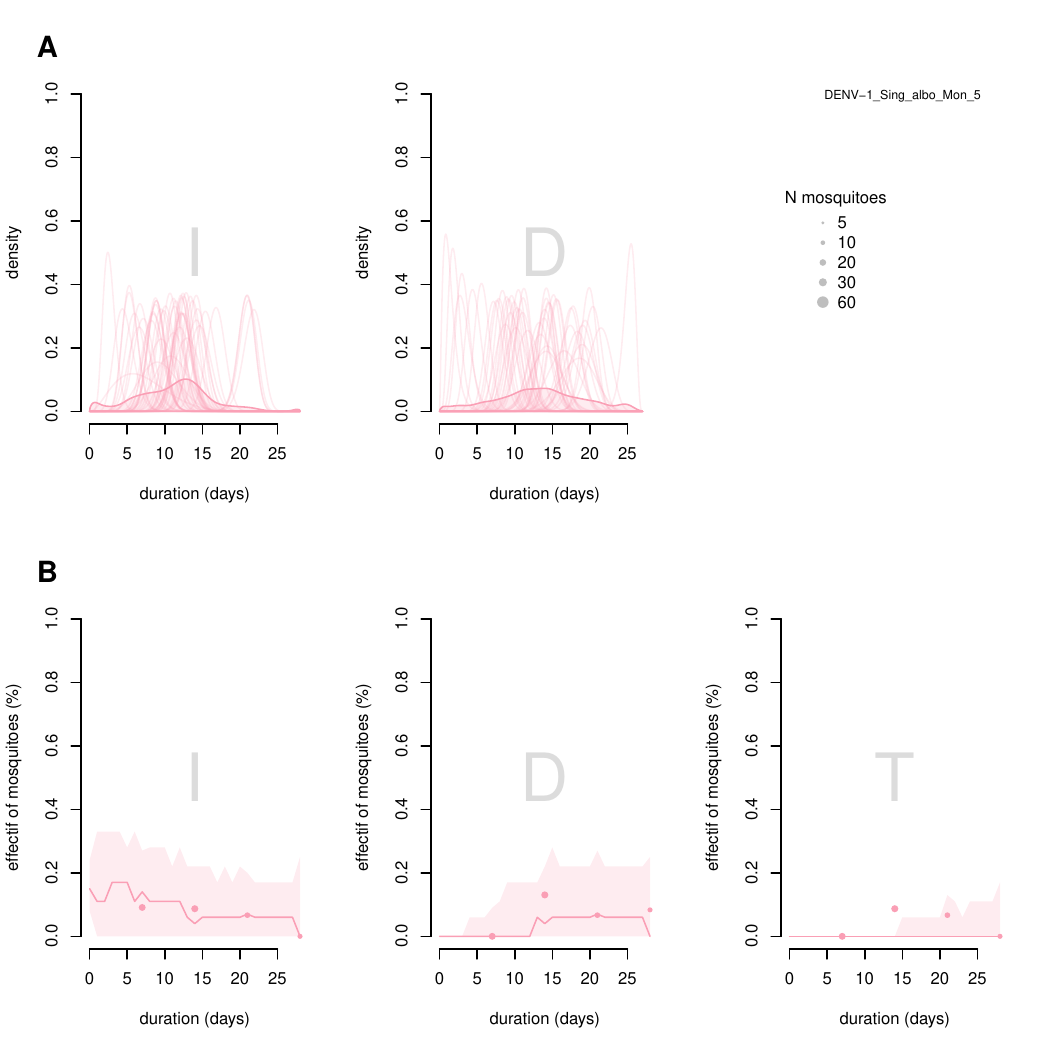}
\caption{Inference results for IVD stages distributions for scenario DENVc5(DENV\-1\_Sing\_albo\_Mon\_5): \textit{Aedes.albopictus} from Montecchio infected by dengue virus from Singapore with an infectious dose of 5 log10 FFU/mL : A) Selected distributions in the infected and disseminated states for the main model selected. The dark line represents the mean of distributions and light lines represent a random sample of 50 distribution among all selected distribution. B) Selected dynamics in the infected (I), disseminated (D), and transmitter (T) states for the main model selected. The dots represent the observed data, the line (mean dynamics), and the uncertainty ribbons (5\%-95\%) represent selected simulated dynamic.}
\label{fig_DENVc5}
\end{figure}

\begin{figure}[H]
\centering
\includegraphics[width=\textwidth]{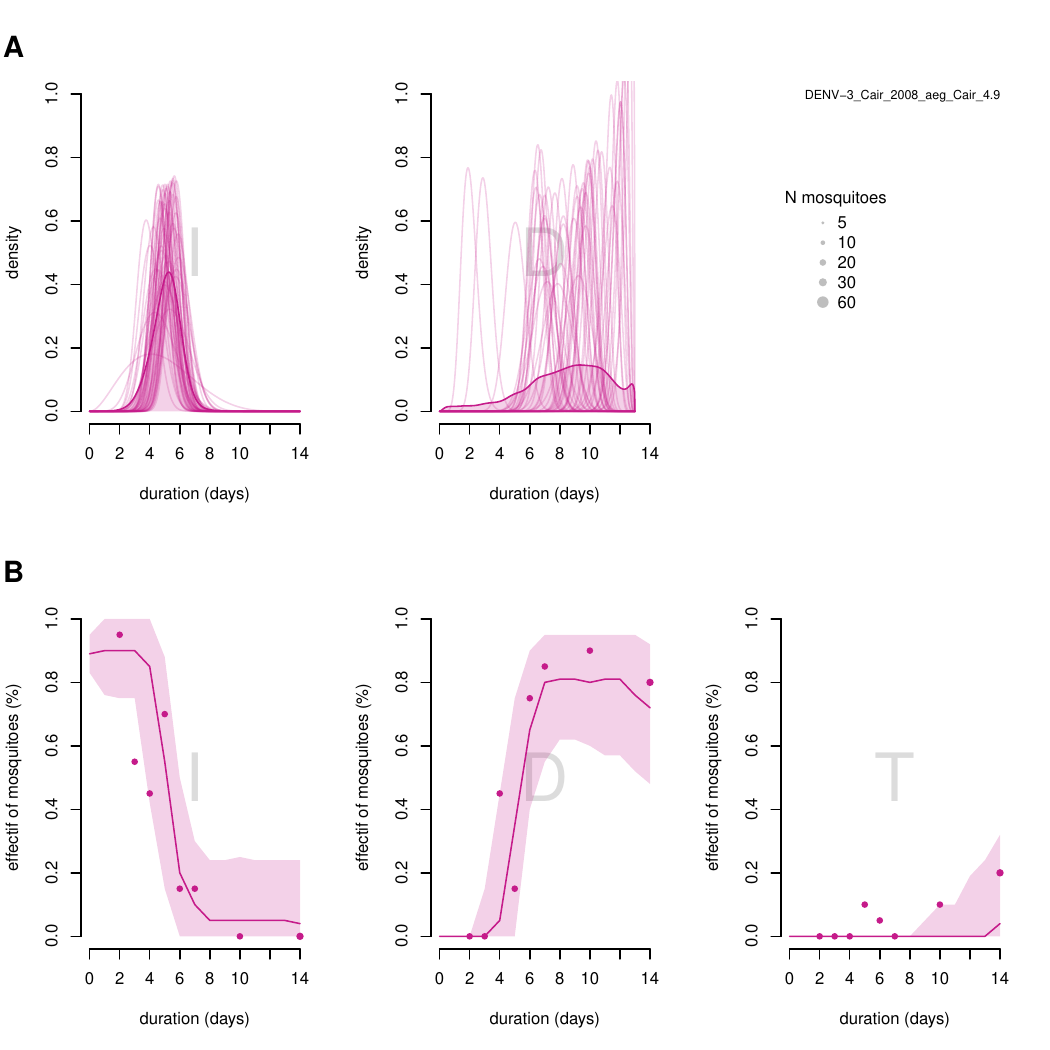}
\caption{Inference results for IVD stages distributions for scenario DENVc6(DENV\-3\_Cair\_2008\_aeg\_Cair\_4.9): \textit{Aedes.aegypti} from Cairns infected by dengue virus from Cairns with an infectious dose of 4.9 log10 FFU/mL : A) Selected distributions in the infected and disseminated states for the main model selected. The dark line represents the mean of distributions and light lines represent a random sample of 50 distribution among all selected distribution. B) Selected dynamics in the infected (I), disseminated (D), and transmitter (T) states for the main model selected. The dots represent the observed data, the line (mean dynamics), and the uncertainty ribbons (5\%-95\%) represent selected simulated dynamic.}
\label{fig_DENVc6}
\end{figure}

\begin{figure}[H]
\centering
\includegraphics[width=\textwidth]{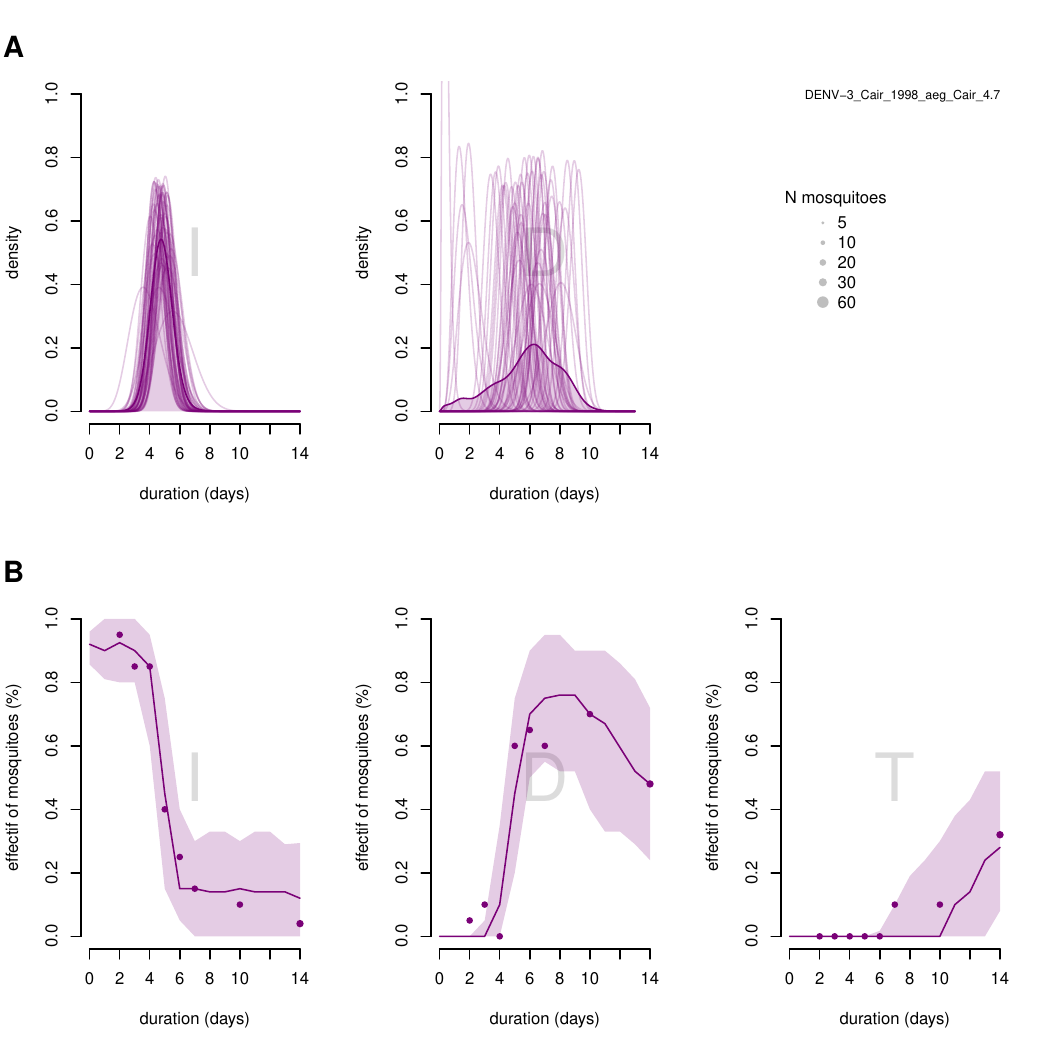}
\caption{Inference results for IVD stages distributions for scenario DENVc7(DENV\-3\_Cair\_1998\_aeg\_Cair\_4.7): \textit{Aedes.aegypti} from Cairns infected by dengue virus from Cairns with an infectious dose of 4.7 log10 FFU/mL : A) Selected distributions in the infected and disseminated states for the main model selected. The dark line represents the mean of distributions and light lines represent a random sample of 50 distribution among all selected distribution. B) Selected dynamics in the infected (I), disseminated (D), and transmitter (T) states for the main model selected. The dots represent the observed data, the line (mean dynamics), and the uncertainty ribbons (5\%-95\%) represent selected simulated dynamic.}
\label{fig_DENVc7}
\end{figure}

\begin{figure}[H]
\centering
\includegraphics[width=\textwidth]{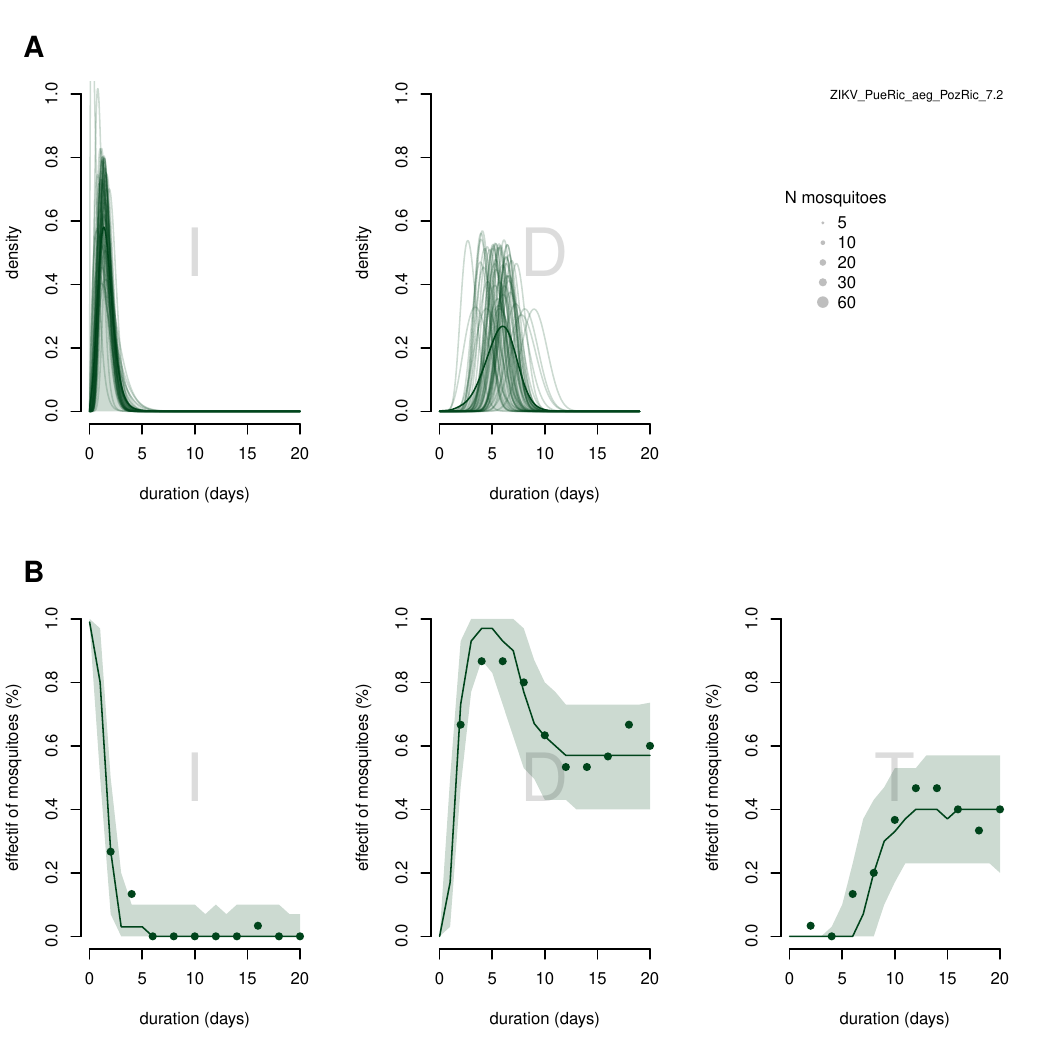}
\caption{Inference results for IVD stages distributions for scenario ZIKVc1(ZIKV\_PueRic\_aeg\_PozRic\_7.2): \textit{Aedes.aegypti} from Poza Rica infected by Zika virus from Puerto Rico with an infectious dose of 7.2 log10 FFU/mL : A) Selected distributions in the infected and disseminated states for the main model selected. The dark line represents the mean of distributions and light lines represent a random sample of 50 distribution among all selected distribution. B) Selected dynamics in the infected (I), disseminated (D), and transmitter (T) states for the main model selected. The dots represent the observed data, the line (mean dynamics), and the uncertainty ribbons (5\%-95\%) represent selected simulated dynamic.}
\label{fig_ZIKVc1}
\end{figure}

\begin{figure}[H]
\centering
\includegraphics[width=\textwidth]{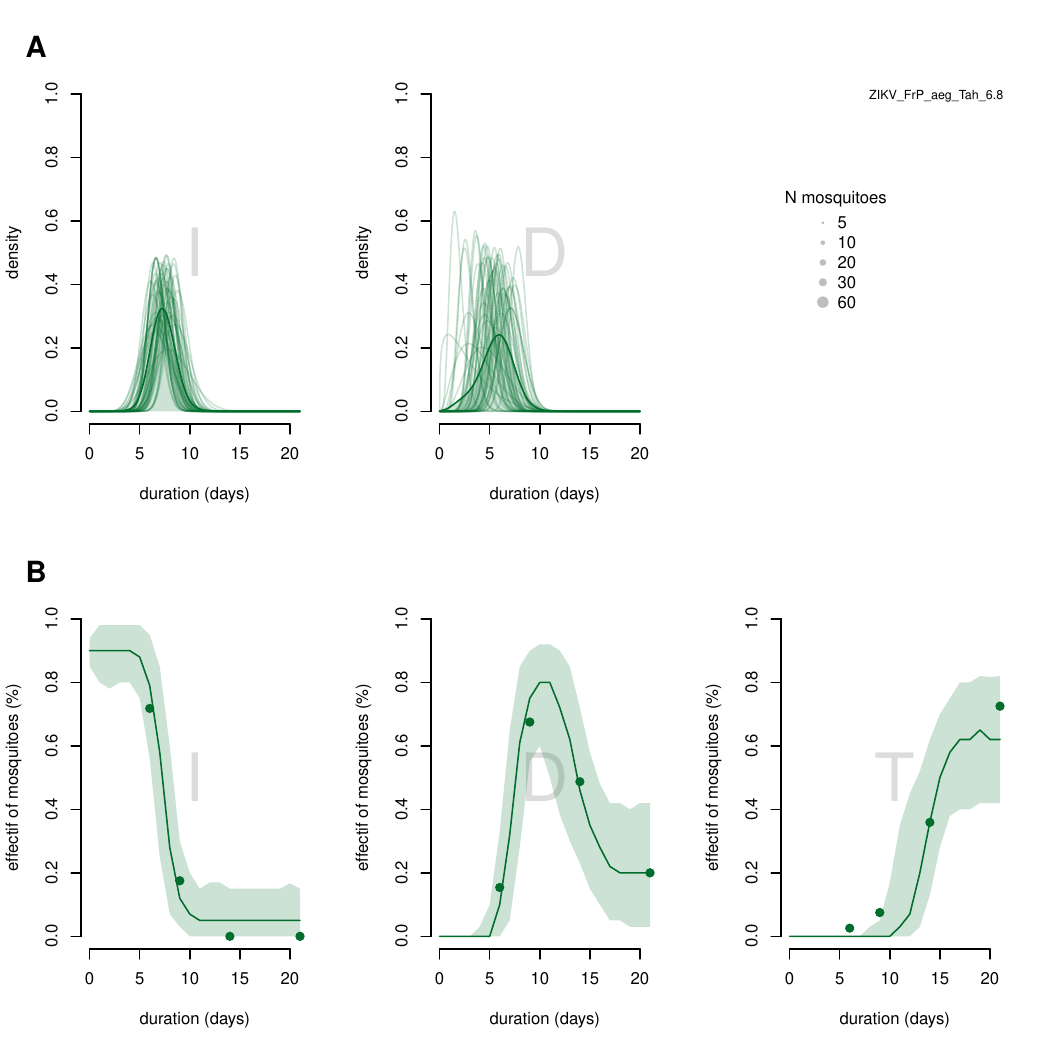}
\caption{Inference results for IVD stages distributions for scenario ZIKVc2(ZIKV\_FrP\_aeg\_Tah\_6.8): \textit{Aedes.aegypti} from Tahiti infected by Zika virus from French Polynesia with an infectious dose of 6.8 log10 FFU/mL : A) Selected distributions in the infected and disseminated states for the main model selected. The dark line represents the mean of distributions and light lines represent a random sample of 50 distribution among all selected distribution. B) Selected dynamics in the infected (I), disseminated (D), and transmitter (T) states for the main model selected. The dots represent the observed data, the line (mean dynamics), and the uncertainty ribbons (5\%-95\%) represent selected simulated dynamic.}
\label{fig_ZIKVc2}
\end{figure}

\begin{figure}[H]
\centering
\includegraphics[width=\textwidth]{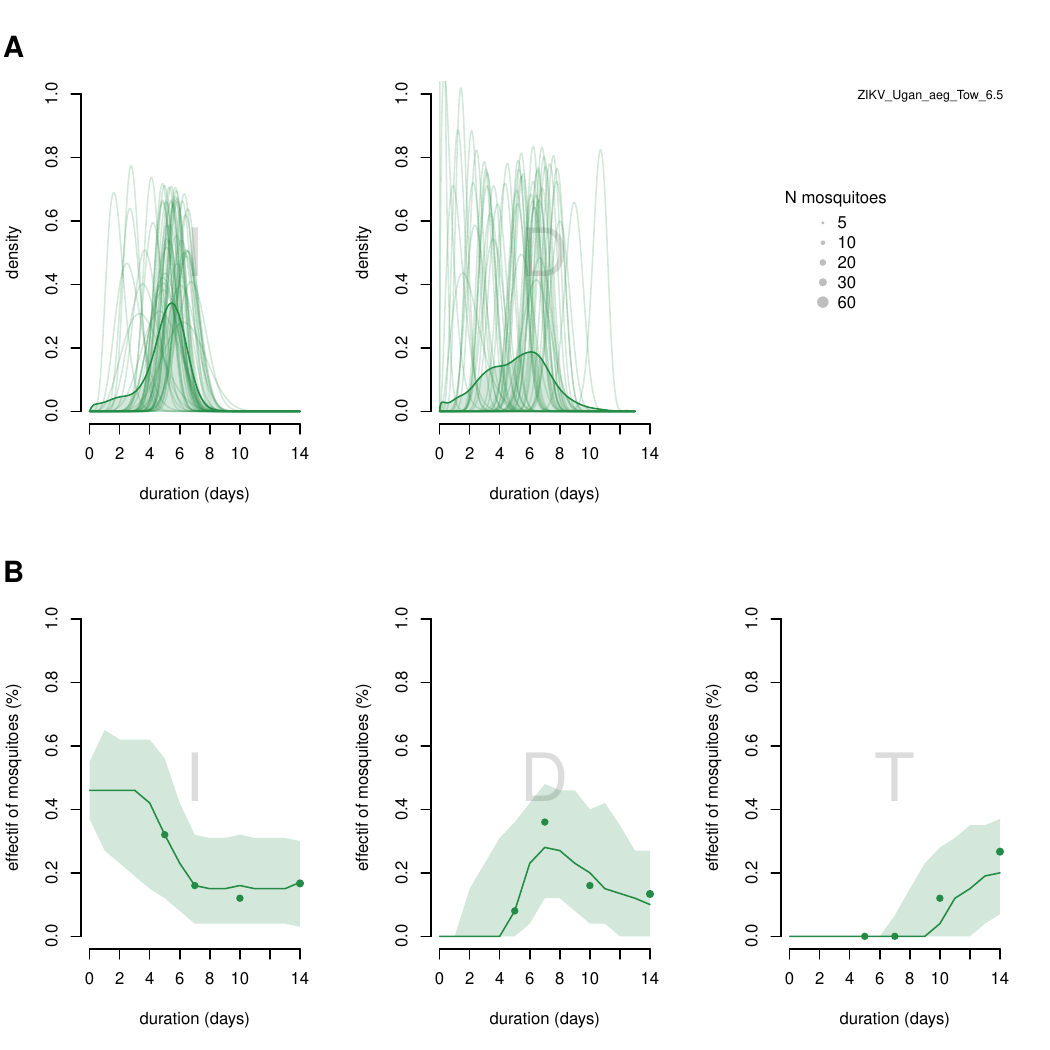}
\caption{Inference results for IVD stages distributions for scenario ZIKVc3(ZIKV\_Ugan\_aeg\_Tow\_6.5): \textit{Aedes.aegypti} from Townsville infected by Zika virus from Uganda with an infectious dose of 6.5 log10 FFU/mL : A) Selected distributions in the infected and disseminated states for the main model selected. The dark line represents the mean of distributions and light lines represent a random sample of 50 distribution among all selected distribution. B) Selected dynamics in the infected (I), disseminated (D), and transmitter (T) states for the main model selected. The dots represent the observed data, the line (mean dynamics), and the uncertainty ribbons (5\%-95\%) represent selected simulated dynamic.}
\label{fig_ZIKVc3}
\end{figure}

\begin{figure}[H]
\centering
\includegraphics[width=\textwidth]{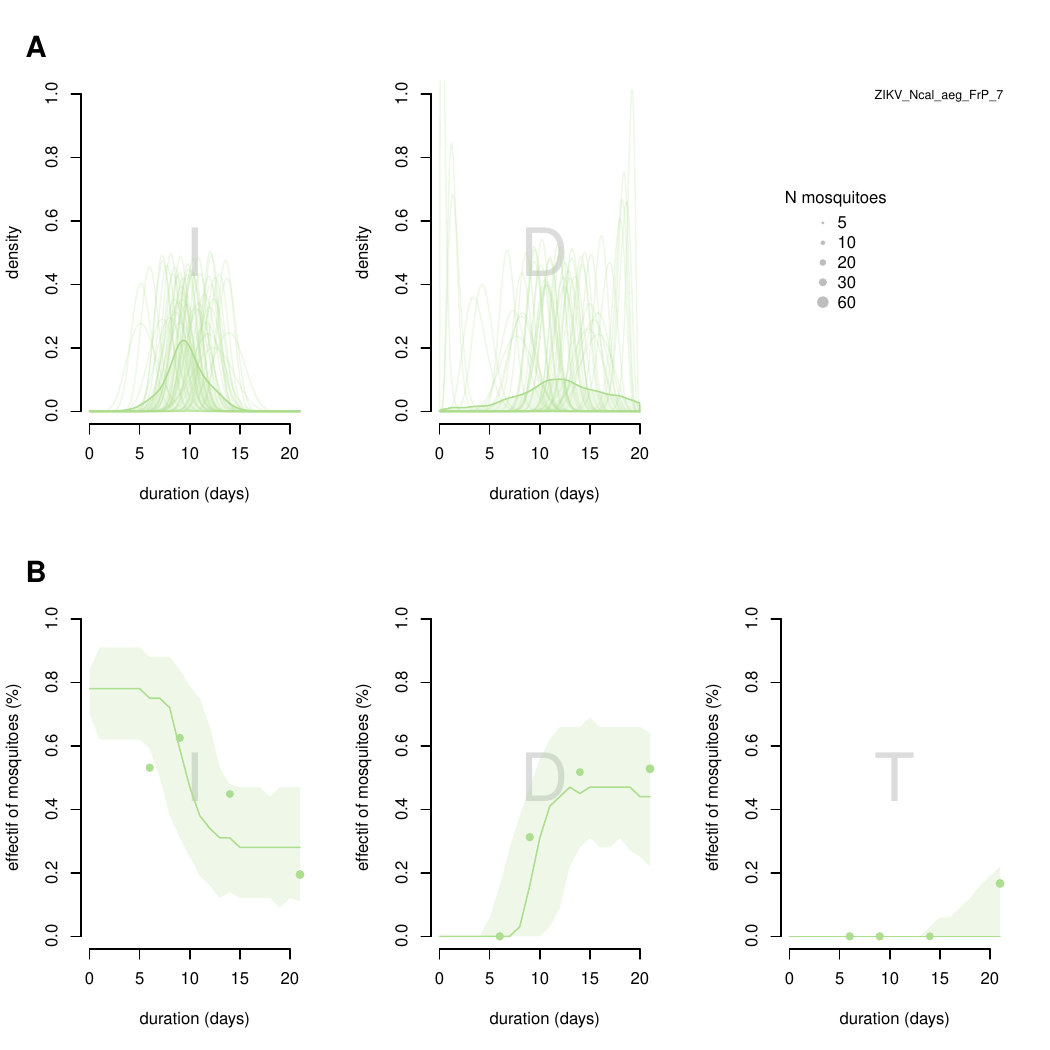}
\caption{Inference results for IVD stages distributions for scenario ZIKVc4(ZIKV\_Ncal\_aeg\_FrP\_7): \textit{Aedes.aegypti} from French Polynesia infected by Zika virus from New Caledonia with an infectious dose of 7 log10 FFU/mL : A) Selected distributions in the infected and disseminated states for the main model selected. The dark line represents the mean of distributions and light lines represent a random sample of 50 distribution among all selected distribution. B) Selected dynamics in the infected (I), disseminated (D), and transmitter (T) states for the main model selected. The dots represent the observed data, the line (mean dynamics), and the uncertainty ribbons (5\%-95\%) represent selected simulated dynamic.}
\label{fig_ZIKVc4}
\end{figure}

\begin{figure}[H]
\centering
\includegraphics[width=\textwidth]{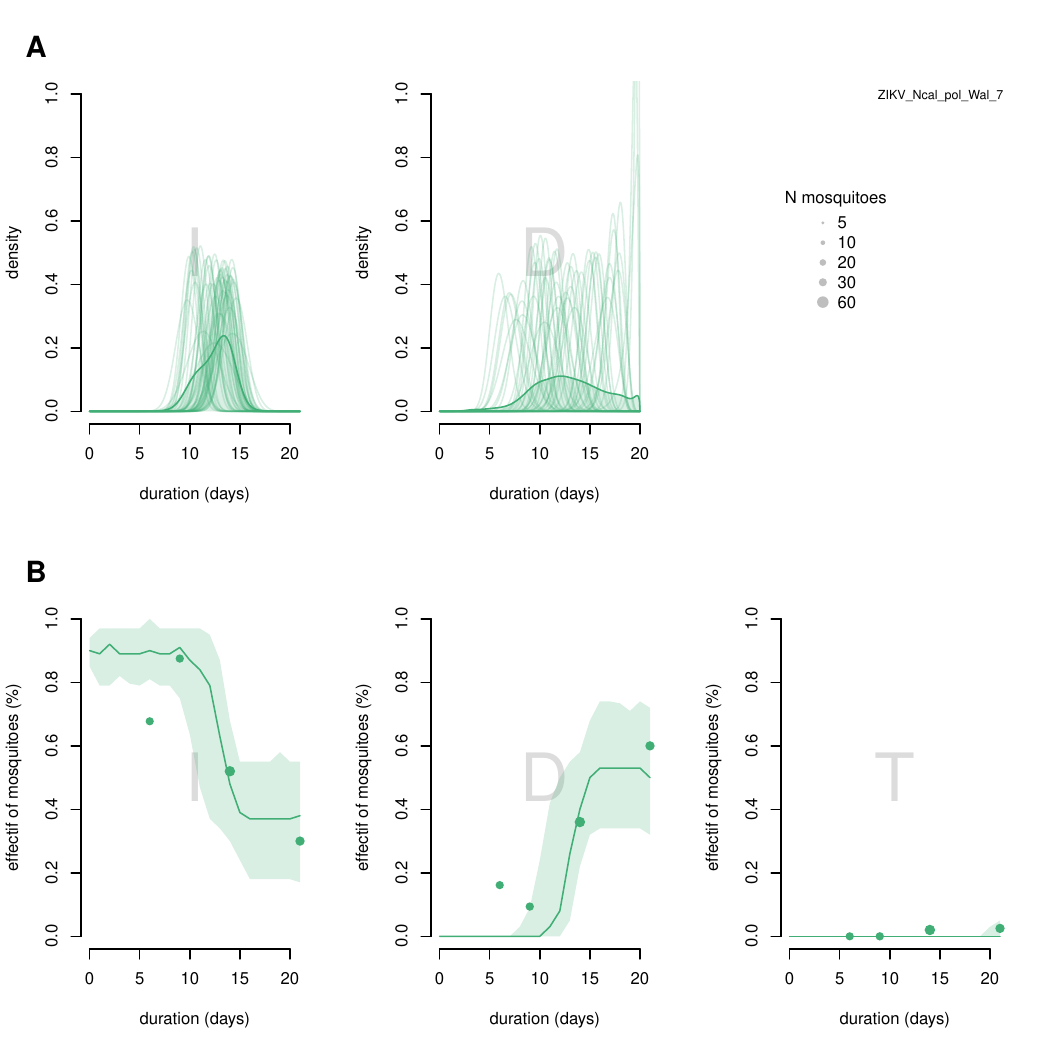}
\caption{Inference results for IVD stages distributions for scenario ZIKVc5(ZIKV\_Ncal\_pol\_Wal\_7): \textit{Aedes.polynesiensis} from Wallis infected by Zika virus from New Caledonia with an infectious dose of 7 log10 FFU/mL : A) Selected distributions in the infected and disseminated states for the main model selected. The dark line represents the mean of distributions and light lines represent a random sample of 50 distribution among all selected distribution. B) Selected dynamics in the infected (I), disseminated (D), and transmitter (T) states for the main model selected. The dots represent the observed data, the line (mean dynamics), and the uncertainty ribbons (5\%-95\%) represent selected simulated dynamic.}
\label{fig_ZIKVc5}
\end{figure}

\begin{figure}[H]
\centering
\includegraphics[width=\textwidth]{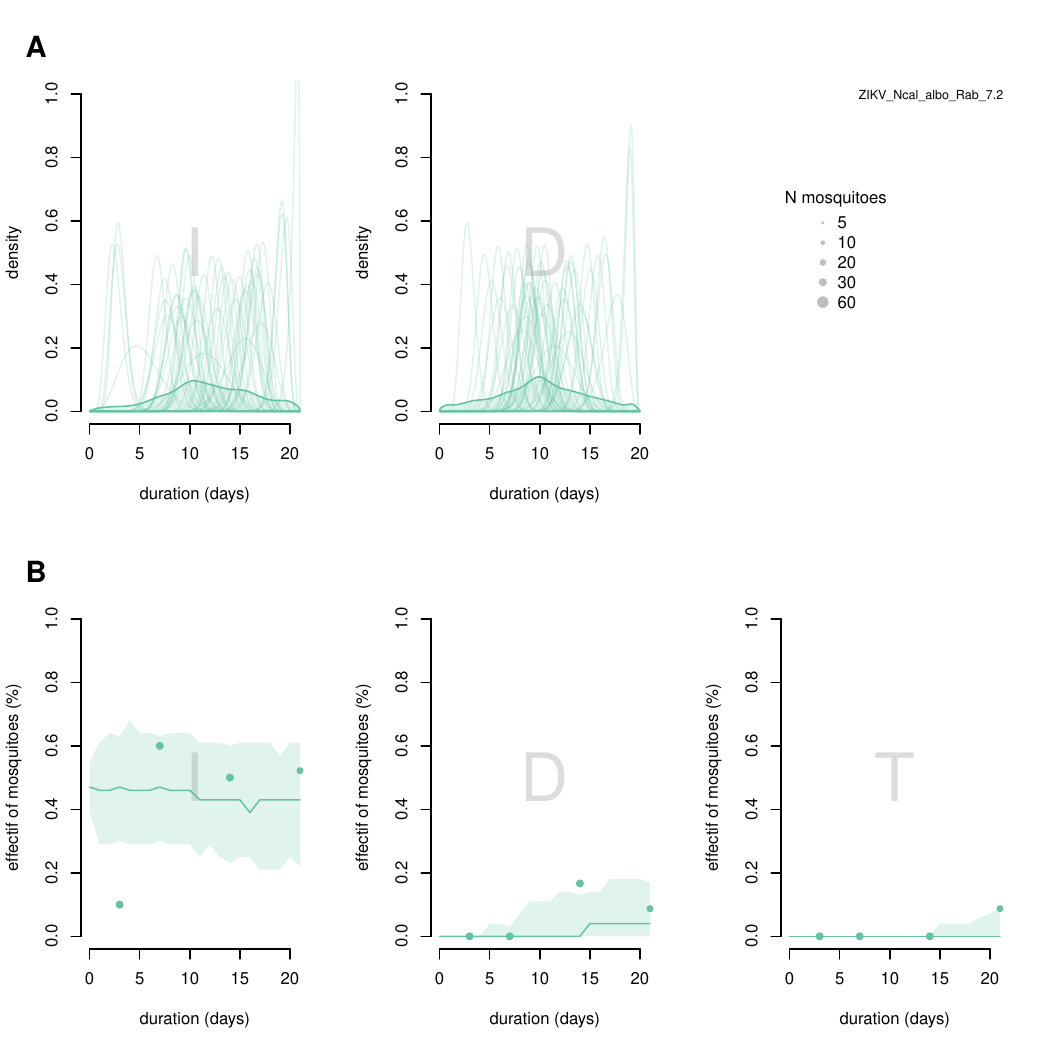}
\caption{Inference results for IVD stages distributions for scenario ZIKVc6(ZIKV\_Ncal\_albo\_Rab\_7.2): \textit{Aedes.albopictus} from Rabat infected by Zika virus from New Caledonia with an infectious dose of 7.2 log10 FFU/mL : A) Selected distributions in the infected and disseminated states for the main model selected. The dark line represents the mean of distributions and light lines represent a random sample of 50 distribution among all selected distribution. B) Selected dynamics in the infected (I), disseminated (D), and transmitter (T) states for the main model selected. The dots represent the observed data, the line (mean dynamics), and the uncertainty ribbons (5\%-95\%) represent selected simulated dynamic.}
\label{fig_ZIKVc6}
\end{figure}

\begin{figure}[H]
\centering
\includegraphics[width=\textwidth]{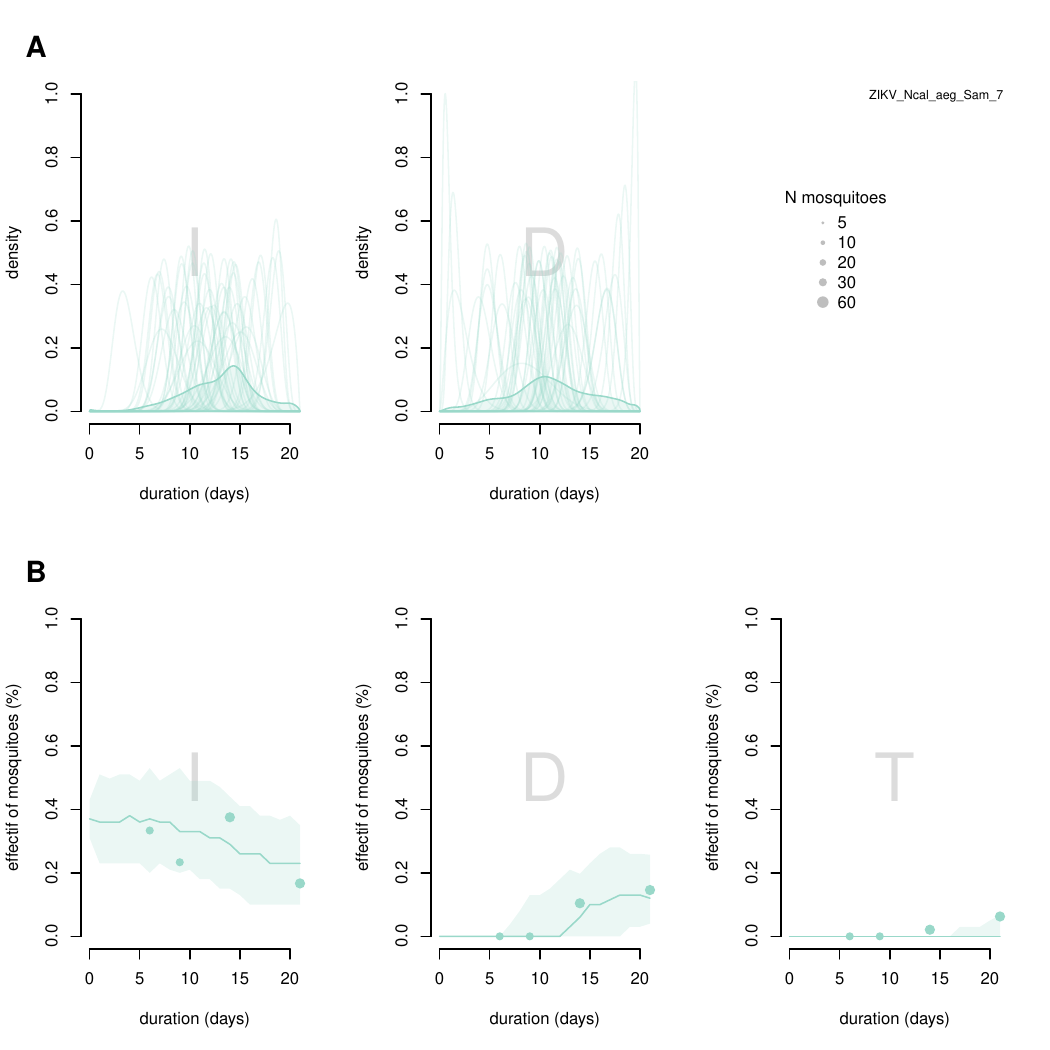}
\caption{Inference results for IVD stages distributions for scenario ZIKVc7(ZIKV\_Ncal\_aeg\_Sam\_7): \textit{Aedes.aegypti} from Samoa infected by Zika virus from New Caledonia with an infectious dose of 7 log10 FFU/mL : A) Selected distributions in the infected and disseminated states for the main model selected. The dark line represents the mean of distributions and light lines represent a random sample of 50 distribution among all selected distribution. B) Selected dynamics in the infected (I), disseminated (D), and transmitter (T) states for the main model selected. The dots represent the observed data, the line (mean dynamics), and the uncertainty ribbons (5\%-95\%) represent selected simulated dynamic.}
\label{fig_ZIKVc7}
\end{figure}

\begin{figure}[H]
\centering
\includegraphics[width=\textwidth]{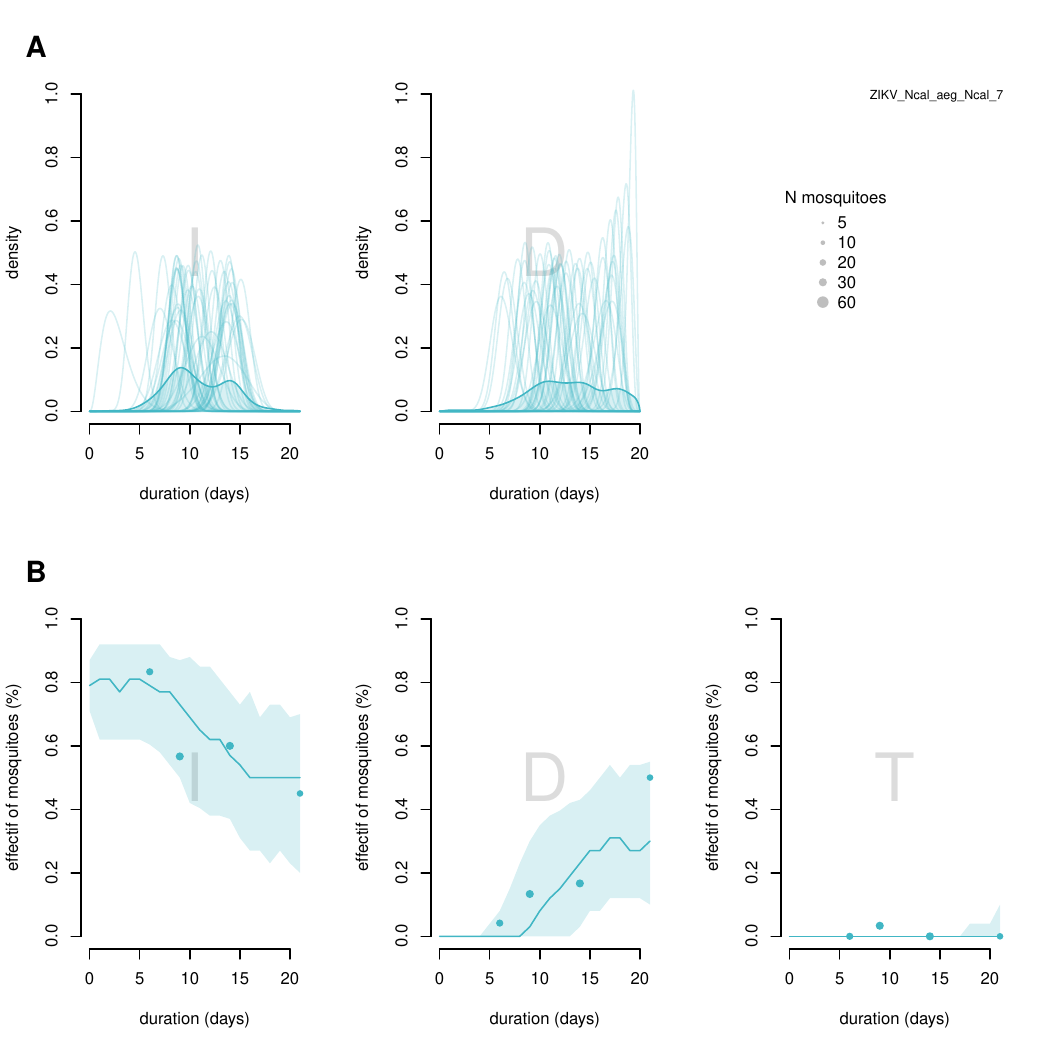}
\caption{Inference results for IVD stages distributions for scenario ZIKVc8(ZIKV\_Ncal\_aeg\_Ncal\_7): \textit{Aedes.aegypti} from New caledonia infected by Zika virus from New Caledonia with an infectious dose of 7 log10 FFU/mL : A) Selected distributions in the infected and disseminated states for the main model selected. The dark line represents the mean of distributions and light lines represent a random sample of 50 distribution among all selected distribution. B) Selected dynamics in the infected (I), disseminated (D), and transmitter (T) states for the main model selected. The dots represent the observed data, the line (mean dynamics), and the uncertainty ribbons (5\%-95\%) represent selected simulated dynamic.}
\label{fig_ZIKVc8}
\end{figure}

\begin{figure}[H]
\centering
\includegraphics[width=\textwidth]{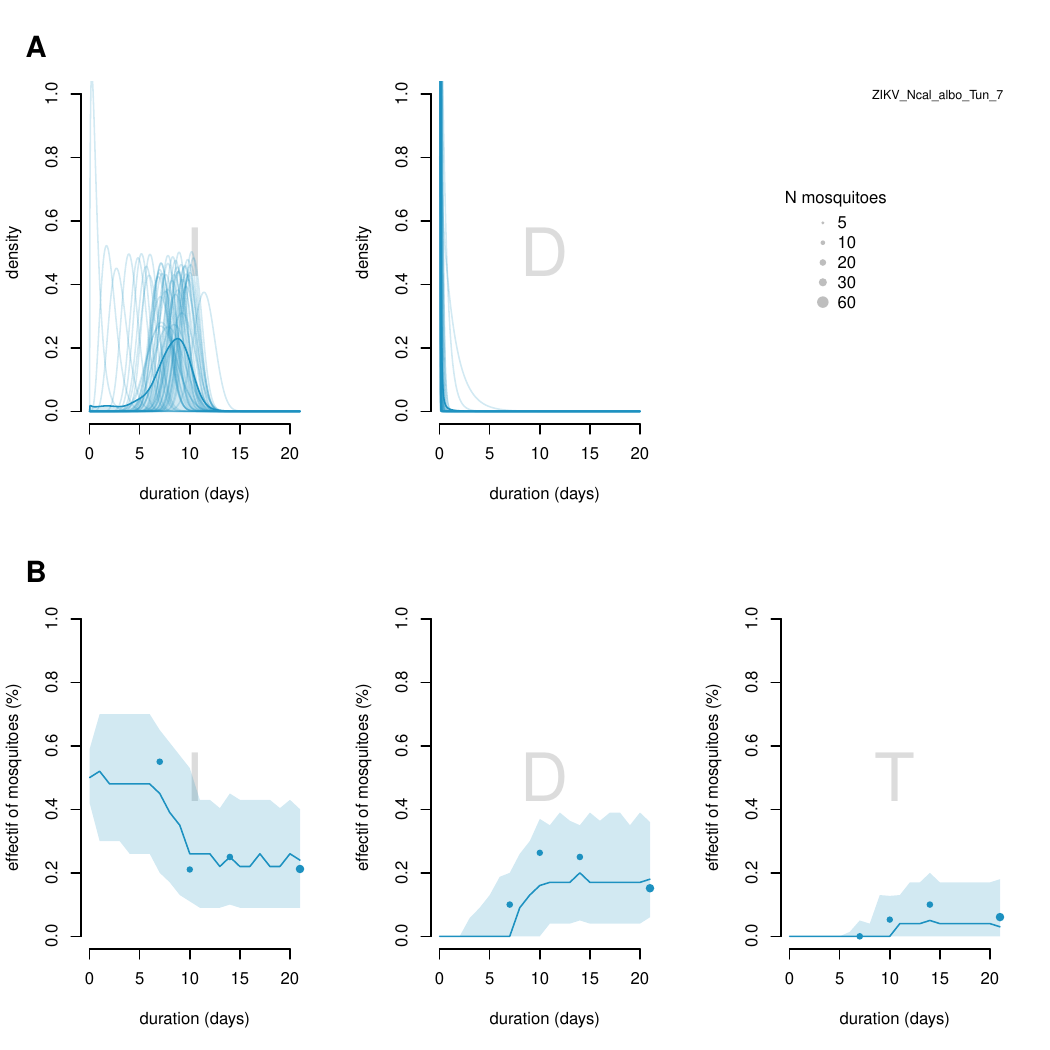}
\caption{Inference results for IVD stages distributions for scenario ZIKVc9(ZIKV\_Ncal\_albo\_Tun\_7): \textit{Aedes.albopictus} from Tunisia infected by Zika virus from New Caledonia with an infectious dose of 7 log10 FFU/mL : A) Selected distributions in the infected and disseminated states for the main model selected. The dark line represents the mean of distributions and light lines represent a random sample of 50 distribution among all selected distribution. B) Selected dynamics in the infected (I), disseminated (D), and transmitter (T) states for the main model selected. The dots represent the observed data, the line (mean dynamics), and the uncertainty ribbons (5\%-95\%) represent selected simulated dynamic.}
\label{fig_ZIKVc9}
\end{figure}

\begin{figure}[H]
\centering
\includegraphics[width=\textwidth]{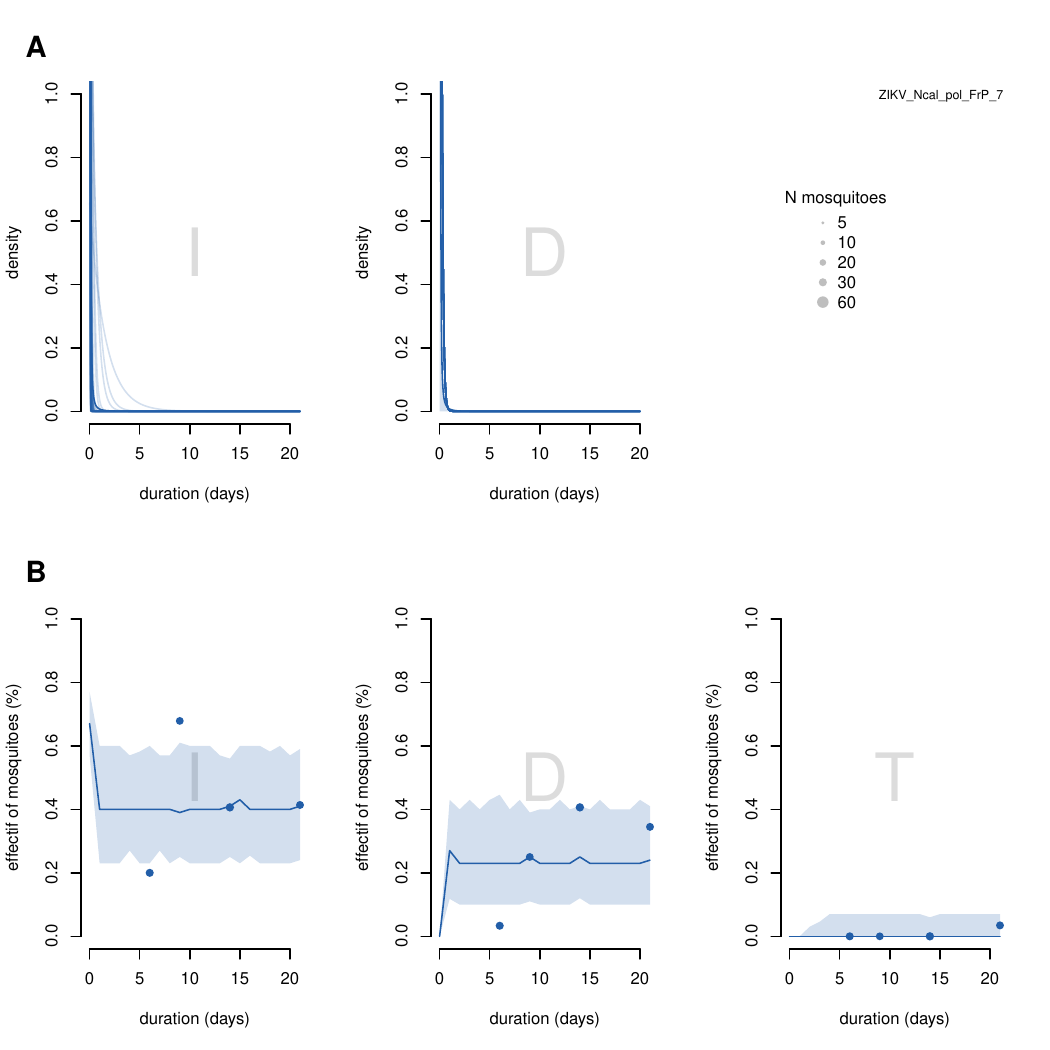}
\caption{Inference results for IVD stages distributions for scenario ZIKVc10(ZIKV\_Ncal\_pol\_FrP\_7): \textit{Aedes.polynesiensis} from French Polynesia infected by Zika virus from New Caledonia with an infectious dose of 7 log10 FFU/mL : A) Selected distributions in the infected and disseminated states for the main model selected. The dark line represents the mean of distributions and light lines represent a random sample of 50 distribution among all selected distribution. B) Selected dynamics in the infected (I), disseminated (D), and transmitter (T) states for the main model selected. The dots represent the observed data, the line (mean dynamics), and the uncertainty ribbons (5\%-95\%) represent selected simulated dynamic.}
\label{fig_ZIKVc10}
\end{figure}

\begin{figure}[H]
\centering
\includegraphics[width=\textwidth]{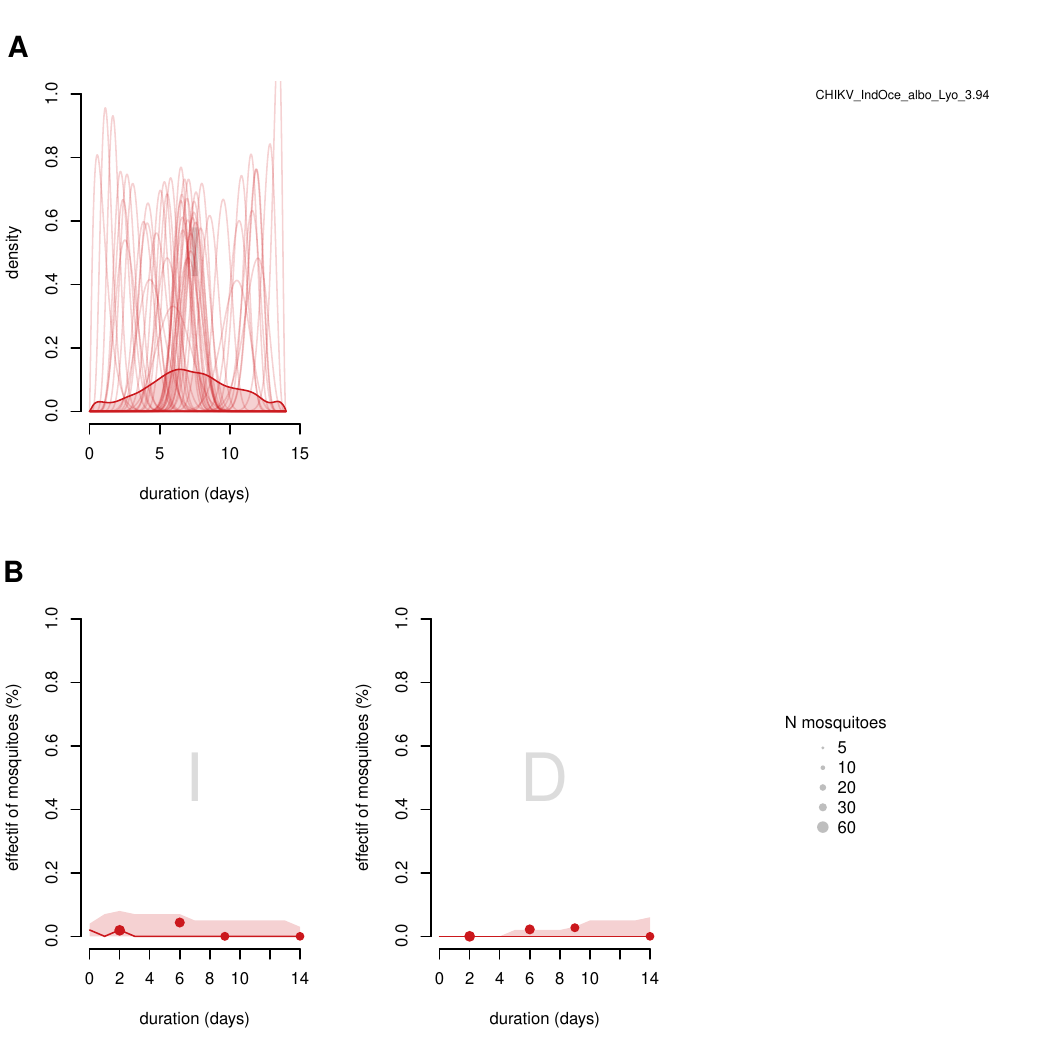}
\caption{Inference results for IVD stages distributions for scenario CHIKVp1(CHIKV\_IndOce\_albo\_Lyo\_3.94): \textit{Aedes.albopictus} from Lyon infected by chikungunya virus from Indian Ocean with an infectious dose of 3.94 log10 FFU/mL : A) Selected distributions in the infected state for the main model selected. The dark line represents the mean of distributions and light lines represent a random sample of 50 distribution among all selected distribution. B) Selected dynamics in the infected (I) and disseminated (D) states for the main model selected. The dots represent the observed data, the line (mean dynamics), and the uncertainty ribbons (5\%-95\%) represent selected simulated dynamic.}
\label{fig_CHIKVp1}
\end{figure}

\begin{figure}[H]
\centering
\includegraphics[width=\textwidth]{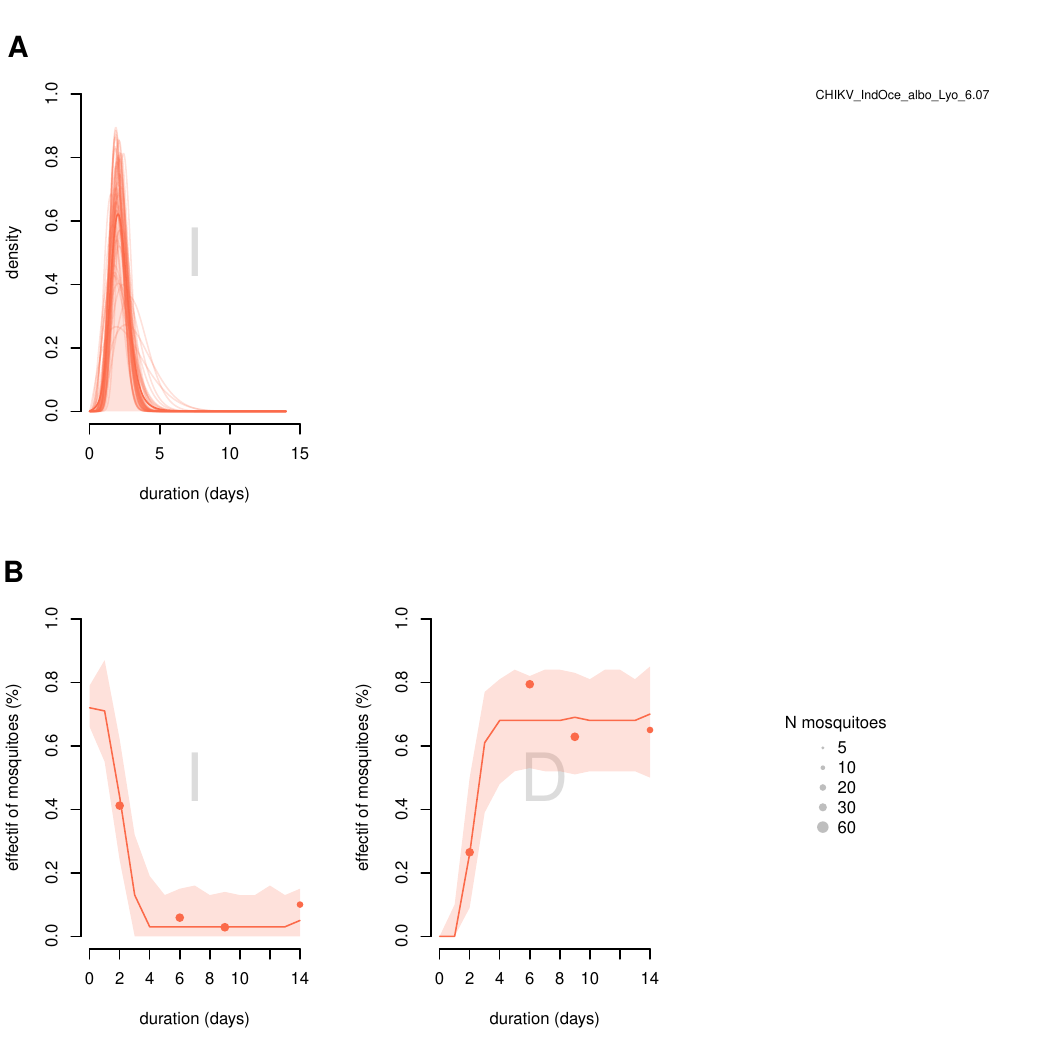}
\caption{Inference results for IVD stages distributions for scenario CHIKVp2(CHIKV\_IndOce\_albo\_Lyo\_6.07): \textit{Aedes.albopictus} from Lyon infected by chikungunya virus from Indian Ocean with an infectious dose of 6.07 log10 FFU/mL : A) Selected distributions in the infected state for the main model selected. The dark line represents the mean of distributions and light lines represent a random sample of 50 distribution among all selected distribution. B) Selected dynamics in the infected (I) and disseminated (D) states for the main model selected. The dots represent the observed data, the line (mean dynamics), and the uncertainty ribbons (5\%-95\%) represent selected simulated dynamic.}
\label{fig_CHIKVp2}
\end{figure}

\begin{figure}[H]
\centering
\includegraphics[width=\textwidth]{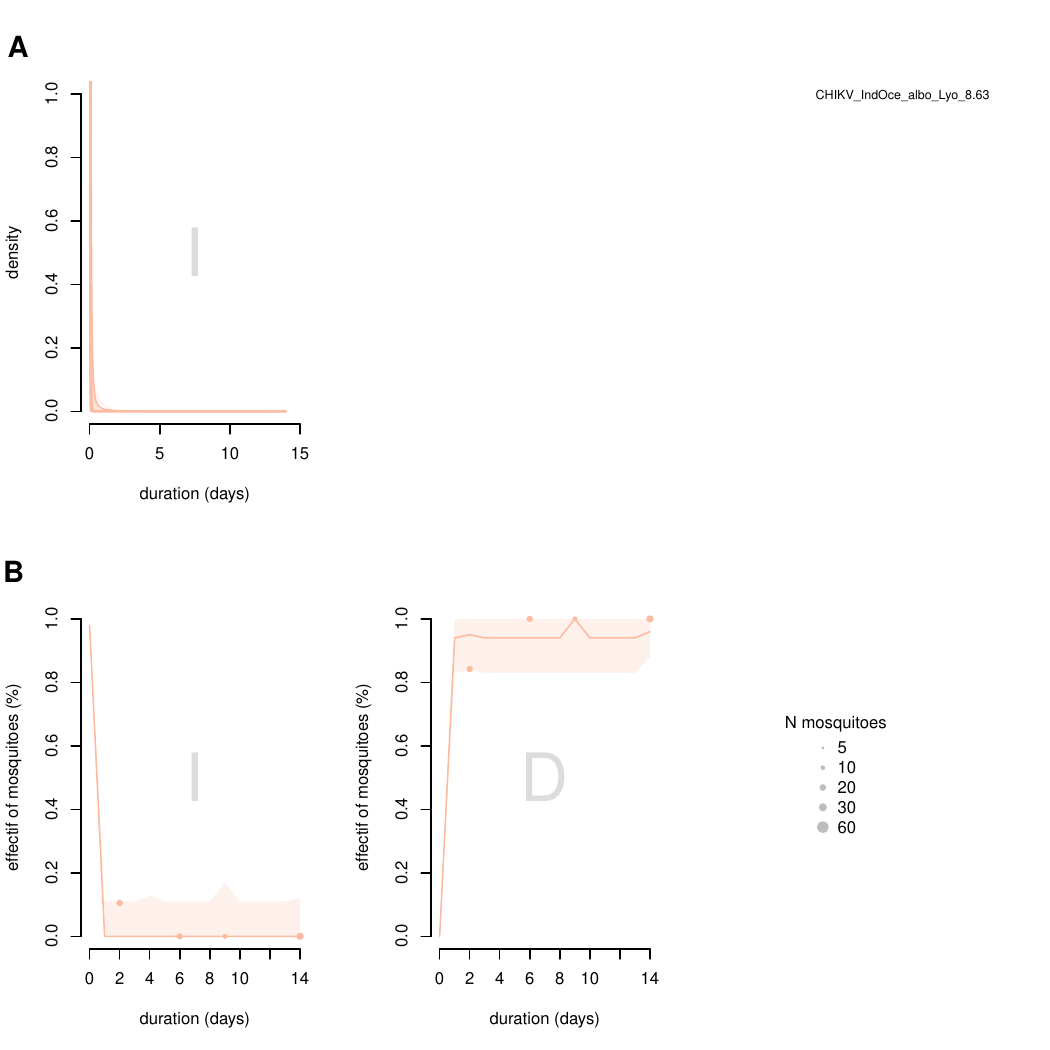}
\caption{Inference results for IVD stages distributions for scenario CHIKVp3(CHIKV\_IndOce\_albo\_Lyo\_8.63): \textit{Aedes.albopictus} from Lyon infected by chikungunya virus from Indian Ocean with an infectious dose of 8.63 log10 FFU/mL : A) Selected distributions in the infected state for the main model selected. The dark line represents the mean of distributions and light lines represent a random sample of 50 distribution among all selected distribution. B) Selected dynamics in the infected (I) and disseminated (D) states for the main model selected. The dots represent the observed data, the line (mean dynamics), and the uncertainty ribbons (5\%-95\%) represent selected simulated dynamic.}
\label{fig_CHIKVp3}
\end{figure}

\begin{figure}[H]
\centering
\includegraphics[width=\textwidth]{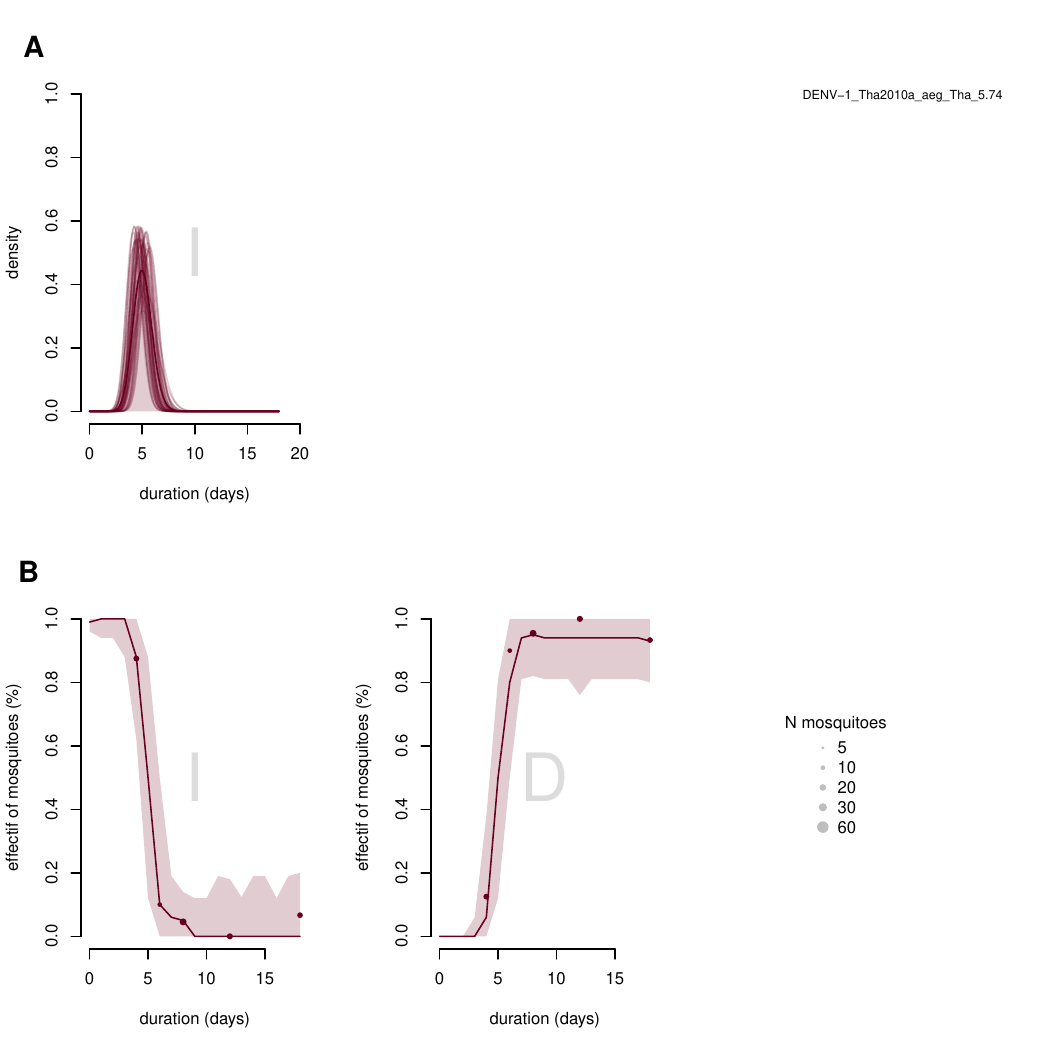}
\caption{Inference results for IVD stages distributions for scenario DENVp1(DENV1\_Tha2010a\_aeg\_Tha\_5.74): \textit{Aedes.aegypti} from Thailand infected by dengue virus from Thailand with an infectious dose of 5.74 log10 FFU/mL : A) Selected distributions in the infected state for the main model selected. The dark line represents the mean of distributions and light lines represent a random sample of 50 distribution among all selected distribution. B) Selected dynamics in the infected (I) and disseminated (D) states for the main model selected. The dots represent the observed data, the line (mean dynamics), and the uncertainty ribbons (5\%-95\%) represent selected simulated dynamic.}
\label{fig_DENVp1}
\end{figure}

\begin{figure}[H]
\centering
\includegraphics[width=\textwidth]{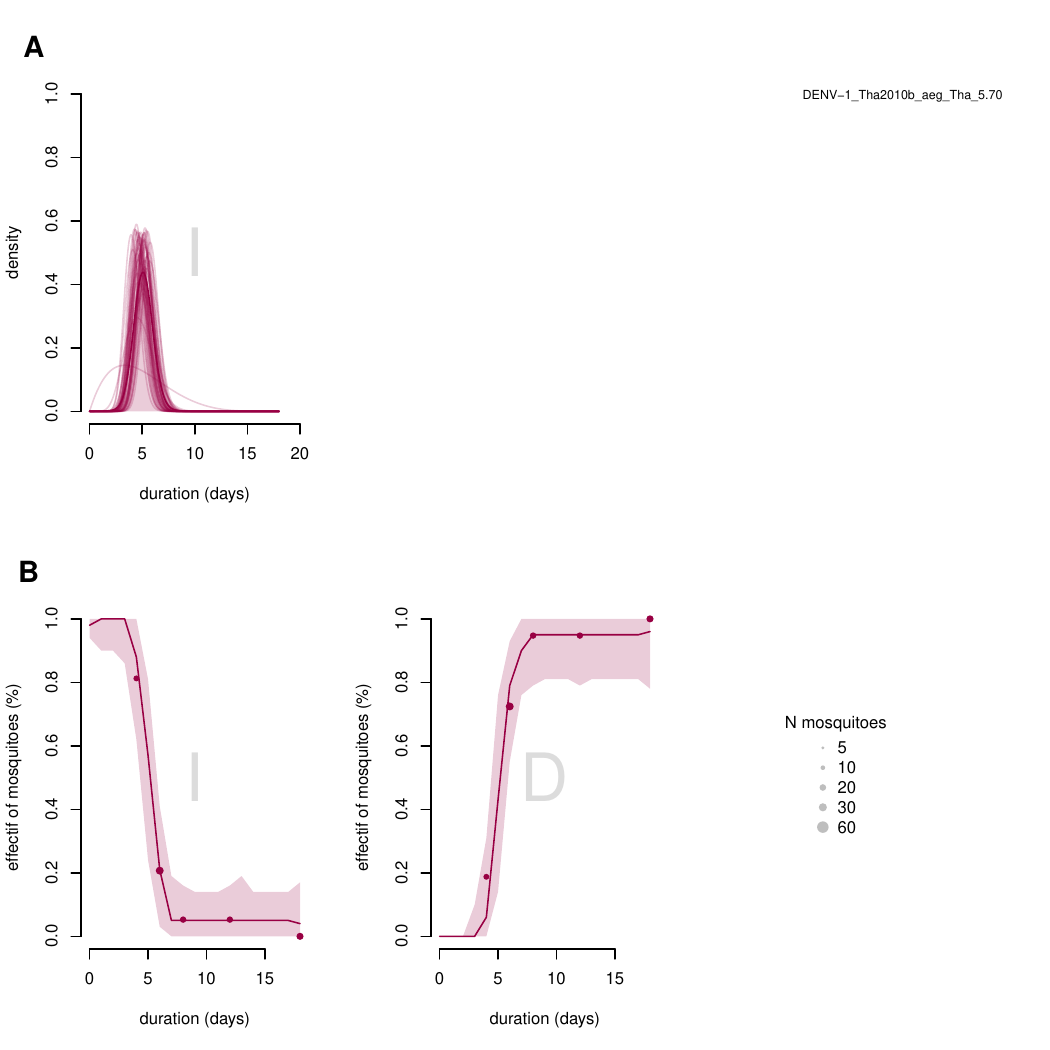}
\caption{Inference results for IVD stages distributions for scenario DENVp2(DENV1\_Tha2010b\_aeg\_Tha\_5.70): \textit{Aedes.aegypti} from Thailand infected by dengue virus from Thailand with an infectious dose of 5.70 log10 FFU/mL : A) Selected distributions in the infected state for the main model selected. The dark line represents the mean of distributions and light lines represent a random sample of 50 distribution among all selected distribution. B) Selected dynamics in the infected (I) and disseminated (D) states for the main model selected. The dots represent the observed data, the line (mean dynamics), and the uncertainty ribbons (5\%-95\%) represent selected simulated dynamic.}
\label{fig_DENVp2}
\end{figure}

\begin{figure}[H]
\centering
\includegraphics[width=\textwidth]{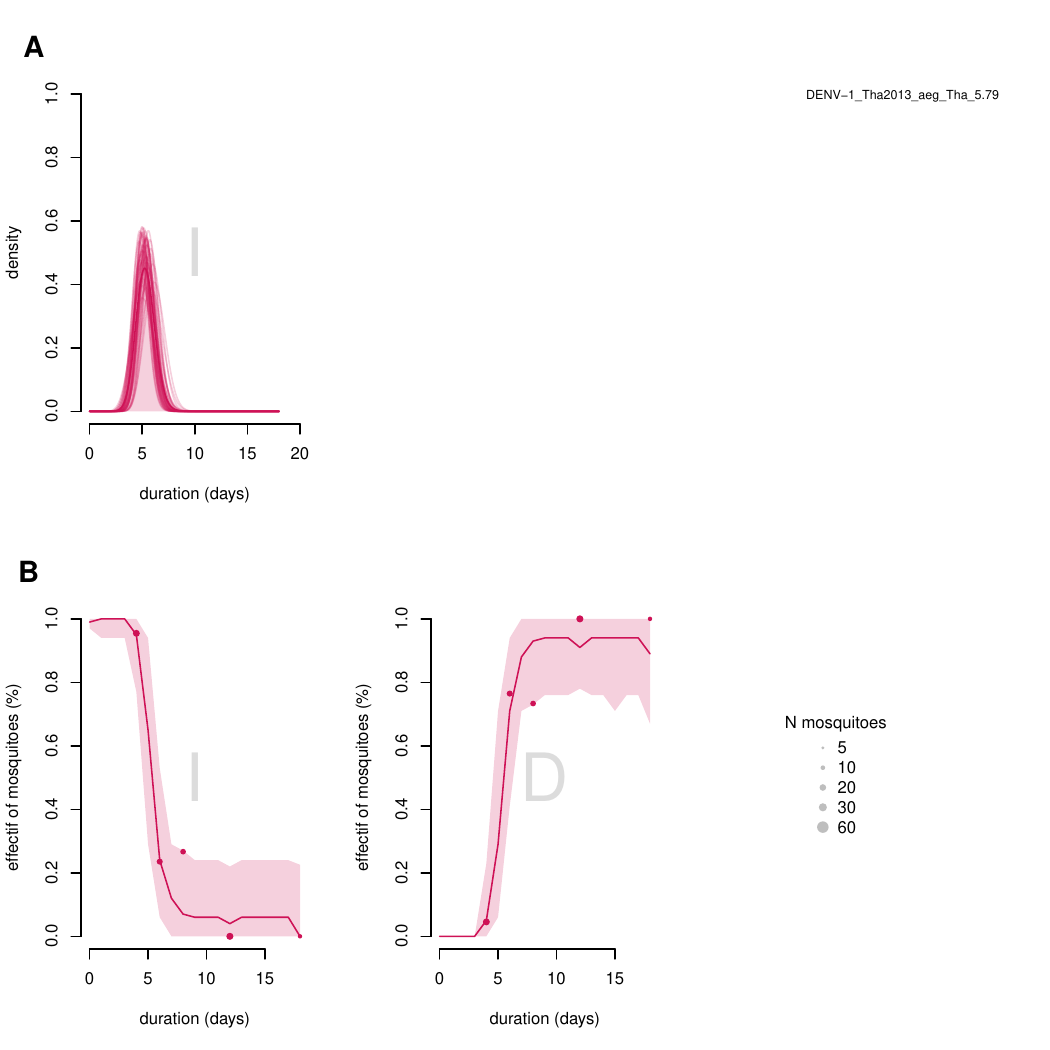}
\caption{Inference results for IVD stages distributions for scenario DENVp3(DENV1\_Tha2013\_aeg\_Tha\_5.79): \textit{Aedes.aegypti} from Thailand infected by dengue virus from Thailand with an infectious dose of 5.79 log10 FFU/mL : A) Selected distributions in the infected state for the main model selected. The dark line represents the mean of distributions and light lines represent a random sample of 50 distribution among all selected distribution. B) Selected dynamics in the infected (I) and disseminated (D) states for the main model selected. The dots represent the observed data, the line (mean dynamics), and the uncertainty ribbons (5\%-95\%) represent selected simulated dynamic.}
\label{fig_DENVp3}
\end{figure}

\begin{figure}[H]
\centering
\includegraphics[width=\textwidth]{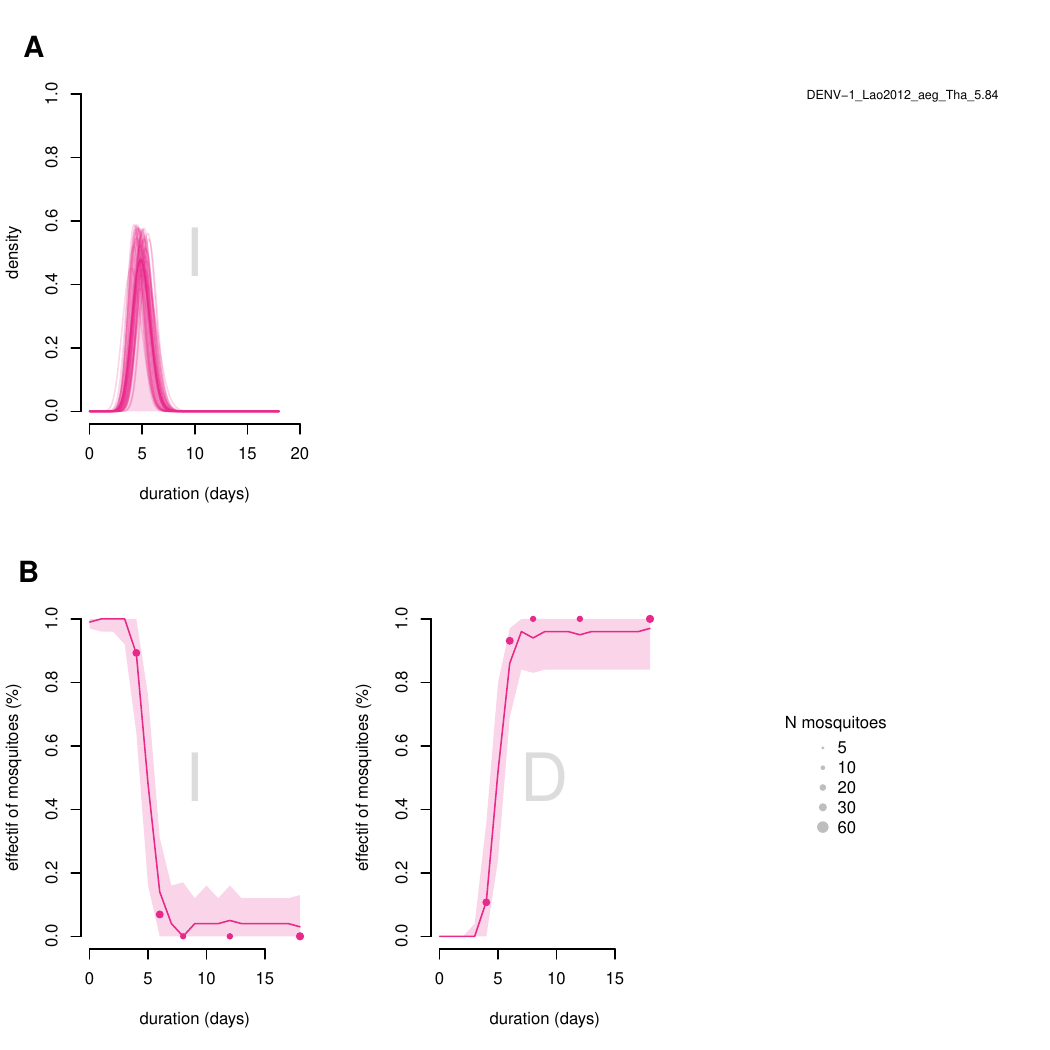}
\caption{Inference results for IVD stages distributions for scenario DENVp4(DENV1\_Lao2012\_aeg\_Tha\_5.84): \textit{Aedes.aegypti} from Laos infected by dengue virus from Thailand with an infectious dose of 5.84 log10 FFU/mL : A) Selected distributions in the infected state for the main model selected. The dark line represents the mean of distributions and light lines represent a random sample of 50 distribution among all selected distribution. B) Selected dynamics in the infected (I) and disseminated (D) states for the main model selected. The dots represent the observed data, the line (mean dynamics), and the uncertainty ribbons (5\%-95\%) represent selected simulated dynamic.}
\label{fig_DENVp4}
\end{figure}

\begin{figure}[H]
\centering
\includegraphics[width=\textwidth]{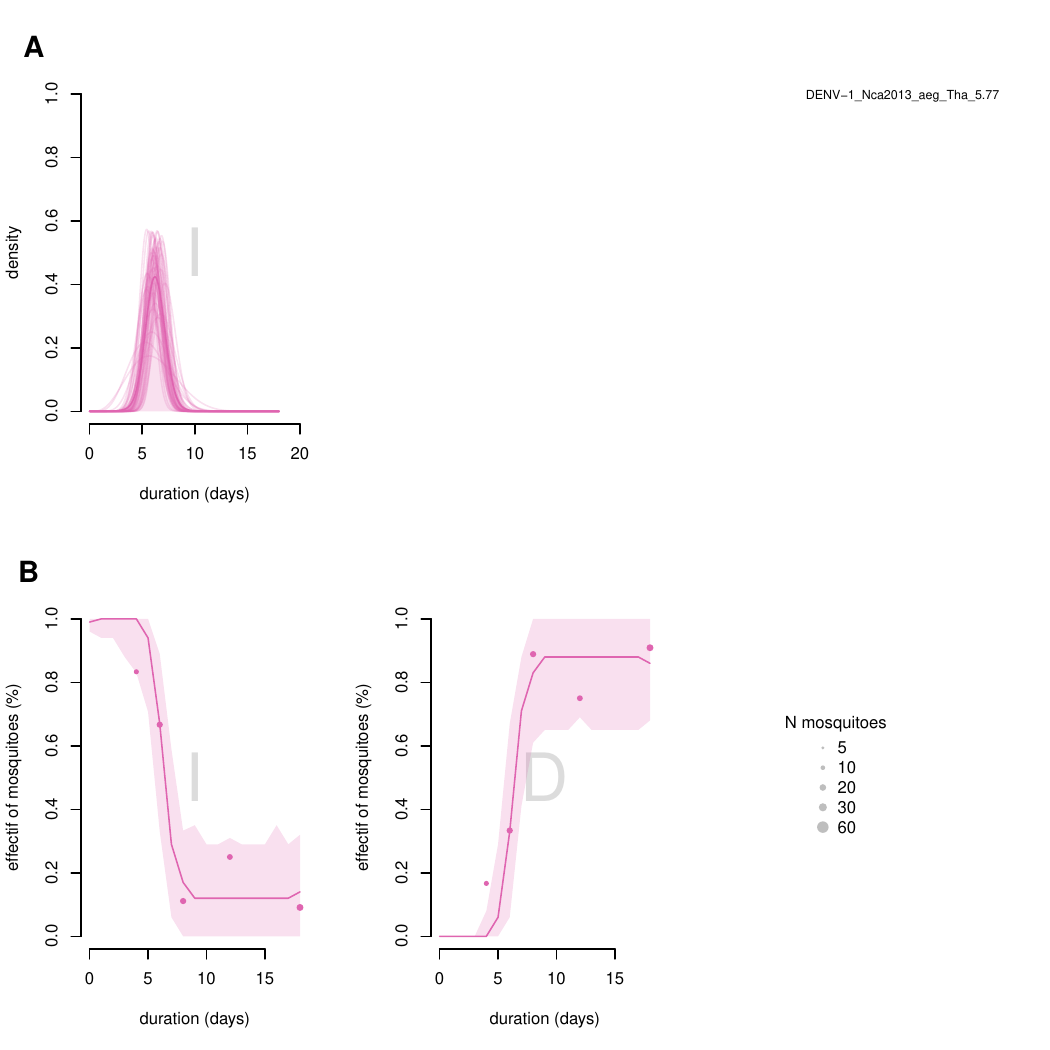}
\caption{Inference results for IVD stages distributions for scenario DENVp5(DENV1\_Nca2013\_aeg\_Tha\_5.77): \textit{Aedes.aegypti} from Thailand infected by dengue virus from New Caledonia with an infectious dose of 5.77 log10 FFU/mL : A) Selected distributions in the infected state for the main model selected. The dark line represents the mean of distributions and light lines represent a random sample of 50 distribution among all selected distribution. B) Selected dynamics in the infected (I) and disseminated (D) states for the main model selected. The dots represent the observed data, the line (mean dynamics), and the uncertainty ribbons (5\%-95\%) represent selected simulated dynamic.}
\label{fig_DENVp5}
\end{figure}

\begin{figure}[H]
\centering
\includegraphics[width=\textwidth]{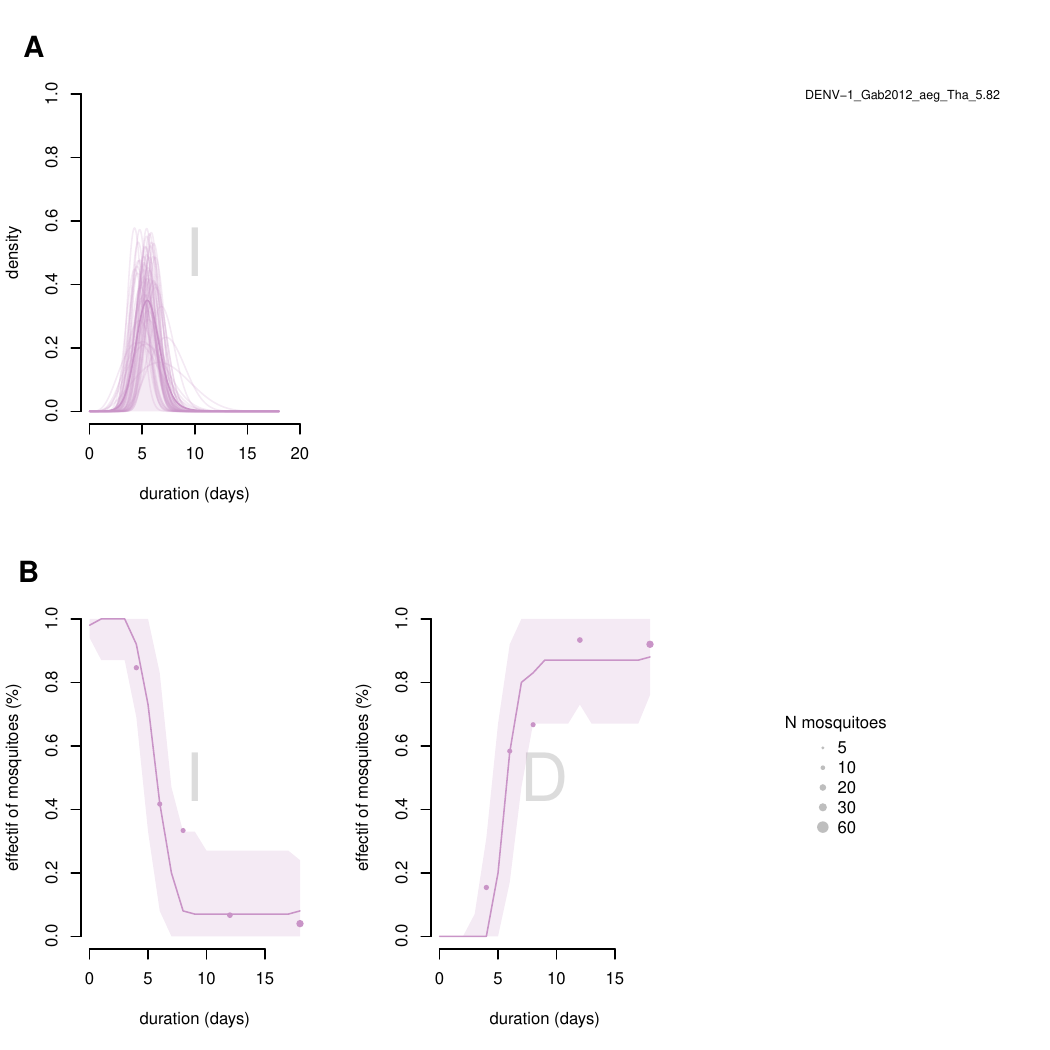}
\caption{Inference results for IVD stages distributions for scenario DENVp6(DENV1\_Gab2012\_aeg\_Tha\_5.82): \textit{Aedes.aegypti} from Thailand infected by dengue virus from Gabon with an infectious dose of 5.82 log10 FFU/mL : A) Selected distributions in the infected state for the main model selected. The dark line represents the mean of distributions and light lines represent a random sample of 50 distribution among all selected distribution. B) Selected dynamics in the infected (I) and disseminated (D) states for the main model selected. The dots represent the observed data, the line (mean dynamics), and the uncertainty ribbons (5\%-95\%) represent selected simulated dynamic.}
\label{fig_DENVp6}
\end{figure}

\begin{figure}[H]
\centering
\includegraphics[width=\textwidth]{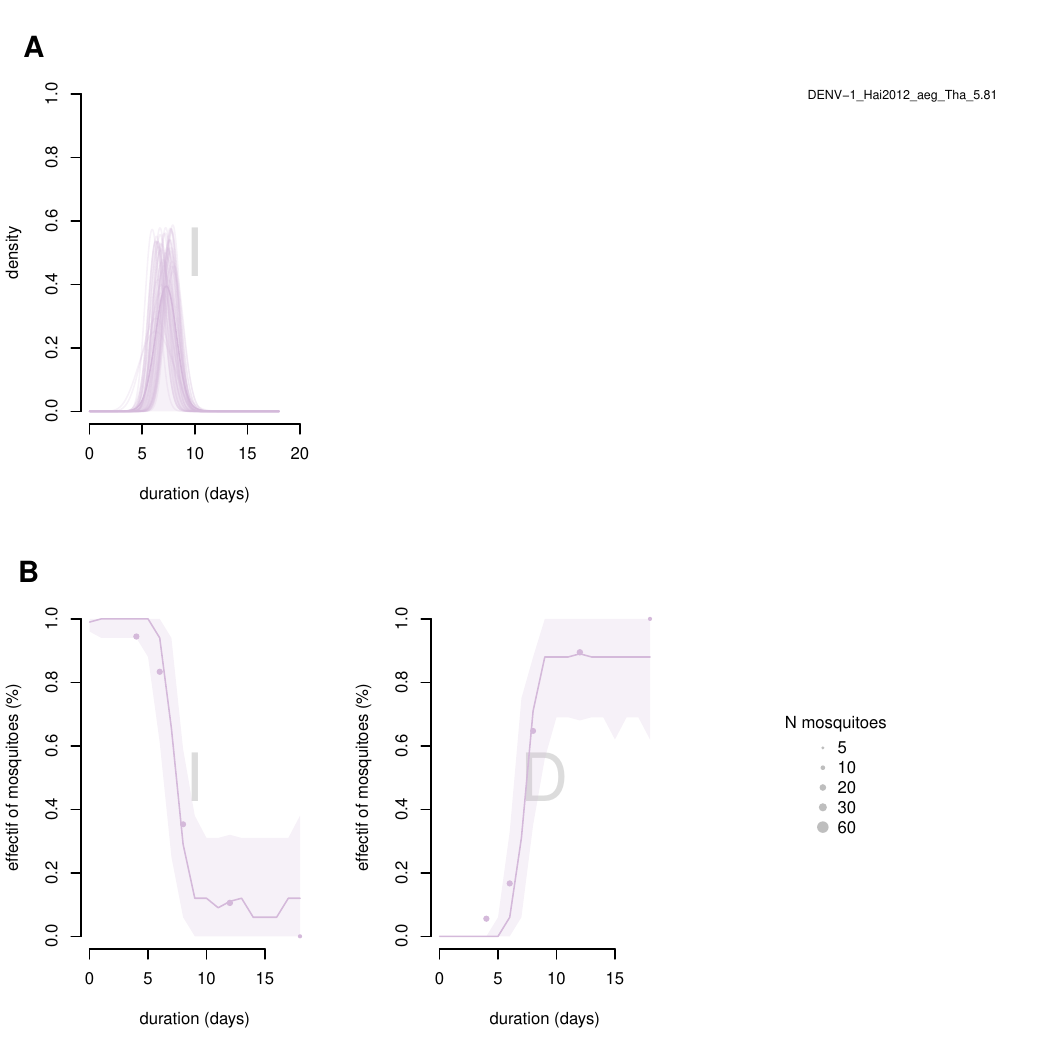}
\caption{Inference results for IVD stages distributions for scenario DENVp7(DENV1\_Hai2012\_aeg\_Tha\_5.81): \textit{Aedes.aegypti} from Thailand infected by dengue virus from Haïti with an infectious dose of 5.81 log10 FFU/mL : A) Selected distributions in the infected state for the main model selected. The dark line represents the mean of distributions and light lines represent a random sample of 50 distribution among all selected distribution. B) Selected dynamics in the infected (I) and disseminated (D) states for the main model selected. The dots represent the observed data, the line (mean dynamics), and the uncertainty ribbons (5\%-95\%) represent selected simulated dynamic.}
\label{fig_DENVp7}
\end{figure}

\begin{figure}[H]
\centering
\includegraphics[width=\textwidth]{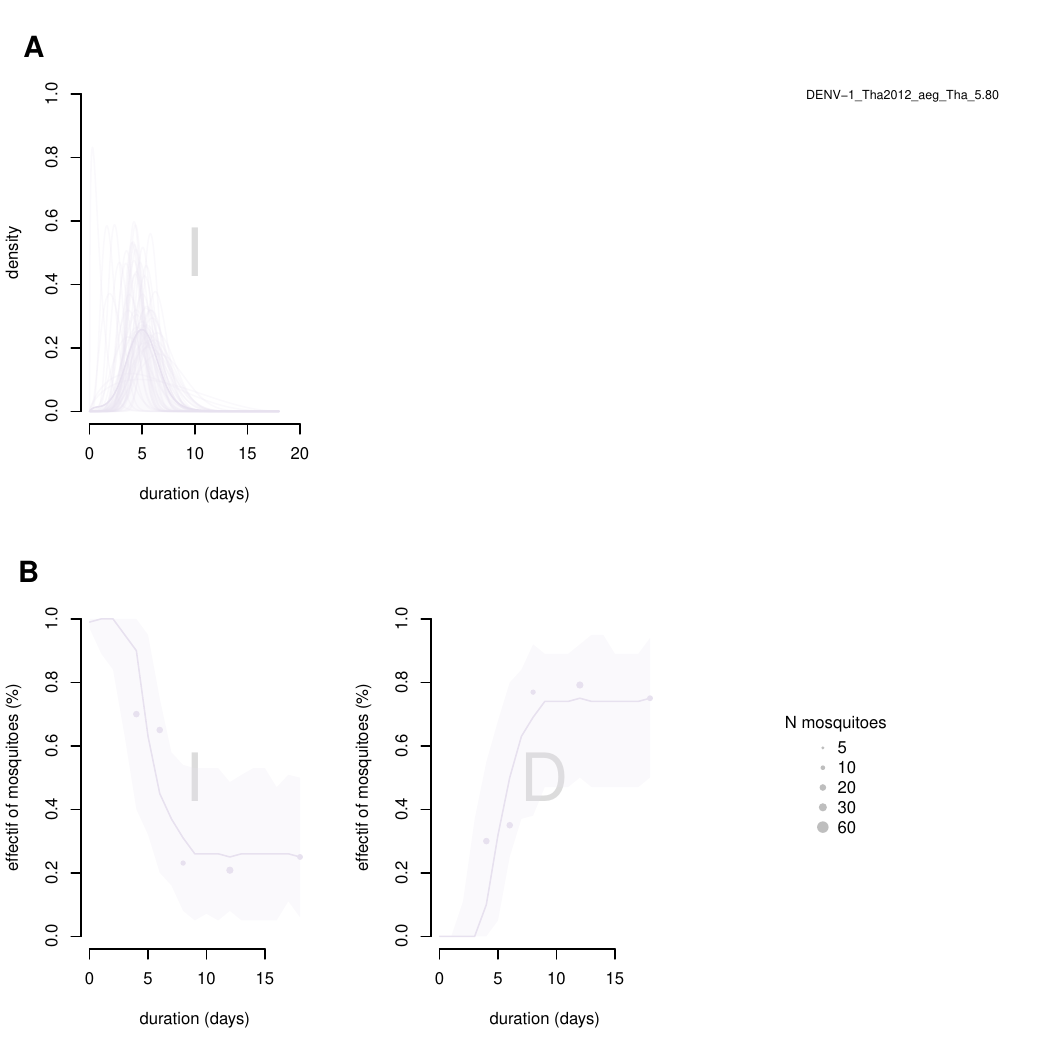}
\caption{Inference results for IVD stages distributions for scenario DENVp8(DENV1\_Tha2012\_aeg\_Tha\_5.80): \textit{Aedes.aegypti} from Thailand infected by dengue virus from Thailand with an infectious dose of 5.80 log10 FFU/mL : A) Selected distributions in the infected state for the main model selected. The dark line represents the mean of distributions and light lines represent a random sample of 50 distribution among all selected distribution. B) Selected dynamics in the infected (I) and disseminated (D) states for the main model selected. The dots represent the observed data, the line (mean dynamics), and the uncertainty ribbons (5\%-95\%) represent selected simulated dynamic.}
\label{fig_DENVp8}
\end{figure}

\begin{figure}[H]
\centering
\includegraphics[width=\textwidth]{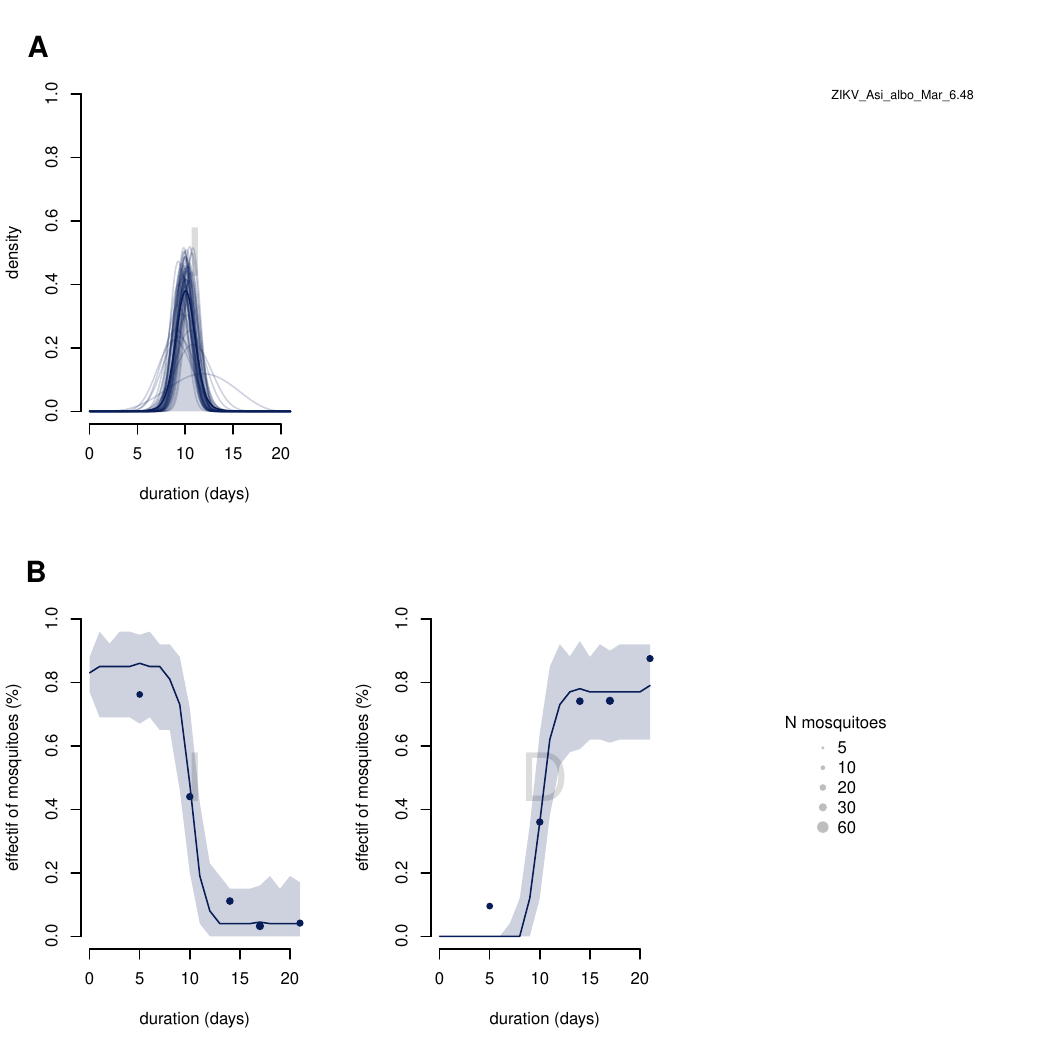}
\caption{Inference results for IVD stages distributions for scenario ZIKVp1(ZIKV\_Asi\_albo\_Mar\_6.48): \textit{Aedes.albopictus} from Marseille infected by Zika virus from Asia with an infectious dose of 6.48 log10 FFU/mL : A) Selected distributions in the infected state for the main model selected. The dark line represents the mean of distributions and light lines represent a random sample of 50 distribution among all selected distribution. B) Selected dynamics in the infected (I) and disseminated (D) states for the main model selected. The dots represent the observed data, the line (mean dynamics), and the uncertainty ribbons (5\%-95\%) represent selected simulated dynamic.}
\label{fig_ZIKVp1}
\end{figure}

\begin{figure}[H]
\centering
\includegraphics[width=\textwidth]{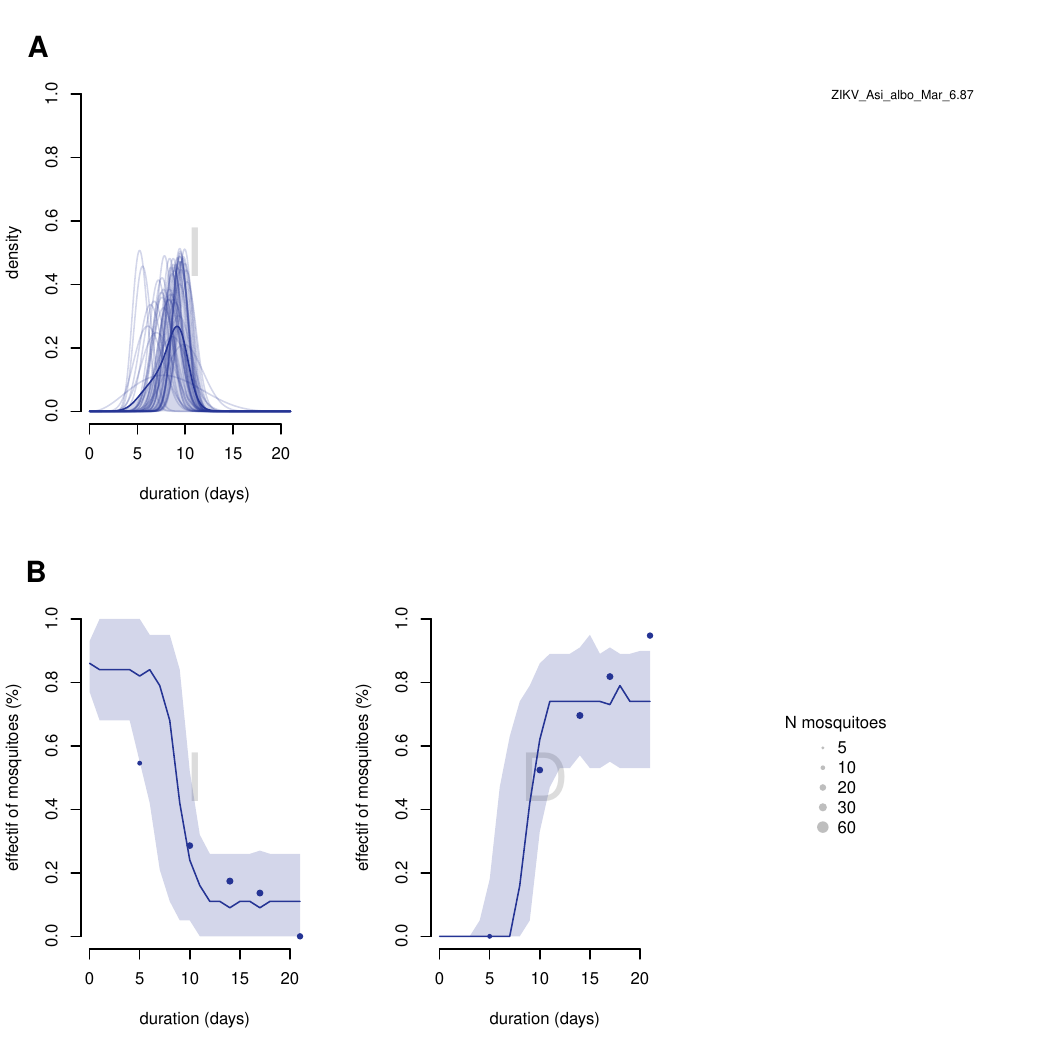}
\caption{Inference results for IVD stages distributions for scenario ZIKVp2(ZIKV\_Asi\_albo\_Mar\_6.87): \textit{Aedes.albopictus} from Marseille infected by Zika virus from Asia with an infectious dose of 6.87 log10 FFU/mL : A) Selected distributions in the infected state for the main model selected. The dark line represents the mean of distributions and light lines represent a random sample of 50 distribution among all selected distribution. B) Selected dynamics in the infected (I) and disseminated (D) states for the main model selected. The dots represent the observed data, the line (mean dynamics), and the uncertainty ribbons (5\%-95\%) represent selected simulated dynamic.}
\label{fig_ZIKVp2}
\end{figure}

\begin{figure}[H]
\centering
\includegraphics[width=\textwidth]{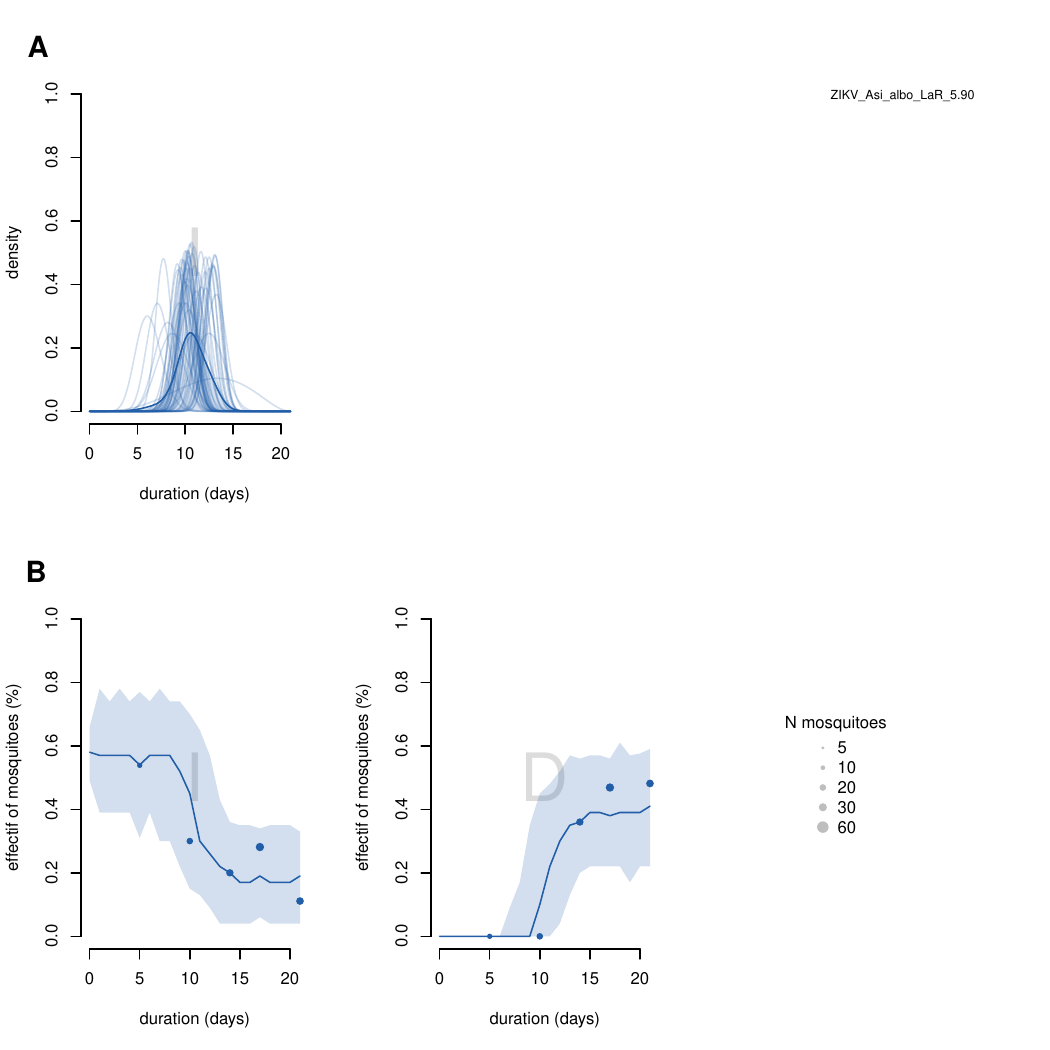}
\caption{Inference results for IVD stages distributions for scenario ZIKVp3(ZIKV\_Asi\_albo\_LaR\_5.90): \textit{Aedes.albopictus} from Reunion Island infected by Zika virus from Asia with an infectious dose of 5.90 log10 FFU/mL : A) Selected distributions in the infected state for the main model selected. The dark line represents the mean of distributions and light lines represent a random sample of 50 distribution among all selected distribution. B) Selected dynamics in the infected (I) and disseminated (D) states for the main model selected. The dots represent the observed data, the line (mean dynamics), and the uncertainty ribbons (5\%-95\%) represent selected simulated dynamic.}
\label{fig_ZIKVp3}
\end{figure}

\begin{figure}[H]
\centering
\includegraphics[width=\textwidth]{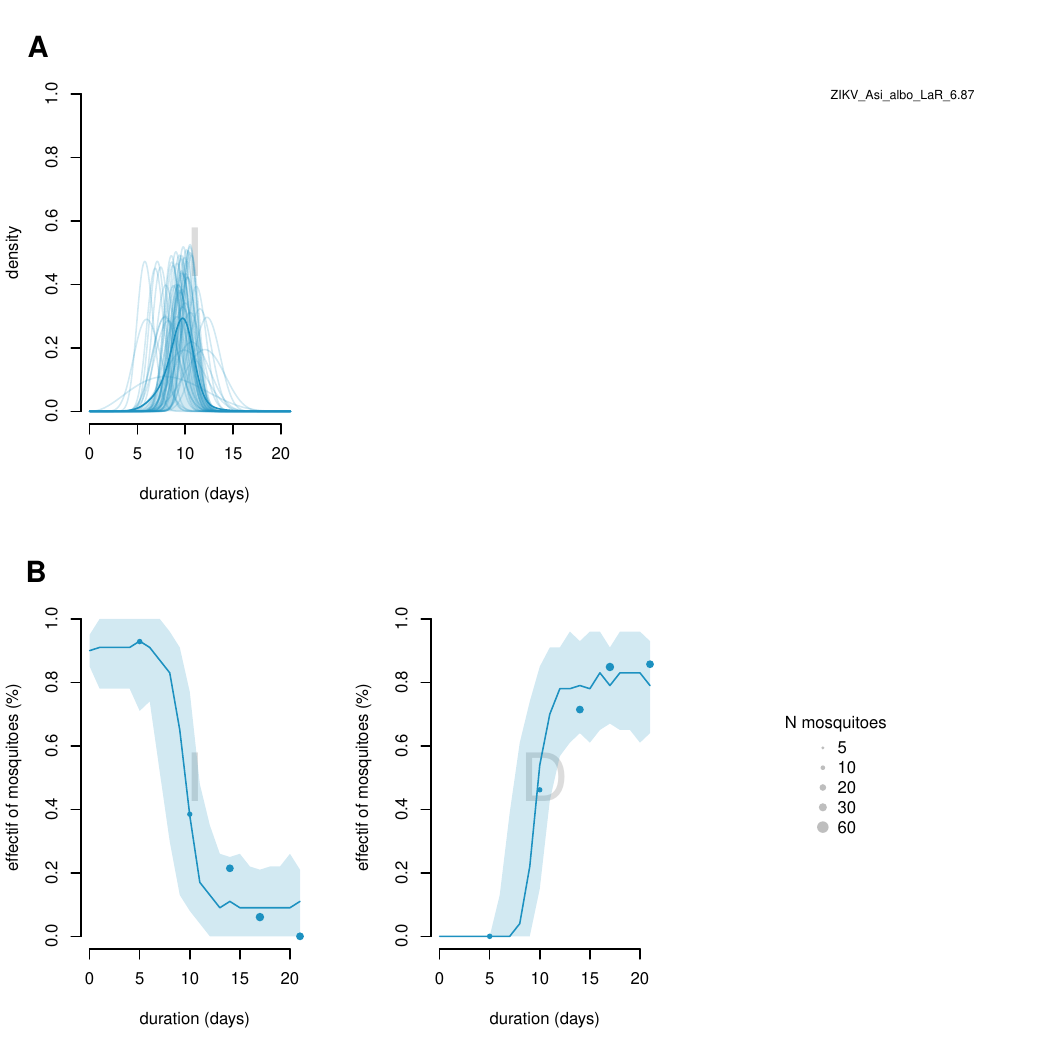}
\caption{Inference results for IVD stages distributions for scenario ZIKVp4(ZIKV\_Asi\_albo\_LaR\_6.87): \textit{Aedes.albopictus} from Reunion Island infected by Zika virus from Asia with an infectious dose of 6.87 log10 FFU/mL : A) Selected distributions in the infected state for the main model selected. The dark line represents the mean of distributions and light lines represent a random sample of 50 distribution among all selected distribution. B) Selected dynamics in the infected (I) and disseminated (D) states for the main model selected. The dots represent the observed data, the line (mean dynamics), and the uncertainty ribbons (5\%-95\%) represent selected simulated dynamic.}
\label{fig_ZIKVp4}
\end{figure}

\begin{figure}[H]
\centering
\includegraphics[width=\textwidth]{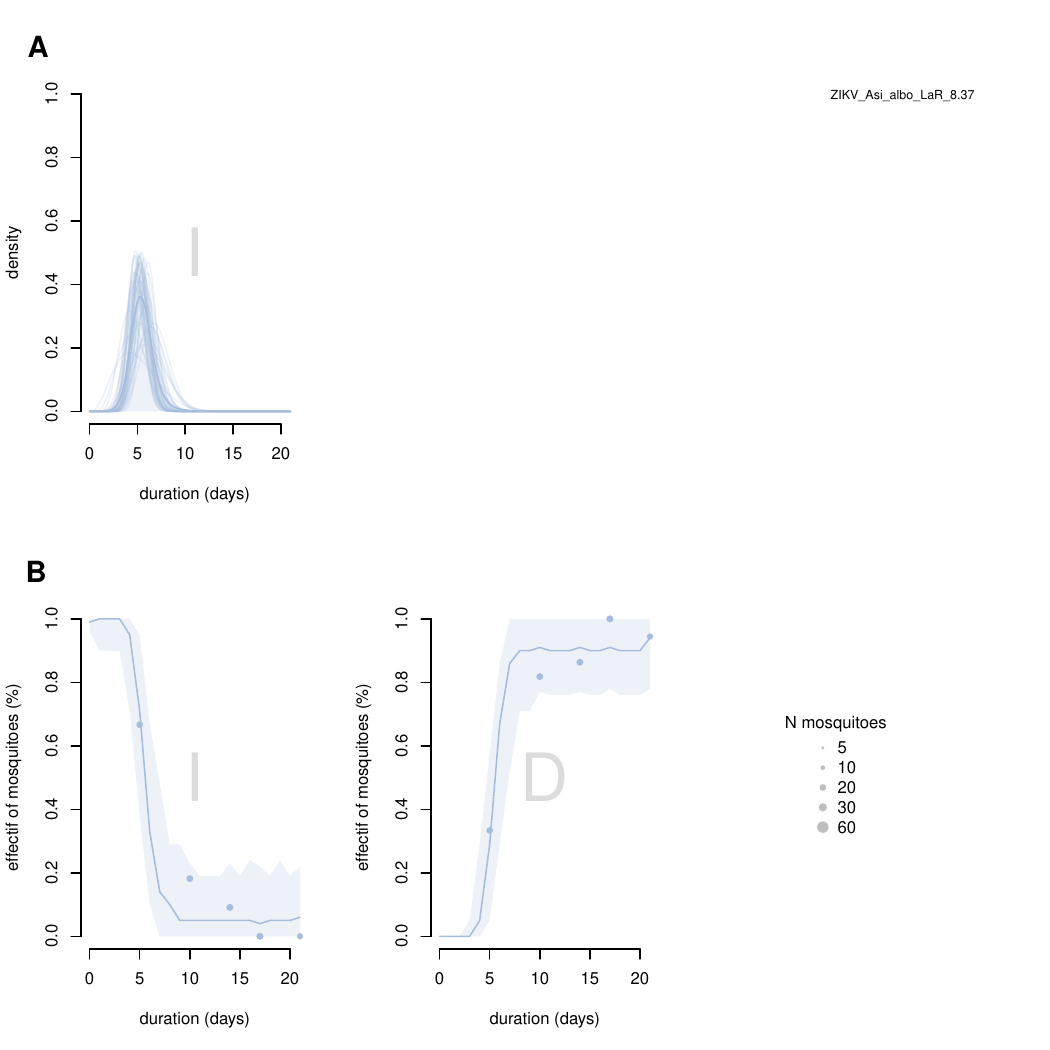}
\caption{Inference results for IVD stages distributions for scenario ZIKVp5(ZIKV\_Asi\_albo\_LaR\_8.37): \textit{Aedes.albopictus} from Reunion Island infected by Zika virus from Asia with an infectious dose of 8.37 log10 FFU/mL : A) Selected distributions in the infected state for the main model selected. The dark line represents the mean of distributions and light lines represent a random sample of 50 distribution among all selected distribution. B) Selected dynamics in the infected (I) and disseminated (D) states for the main model selected. The dots represent the observed data, the line (mean dynamics), and the uncertainty ribbons (5\%-95\%) represent selected simulated dynamic.}
\label{fig_ZIKVp5}
\end{figure}

\begin{figure}[H]
\centering
\includegraphics[width=\textwidth]{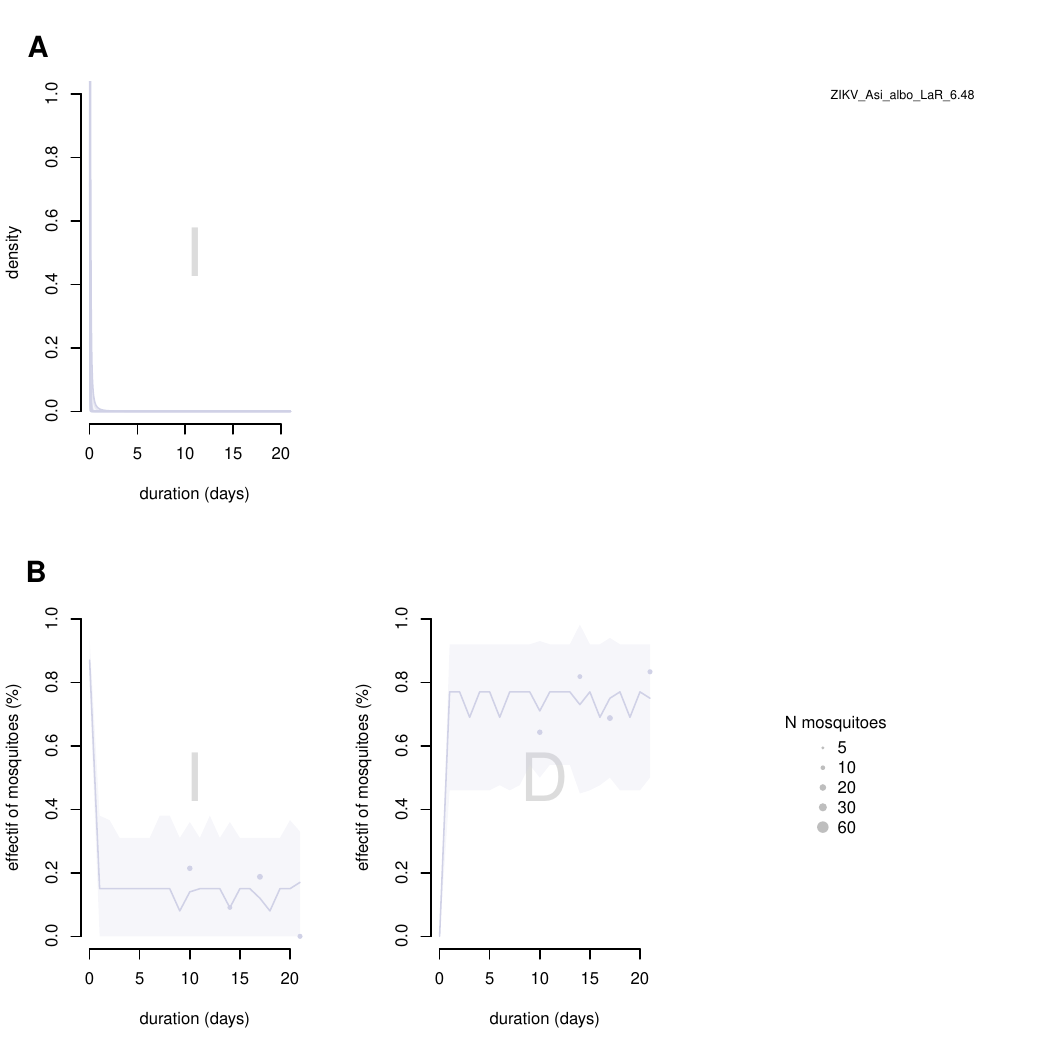}
\caption{Inference results for IVD stages distributions for scenario ZIKVp6(ZIKV\_Asi\_albo\_LaR\_6.48): \textit{Aedes.albopictus} from Reunion Island infected by Zika virus from Asia with an infectious dose of 6.48 log10 FFU/mL : A) Selected distributions in the infected state for the main model selected. The dark line represents the mean of distributions and light lines represent a random sample of 50 distribution among all selected distribution. B) Selected dynamics in the infected (I) and disseminated (D) states for the main model selected. The dots represent the observed data, the line (mean dynamics), and the uncertainty ribbons (5\%-95\%) represent selected simulated dynamic.}
\label{fig_ZIKVp6}
\end{figure}

\clearpage

\subsection{Statistical Analysis of Results} 
\begin{center}
    \vspace*{\fill} 
    {\Huge \bfseries Statistical Analysis of Results} 
    \vspace*{\fill} 
\end{center}

\clearpage

\begin{sidewaystable}[!hp]
\centering
\caption{Results of inference for each crossing barriers parameters for each scenario with the SEIDT model. Scenarios with IC 90\% < 0,15  are shown in bold (16/26 for \(\gamma_I\), 8/26 for \(\gamma_D\) and 2/26 for \(\gamma_T\)). } 
\resizebox{\textwidth}{!}{%
\begin{tabular}{@{}llllllllll@{}}
\toprule
Short name & Mode \(\gamma_I\) {[}IC 90\%{]} & Size IC 90\% & \(\gamma_I\) from litterature & Mode \(\gamma_D\) {[}IC 90\%{]} & Size IC 90\% & \(\gamma_D\) from litterature & Mode \(\gamma_T\) {[}IC 90\%{]} & Size IC 90\% & \(\gamma_T\) from litterature \\
\midrule
CHIKVc1 & 0.84 {[} 0.76 , 0.94 {]} & 0.18 & 0.93 & 0.68 {[} 0.54 , 0.85 {]} & 0.32 & 0.58 & 0.24 {[} 0.12 , 0.48 {]} & 0.36 & 0.37 \\
CHIKVc2 & 0.99 {[} 0.97 , 1 {]} & \textbf{0.02} & 0.99 & 0.94 {[} 0.88 , 0.99 {]} & \textbf{0.11} & 0.85 & 0.25 {[} 0.17 , 0.33 {]} & 0.15 & 0.20 \\
CHIKVc3 & 0.98 {[} 0.95 , 1 {]} & \textbf{0.04} & 1.00 & 0.82 {[} 0.73 , 0.91 {]} & 0.18 & 0.61 & 0.62 {[} 0.52 , 0.75 {]} & 0.22 & 0.54 \\
CHIKVc4* & 0.85 {[} 0.76 , 0.9 {]} & \textbf{0.14} & 0.83 & 0.75 {[} 0.65 , 0.83 {]} & 0.17 & 0.76 & 0.6 {[} 0.51 , 0.7 {]} & 0.19 & 0.61 \\
CHIKVc5* & 0.8 {[} 0.74 , 0.86 {]} & \textbf{0.12} & 0.81 & 0.92 {[} 0.86 , 0.97 {]} & \textbf{0.11} & 0.91 & 0.45 {[} 0.35 , 0.55 {]} & 0.20 & 0.45 \\
CHIKVc6* & 0.98 {[} 0.94 , 1 {]} & \textbf{0.06} & 0.97 & 0.99 {[} 0.95 , 1 {]} & \textbf{0.04} & 0.98 & 0.33 {[} 0.25 , 0.42 {]} & 0.17 & 0.33 \\
CHIKVc7* & 0.99 {[} 0.94 , 1 {]} & \textbf{0.06} & 0.98 & 0.96 {[} 0.88 , 0.99 {]} & \textbf{0.11} & 0.94 & 0.4 {[} 0.32 , 0.49 {]} & 0.17 & 0.40 \\
CHIKVc8* & 0.99 {[} 0.96 , 1 {]} & \textbf{0.03} & 1.00 & 0.99 {[} 0.94 , 1 {]} & \textbf{0.06} & 0.98 & 0.39 {[} 0.3 , 0.5 {]} & 0.20 & 0.40 \\
CHIKVc9* & 1 {[} 0.98 , 1 {]} & \textbf{0.02} & 0.99 & 1 {[} 0.99 , 1 {]} & \textbf{0.01} & 1 & 0.1 {[} 0.06 , 0.13 {]} & \textbf{0.08} & 0.09 \\
DENVc1 & 0.53 {[} 0.45 , 0.61 {]} & 0.17 & 0.65 & 0.78 {[} 0.56 , 0.92 {]} & 0.36 & 0.53 & 0.21 {[} 0.04 , 0.88 {]} & 0.84 & 0.18 \\
DENVc2 & 0.23 {[} 0.19 , 0.3 {]} & \textbf{0.11} & 0.26 & 0.84 {[} 0.53 , 0.98 {]} & 0.45 & 0.49 & 0.23 {[} 0.05 , 0.91 {]} & 0.86 & 0.17 \\
DENVc3* & 0.18 {[} 0.1 , 0.26 {]} & 0.16 & 0.22 & 0.73 {[} 0.14 , 0.96 {]} & 0.82 & 0.60 & 0.73 {[} 0.05 , 0.92 {]} & 0.87 & 0.44 \\
DENVc4 & 0.92 {[} 0.86 , 0.96 {]} & \textbf{0.11} & 0.91 & 0.86 {[} 0.73 , 0.95 {]} & 0.22 & 0.52 & 0.39 {[} 0.17 , 0.78 {]} & 0.61 & 0.15 \\
DENVc5 & 0.9 {[} 0.83 , 0.95 {]} & \textbf{0.12} & 0.92 & 0.95 {[} 0.77 , 0.99 {]} & 0.23 & 0.61 & 0.17 {[} 0.03 , 0.87 {]} & 0.84 & 0.11 \\
DENVc6 & 0.28 {[} 0.21 , 0.39 {]} & 0.18 & 0.30 & 0.79 {[} 0.52 , 0.96 {]} & 0.44 & 0.48 & 0.18 {[} 0.04 , 0.61 {]} & 0.57 & 0.30 \\
DENVc7* & 0.15 {[} 0.08 , 0.24 {]} & 0.15 & 0.18 & 0.54 {[} 0.24 , 0.93 {]} & 0.69 & 0.62 & 0.29 {[} 0.07 , 0.95 {]} & 0.87 & 0.38 \\
ZIKVc1 & 0.99 {[} 0.98 , 1 {]} & \textbf{0.02} & 0.99 & 0.99 {[} 0.95 , 1 {]} & \textbf{0.05} & 0.96 & 0.39 {[} 0.32 , 0.49 {]} & 0.17 & 0.29 \\
ZIKVc2 & 0.9 {[} 0.85 , 0.94 {]} & \textbf{0.09} & 0.90 & 0.96 {[} 0.86 , 0.99 {]} & \textbf{0.13} & 0.75 & 0.76 {[} 0.56 , 0.94 {]} & 0.38 & 0.44 \\
ZIKVc3 & 0.45 {[} 0.37 , 0.55 {]} & 0.18 & 0.48 & 0.66 {[} 0.51 , 0.82 {]} & 0.31 & 0.60 & 0.74 {[} 0.43 , 0.95 {]} & 0.53 & 0.37 \\
ZIKVc4 & 0.78 {[} 0.7 , 0.84 {]} & \textbf{0.14} & 0.83 & 0.65 {[} 0.49 , 0.78 {]} & 0.29 & 0.47 & 0.21 {[} 0.03 , 0.9 {]} & 0.87 & 0.12 \\
ZIKVc5 & 0.9 {[} 0.85 , 0.94 {]} & \textbf{0.09} & 0.91 & 0.6 {[} 0.45 , 0.75 {]} & 0.31 & 0.37 & 0.25 {[} 0.03 , 0.89 {]} & 0.86 & 0.04 \\
ZIKVc6* & 0.47 {[} 0.39 , 0.55 {]} & 0.17 & 0.50 & 0.08 {[} 0.01 , 0.33 {]} & 0.32 & 0.16 & 0.27 {[} 0.05 , 0.93 {]} & 0.88 & 0.22 \\
ZIKVc7* & 0.37 {[} 0.31 , 0.43 {]} & \textbf{0.12} & 0.38 & 0.35 {[} 0.21 , 0.63 {]} & 0.42 & 0.27 & 0.36 {[} 0.05 , 0.92 {]} & 0.88 & 0.25 \\
ZIKVc8* & 0.78 {[} 0.71 , 0.87 {]} & 0.15 & 0.82 & 0.39 {[} 0.22 , 0.63 {]} & 0.41 & 0.25 & 0.39 {[} 0.03 , 0.94 {]} & 0.90 & 0.05 \\
ZIKVc9* & 0.5 {[} 0.42 , 0.59 {]} & 0.17 & 0.53 & 0.46 {[} 0.34 , 0.71 {]} & 0.37 & 0.45 & 0.18 {[} 0.03 , 0.44 {]} & 0.41 & 0.23 \\
ZIKVc10* & 0.69 {[} 0.57 , 0.77 {]} & 0.20 & 0.69 & 0.39 {[} 0.27 , 0.51 {]} & 0.24 & 0.39 & 0.03 {[} 0.01 , 0.11 {]} & \textbf{0.11} & 0.03 \\
\bottomrule \\
\end{tabular}

}
* scenarios for which the infection, dissemination and transmission rates, taken from articles, were within the 90\% CI obtained for the barrier crossing parameters \(\gamma_I\), \(\gamma_D\), and \(\gamma_T\)

\label{table_params_SEIDT}
\end{sidewaystable}

\FloatBarrier

\begin{table}[!hp]
\centering
\caption{Results of inference for each crossing barriers parameters for each scenario with the SEID model. Scenarios with IC 90\% < 0,15 are shown in bold ( 14/17 for \(\gamma_I\) and 7/17 for \(\gamma_D\)).} 
\begin{tabular}{@{}lllll@{}}
\toprule
Short name  & Mode \(\gamma_I\) {[}IC 90\%{]} & Size IC 90\% & Mode \(\gamma_D\) {[}IC 90\%{]} & Size IC 90\%\\
\midrule
CHIKVp1 & 0.02 {[} 0 , 0.04 {]} & \textbf{0.04} & 0.6 {[} 0.07 , 0.93 {]} & 0.86 \\
CHIKVp2 & 0.72 {[} 0.66 , 0.79 {]} & \textbf{0.13} & 0.96 {[} 0.86 , 0.99 {]} & \textbf{0.13} \\
CHIKVp3 & 0.99 {[} 0.95 , 1 {]} & \textbf{0.05} & 0.99 {[} 0.92 , 1 {]} & \textbf{0.07} \\
DENVp1 & 0.99 {[} 0.96 , 1 {]} & \textbf{0.06} & 0.98 {[} 0.9 , 1 {]} & \textbf{0.10} \\
DENVp2 & 0.98 {[} 0.94 , 1 {]} & \textbf{0.07} & 0.98 {[} 0.89 , 1 {]} & \textbf{0.11} \\
DENVp3 & 1 {[} 0.97 , 1 {]} & \textbf{0.08} & 0.94 {[} 0.83 , 0.99 {]} & 0.16 \\
DENVp4 & 0.99 {[} 0.97 , 1 {]} & \textbf{0.09} & 0.97 {[} 0.91 , 1 {]} & \textbf{0.09} \\
DENVp5 & 0.99 {[} 0.96 , 1 {]} & \textbf{0.10} & 0.89 {[} 0.76 , 0.95 {]} & 0.19 \\
DENVp6 & 0.98 {[} 0.94 , 1 {]} & \textbf{0.11} & 0.91 {[} 0.8 , 0.98 {]} & 0.18 \\
DENVp7 & 0.99 {[} 0.96 , 1 {]} & \textbf{0.12} & 0.93 {[} 0.75 , 0.99 {]} & 0.23 \\
DENVp8 & 0.99 {[} 0.97 , 1 {]} & \textbf{0.13} & 0.74 {[} 0.58 , 0.87 {]} & 0.29 \\
ZIKVp1 & 0.83 {[} 0.77 , 0.88 {]} & \textbf{0.14} & 0.96 {[} 0.85 , 0.99 {]} & \textbf{0.13} \\
ZIKVp2 & 0.87 {[} 0.77 , 0.93 {]} & 0.15 & 0.88 {[} 0.77 , 0.95 {]} & 0.18 \\
ZIKVp3 & 0.57 {[} 0.49 , 0.66 {]} & 0.17 & 0.67 {[} 0.56 , 0.8 {]} & 0.24 \\
ZIKVp4 & 0.91 {[} 0.85 , 0.95 {]} & \textbf{0.10} & 0.89 {[} 0.81 , 0.96 {]} & \textbf{0.14} \\
ZIKVp5 & 0.99 {[} 0.96 , 1 {]} & \textbf{0.04} & 0.93 {[} 0.84 , 0.99 {]} & 0.15 \\
ZIKVp6 & 0.86 {[} 0.79 , 0.94 {]} & 0.16 & 0.84 {[} 0.74 , 0.93 {]} & 0.19 \\
\bottomrule \\
\end{tabular}
\label{table_params_SEID}
\end{table}
\FloatBarrier

\centering
\begin{table}[!hp]
\caption{ Results of Wilcoxon test to study superiority to 0.9 of \(\gamma_I\), \(\gamma_D\) and \(\gamma_T\) values (SEIDT model). P-values < 0,05 corresponding to scenarios for which parameters values are statistically > 0.9  are shown in bold ( 8/26 for \(\gamma_I\) 9/26 for \(\gamma_D\) and 0/26 for \(\gamma_T\)) .} 
\begin{tabular}{llll}
\toprule
scenario & \(\gamma_I\)pvalue  & \(\gamma_D\)pvalue  & \(\gamma_T\)pvalue  \\
\midrule 
CHIKV\_LaR\_albo\_Rab\_7 & 1 & 1 & 1 \\
CHIKV\_FrCarIs\_aeg\_Tha\_6 & \textbf{3.1947121779714e-117} & \textbf{1.61599883470573e-94} & 1 \\
CHIKV\_Ind\_gen\_Tir\_8 & \textbf{1.79395945393212e-105} & 1 & 1 \\
CHIKV\_Tah\_aeg\_Tah\_7 & 1 & 1 & 1 \\
CHIKV\_LaR\_albo\_Tun\_7 & 1 & \textbf{8.30420846567072e-14} & 1 \\
CHIKV\_Ind\_albo\_Tir\_8 & \textbf{6.96294850737994e-86} & \textbf{6.96294850737994e-86} & 1 \\
CHIKV\_Ncal\_aeg\_Mad\_Fun\_7.3 & \textbf{1.95693036570352e-91} & \textbf{3.39955867927649e-77} & 1 \\
CHIKV\_Ncal\_aeg\_Mad\_Pa\_do\_Ma\_7.3 & \textbf{1.4224331828238e-108} & \textbf{1.71069404376581e-108} & 1 \\
CHIKV\_BriVirIsl\_aeg\_PozRic\_6.9 & \textbf{1.12833022090229e-111} & \textbf{1.12833022090229e-111} & 1 \\
DENV-2\_Bang\_albo\_Rab\_7 & 1 & 1 & 1 \\
DENV-2\_Bang\_albo\_Tun\_7 & 1 & 1 & 1 \\
DENV-1\_Sing\_albo\_Rey\_5 & 1 & 1 & 1 \\
DENV-3\_Cair\_1998\_aeg\_Cair\_4.7 & \textbf{6.6603154179305e-17} & 1 & 1 \\
DENV-3\_Cair\_2008\_aeg\_Cair\_4.9 & 0.999968405140845 & \textbf{0.000121915611642587} & 1 \\
DENV-1\_Sing\_albo\_Rom\_5 & 1 & 1 & 1 \\
DENV-1\_Sing\_albo\_Mon\_5 & 1 & 1 & 1 \\
ZIKV\_PueRic\_aeg\_PozRic\_7.2 & \textbf{1.4734456821593e-132} & \textbf{1.4790011082154e-132} & 1 \\
ZIKV\_FrP\_aeg\_Tah\_6.8 & 0.997989130148858 & \textbf{1.54124620229698e-76} & 1 \\
ZIKV\_Ugan\_aeg\_Tow\_6.5 & 1 & 1 & 1 \\
ZIKV\_Ncal\_aeg\_FrP\_7 & 1 & 1 & 1 \\
ZIKV\_Ncal\_pol\_Wal\_7 & 0.602811501831589 & 1 & 1 \\
ZIKV\_Ncal\_albo\_Rab\_7.2 & 1 & 1 & 1 \\
ZIKV\_Ncal\_aeg\_Sam\_7 & 1 & 1 & 1 \\
ZIKV\_Ncal\_aeg\_Ncal\_7 & 1 & 1 & 1 \\
ZIKV\_Ncal\_albo\_Tun\_7 & 1 & 1 & 1 \\
ZIKV\_Ncal\_pol\_FrP\_7 & 1 & 1 & 1 \\
\bottomrule \\
\end{tabular}
\label{table_wilcox_0.9_SEIDT}
\end{table}

\FloatBarrier

\begin{table}[!hp]
\centering
\caption{Results of Wilcoxon test to study superiority to 0.9 of \(\gamma_I\) and \(\gamma_D\) values (SEID model). P-values < 0,05 corresponding to scenarios for which parameters values are statistically > 0.9  are shown in bold ( 10/17 for \(\gamma_I\) and 8/17 for \(\gamma_D\)).} 
\begin{tabular}{lll}
\toprule
scenario & \(\gamma_I\)pvalue  & \(\gamma_D\)pvalue \\
\midrule 
CHIKV\_IndOce\_albo\_Lyo\_3.94 & 1 & 1 \\
CHIKV\_IndOce\_albo\_Lyo\_6.07 & 1 & \textbf{1.55002179699219e-60} \\
CHIKV\_IndOce\_albo\_Lyo\_8.63 & \textbf{6.64172503377505e-93} & \textbf{1.44714556069923e-92} \\
DENV-1\_Tha2010a\_aeg\_Tha\_5.74 & \textbf{2.99661165023998e-100} & \textbf{6.12994676643568e-93} \\
DENV-1\_Tha2010b\_aeg\_Tha\_5.70 & \textbf{2.99661165023998e-100} & \textbf{4.42176195492722e-90} \\
DENV-1\_Tha2013\_aeg\_Tha\_5.79 & \textbf{2.99661165023998e-100} & \textbf{4.57944374121786e-27} \\
DENV-1\_Lao2012\_aeg\_Tha\_5.84 & \textbf{2.99661165023998e-100} & \textbf{3.3115204412706e-96} \\
DENV-1\_Nca2013\_aeg\_Tha\_5.77 & \textbf{2.99661165023998e-100} & 1 \\
DENV-1\_Gab2012\_aeg\_Tha\_5.82 & \textbf{3.04196141838421e-100} & 0.900022357291413 \\
DENV-1\_Hai2012\_aeg\_Tha\_5.81 & \textbf{2.99661165023998e-100} & 0.998686850092147 \\
DENV-1\_Tha2012\_aeg\_Tha\_5.80 & \textbf{2.7248529270144e-98} & 1 \\
ZIKV\_Asi\_albo\_Mar\_6.48 & 1 & \textbf{1.5139955770437e-47} \\
ZIKV\_Asi\_albo\_Mar\_6.87 & 1 & 1 \\
ZIKV\_Asi\_albo\_LaR\_5.90 & 1 & 1 \\
ZIKV\_Asi\_albo\_LaR\_6.87 & 0.519010427774142 & 0.999999926511964 \\
ZIKV\_Asi\_albo\_LaR\_8.37 & \textbf{2.99661165023998e-100} & \textbf{1.3348255131e-34} \\
ZIKV\_Asi\_albo\_LaR\_6.48 & 1 & 1 \\
\bottomrule \\
\end{tabular}%
\label{table_wilcox_0.9_SEID}
\end{table}
\FloatBarrier

\begin{table}[!hp]
\centering
\caption{ Results of Wilcoxon test to study inferiority to 0.5 of \(\gamma_I\), \(\gamma_D\) and \(\gamma_T\) values (SEIDT model). P-values < 0,05 corresponding to scenarios for which parameters values are statistically < 0.5  are shown in bold ( 7/26 for \(\gamma_I\) 4/26 for \(\gamma_D\) and 19/26 for \(\gamma_T\) .}
\begin{tabular}{llll}
\toprule
scenario & \(\gamma_I\)pvalue & \(\gamma_D\)pvalue  & \(\gamma_T\)pvalue  \\
\midrule 
CHIKV\_LaR\_albo\_Rab\_7 & 1 & 1 & \textbf{2.56818261412697e-125} \\
CHIKV\_FrCarIs\_aeg\_Tha\_6 & 1 & 1 & \textbf{3.1947121779714e-117} \\
CHIKV\_Ind\_gen\_Tir\_8 & 1 & 1 & 1 \\
CHIKV\_Tah\_aeg\_Tah\_7 & 1 & 1 & 1 \\
CHIKV\_LaR\_albo\_Tun\_7 & 1 & 1 & \textbf{5.05604664244194e-36} \\
CHIKV\_Ind\_albo\_Tir\_8 & 1 & 1 & \textbf{6.96294850737994e-86} \\
CHIKV\_Ncal\_aeg\_Mad\_Fun\_7.3 & 1 & 1 & \textbf{2.86977111117595e-89} \\
CHIKV\_Ncal\_aeg\_Mad\_Pa\_do\_Ma\_7.3 & 1 & 1 & \textbf{1.0642352774086e-105} \\
CHIKV\_BriVirIsl\_aeg\_PozRic\_6.9 & 1 & 1 & \textbf{1.12833022090229e-111} \\
DENV-2\_Bang\_albo\_Rab\_7 & 1 & 1 & \textbf{1.39847279239048e-18} \\
DENV-2\_Bang\_albo\_Tun\_7 & \textbf{2.90061772234249e-115} & 1 & \textbf{8.2524889109902e-07} \\
DENV-1\_Sing\_albo\_Rey\_5 & \textbf{2.85400805651275e-83} & 0.999999999996661 & 0.716416349037314 \\
DENV-3\_Cair\_1998\_aeg\_Cair\_4.7 & 1 & 1 & \textbf{2.45691916468642e-19} \\
DENV-3\_Cair\_2008\_aeg\_Cair\_4.9 & 1 & 1 & \textbf{1.22262900221834e-22} \\
DENV-1\_Sing\_albo\_Rom\_5 & \textbf{8.87668483005163e-75} & 1 & \textbf{5.16473647104948e-57} \\
DENV-1\_Sing\_albo\_Mon\_5 & \textbf{3.92523870829146e-48} & 0.99999999999926 & 0.426393258903374 \\
ZIKV\_PueRic\_aeg\_PozRic\_7.2 & 1 & 1 & \textbf{1.28122608522439e-130} \\
ZIKV\_FrP\_aeg\_Tah\_6.8 & 1 & 1 & 1 \\
ZIKV\_Ugan\_aeg\_Tow\_6.5 & \textbf{4.37943933207197e-68} & 1 & 1 \\
ZIKV\_Ncal\_aeg\_FrP\_7 & 1 & 1 & \textbf{8.09665172169328e-22} \\
ZIKV\_Ncal\_pol\_Wal\_7 & 1 & 1 & \textbf{9.63666290727866e-10} \\
ZIKV\_Ncal\_albo\_Rab\_7.2 & \textbf{7.13051578275215e-36} & \textbf{1.17446401288081e-104} & 0.579774330713689 \\
ZIKV\_Ncal\_aeg\_Sam\_7 & \textbf{9.67237173105063e-93} & \textbf{1.09701260596836e-51} & \textbf{2.35623108497786e-05} \\
ZIKV\_Ncal\_aeg\_Ncal\_7 & 1 & \textbf{2.88279836012155e-51} & \textbf{5.00990688705673e-13} \\
ZIKV\_Ncal\_albo\_Tun\_7 & 0.745710435635756 & 0.591847563676192 & \textbf{4.26470649762686e-63} \\
ZIKV\_Ncal\_pol\_FrP\_7 & 1 & \textbf{1.27805340192947e-72} & \textbf{3.00710473305687e-76} \\
\bottomrule \\
\end{tabular}%
\label{table_wilcox_0.5_SEIDT}
\end{table}
\FloatBarrier

\begin{sidewaystable}[!hp]
\centering
\caption{Visual fit between observation and simulation for SEIDT model. Total of scenarios with good to very good quality is equal to 19/26 } 
\resizebox{\textwidth}{!}{%
\begin{tabular}{llllllllll}
\toprule
Dpe & Number of Dpe & Mosquito number (mean by Dpe) & ID scenario & dynI\_dynD & Quality of visual fit between ssObs and ssSim &  &  & Scoring visual fit between  obs vs sim  (/6) \\
 &  &  &  &  & I & D & T &  \\
 \midrule
3,7,14,21 & 4 & 30 & CHIKVc1 & Beta-Expo & ++ & + & - & 3 \\
3,6,9,12 & 4 & 48 & CHIKVc2 & Beta-Expo & ++ & + & + & 4 \\
3,5,7,10,12,14,20 & 7 & 19 & CHIKVc3 & Beta-Expo & - & + & ++ & 3 \\
6,9,14,21 & 4 & 39 & CHIKVc4 & Expo-Expo & - & ++ & + & 3 \\
3,7,10,14,21 & 5 & 27 & CHIKVc5 & Expo-Expo & ++ & + & + & 4 \\
3,5,7,10,12,14,20 & 7 & 18 & CHIKVc6 & Expo-Expo & ++ & + & + & 4 \\
3,6,9,14 & 4 & 20 & CHIKVc7 & Expo-Expo & ++ & ++ & ++ & 6 \\
3,6,9,14 & 4 & 20 & CHIKVc8 & Expo-Expo & ++ & ++ & ++ & 6 \\
2,4,6,8,10,12,14,16,18,20 & 10 & 60 & CHIKVc9 & Expo-Expo & ++ & - & - & 2 \\
3,7,14,21 & 4 & 28 & DENVc1 & Beta-Beta & + & + & - & 2 \\
3,7,10,14,21 & 5 & 28 & DENVc2 & Beta-Beta & + & + & + & 3 \\
7,14,21,28 & 4 & 17 & DENVc3 & Beta-Beta & ++ & ++ & + & 5 \\
2,3,4,5,6,7,10,14 & 8 & 21 & DENVc4 & Beta-Beta & ++ & + & ++ & 5 \\
2,3,4,5,6,7,10,14 & 8 & 21 & DENVc5 & Beta-Beta & + & ++ & - & 3 \\
7,14,21,28 & 4 & 17 & DENVc6 & Beta-Expo & ++ & ++ & ++ & 6 \\
7,14,21,28 & 4 & 18 & DENVc7 & Beta-Beta & ++ & ++ & + & 5 \\
2,4,6,8,10,12,14,16,18,20 & 10 & 30 & ZIKVc1 & Beta-Beta & + & ++ & + & 4 \\
6,9,14,21 & 4 & 39 & ZIKVc2 & Beta-Beta & ++ & ++ & + & 5 \\
5,7,10,14 & 4 & 26 & ZIKVc3 & Beta-Beta & ++ & ++ & ++ & 6 \\
6,9,14,21 & 4 & 32 & ZIKVc4 & Beta-Beta & + & ++ & ++ & 5 \\
6,9,14,21 & 4 & 38 & ZIKVc5 & Beta-Beta & + & + & + & 3 \\
3,7,14,21 & 4 & 30 & ZIKVc6 & Beta-Beta & + & + & ++ & 4 \\
6,9,14,21 & 4 & 39 & ZIKVc7 & Beta-Beta & ++ & ++ & + & 4 \\
6,9,14,21 & 4 & 26 & ZIKVc8 & Beta-Beta & ++ & ++ & + & 5 \\
7,10,14,21 & 4 & 23 & ZIKVc9 & Beta-Expo & ++ & ++ & ++ & 6 \\
6,9,14,21 & 4 & 30 & ZIKVc10 & Expo-Expo & - & + & ++ & 3\\
\bottomrule \\
\end{tabular}
}

(-): poor quality (at least two points outside the envelope/high dispersion of distributions around the mean),\\
(+): good quality (one point outside the envelope/slight dispersion of distributions around the mean),\\(++): very good quality (no points outside the envelope, distributions strongly clustered around the mean).
\label{table_visual_fit_SEIDT}
\end{sidewaystable}
\FloatBarrier

\begin{sidewaystable}[!hp]
\centering
\caption{Visual fit between observation and simulation for SEID model. Total of scenarios with good to very good quality is equal to 17/17 } 
\begin{tabular}{lllllll}
\toprule
Dpe & Mosquito number (mean by Dpe) & ID scenario short & dynI & Quality of visual fit between ssObs and ssSim &  & Scoring visual fit between  obs vs sim (/4) \\
 &  &  &  & I & D &  \\
\midrule 
2,6,9,14 & 43 & CHIKVp1 & Beta & ++ & ++ & 4 \\
 & 31 & CHIKVp2 & Beta & ++ & ++ & 4 \\
 & 19 & CHIKVp3 & Expo & ++ & ++ & 4 \\
4,6,8,12,18 & 16 & DENVp1 & Beta & ++ & ++ & 4 \\
 & 21 & DENVp2 & Beta & ++ & ++ & 4 \\
 & 17 & DENVp3 & Beta & ++ & ++ & 4 \\
 & 25 & DENVp4 & Beta & ++ & ++ & 4 \\
 & 17 & DENVp5 & Beta & ++ & + & 3 \\
 & 15 & DENVp6 & Beta & ++ & ++ & 4 \\
 & 19 & DENVp7 & Beta & ++ & + & 3 \\
 & 19 & DENVp8 & Beta & ++ & ++ & 4 \\
5,10,14,17,21 & 26 & ZIKVp1 & Beta & ++ & + & 3 \\
 & 19 & ZIKVp2 & Beta & ++ & + & 3 \\
 & 23 & ZIKVp3 & Beta & ++ & ++ & 4 \\
 & 23 & ZIKVp4 & Beta & ++ & ++ & 2 \\
 & 21 & ZIKVp5 & Beta & ++ & ++ & 4 \\
 & 13 & ZIKVp6 & Expo & ++ & ++ & 4\\
 \bottomrule \\
\end{tabular}

(-): poor quality (at least two points outside the envelope/high dispersion of distributions around the mean),\\
(+): good quality (one point outside the envelope/slight dispersion of distributions around the mean),\\(++): very good quality (no points outside the envelope, distributions strongly clustered around the mean).
\label{table_visual_fit_SEID}
\end{sidewaystable}
\FloatBarrier

\begin{table}[H]
\centering
\caption{Mean of RMSE for selected dynamics in each states for SEIDT model. Total of scenarios with a mean RMSE lower than 5 is equal to 19/26, 19/26, and 21/26 for the infected, disseminated, and transmitter states, respectively.} 
\begin{tabular}{llll}
\toprule
scenario & mean rmse I & mean rmse D & mean rmse T \\
\midrule 
CHIKV\_LaR\_albo\_Rab\_7 & 5 & 5 & 5 \\
CHIKV\_FrCarIs\_aeg\_Tha\_6 & 3 & 6 & 6 \\
CHIKV\_Ind\_gen\_Tir\_8 & 3 & 3 & 3 \\
CHIKV\_Tah\_aeg\_Tah\_7 & 6 & 4 & 6 \\
CHIKV\_LaR\_albo\_Tun\_7 & 2 & 4 & 5 \\
CHIKV\_Ind\_albo\_Tir\_8 & 1 & 3 & 3 \\
CHIKV\_Ncal\_aeg\_Mad\_Fun\_7,3 & 2 & 3 & 3 \\
CHIKV\_Ncal\_aeg\_Mad\_Pa\_do\_Ma\_7,3 & 1 & 3 & 4 \\
CHIKV\_BriVirIsl\_aeg\_PozRic\_6,9 & 1 & 7 & 7 \\
DENV-2\_Bang\_albo\_Rab\_7 & 6 & 4 & 2 \\
DENV-2\_Bang\_albo\_Tun\_7 & 3 & 2 & 1 \\
DENV-1\_Sing\_albo\_Rey\_5 & 2 & 1 & 1 \\
DENV-3\_Cair\_1998\_aeg\_Cair\_4,7 & 2 & 3 & 2 \\
DENV-3\_Cair\_2008\_aeg\_Cair\_4,9 & 4 & 4 & 2 \\
DENV-1\_Sing\_albo\_Rom\_5 & 2 & 2 & 1 \\
DENV-1\_Sing\_albo\_Mon\_5 & 1 & 1 & 1 \\
ZIKV\_PueRic\_aeg\_PozRic\_7,2 & 2 & 3 & 3 \\
ZIKV\_FrP\_aeg\_Tah\_6,8 & 4 & 5 & 4 \\
ZIKV\_Ugan\_aeg\_Tow\_6,5 & 3 & 3 & 2 \\
ZIKV\_Ncal\_aeg\_FrP\_7 & 6 & 5 & 2 \\
ZIKV\_Ncal\_pol\_Wal\_7 & 5 & 5 & 1 \\
ZIKV\_Ncal\_albo\_Rab\_7,2 & 7 & 2 & 1 \\
ZIKV\_Ncal\_aeg\_Sam\_7 & 4 & 2 & 1 \\
ZIKV\_Ncal\_aeg\_Ncal\_7 & 4 & 4 & 1 \\
ZIKV\_Ncal\_albo\_Tun\_7 & 3 & 2 & 1 \\
ZIKV\_Ncal\_pol\_FrP\_7 & 6 & 5 & 1 \\
\bottomrule \\
\end{tabular}%
\label{table_rmse_SEIDT}
\end{table}
\FloatBarrier

\begin{table}[H]
\centering
\caption{Mean of RMSE for selected dynamics in each states for SEID model.} 
\begin{tabular}{@{}lll@{}}
\toprule
scenario & mean rmse I & mean rmse D \\
\midrule
CHIKV\_IndOce\_albo\_Lyo\_3,94 & 1 & 1 \\
CHIKV\_IndOce\_albo\_Lyo\_6,07 & 2 & 4 \\
CHIKV\_IndOce\_albo\_Lyo\_8,63 & 1 & 2 \\
DENV-1\_Tha2010a\_aeg\_Tha\_5,74 & 1 & 1 \\
DENV-1\_Tha2010b\_aeg\_Tha\_5,70 & 2 & 2 \\
DENV-1\_Tha2013\_aeg\_Tha\_5,79 & 2 & 2 \\
DENV-1\_Lao2012\_aeg\_Tha\_5,84 & 2 & 2 \\
DENV-1\_Nca2013\_aeg\_Tha\_5,77 & 2 & 2 \\
DENV-1\_Gab2012\_aeg\_Tha\_5,82 & 2 & 2 \\
DENV-1\_Hai2012\_aeg\_Tha\_5,81 & 2 & 2 \\
DENV-1\_Tha2012\_aeg\_Tha\_5,80 & 3 & 3 \\
ZIKV\_Asi\_albo\_Mar\_6,48 & 2 & 3 \\
ZIKV\_Asi\_albo\_Mar\_6,87 & 3 & 3 \\
ZIKV\_Asi\_albo\_LaR\_5,90 & 3 & 3 \\
ZIKV\_Asi\_albo\_LaR\_6,87 & 3 & 3 \\
ZIKV\_Asi\_albo\_LaR\_8,37 & 2 & 2 \\
ZIKV\_Asi\_albo\_LaR\_6,48 & 2 & 2 \\
\bottomrule \\
\end{tabular}%
\label{table_rmse_SEID}
\end{table}
\FloatBarrier

\begin{sidewaystable}[!hp]
\centering
\caption{Analysis of the dynamics in each state for each scenario for SEIDT model.} 
\resizebox{\textwidth}{!}{%
\begin{tabular}{@{}lllllllllllllll@{}}
\toprule
& \multicolumn{5}{l}{Dynamics in state I} & \multicolumn{5}{l}{Dynamics in state D} & \multicolumn{4}{l}{Dynamics in state T} \\
scenario & increasing & decreasing & increasing then decreasing & decreasing then increasing & constant & increasing & decreasing & increasing then decreasing & increasing then decreasing then increasing & decreasing & increasing & decreasing & increasing then decreasing & constant \\
\midrule
\rowcolor[HTML]{EFEFEF} 
CHIKVc1 &  & 1 &  &  &  & 1 &  &  &  &  &  &  & 1 &  \\
CHIKVc2 &  & 1 &  &  &  & 1 &  &  &  &  &  &  & 1 &  \\
\rowcolor[HTML]{EFEFEF} 
CHIKVc3 &  & 1 &  &  &  &  &  &  & 1 &  & 1 &  &  &  \\
CHIKVc4 &  & 1 &  &  &  &  &  & 1 &  &  & 1 &  &  &  \\
\rowcolor[HTML]{EFEFEF} 
CHIKVc5 &  & 1 &  &  &  &  &  &  &  & 1 &  &  & 1 &  \\
CHIKVc6 &  & 1 &  &  &  &  &  &  &  & 1 &  &  & 1 &  \\
\rowcolor[HTML]{EFEFEF} 
CHIKVc7 &  & 1 &  &  &  &  &  &  &  & 1 &  &  & 1 &  \\
CHIKVc8 &  & 1 &  &  &  &  &  &  &  & 1 &  &  & 1 &  \\
\rowcolor[HTML]{EFEFEF} 
CHIKVc9 &  &  &  &  & 1 &  &  &  &  & 1 &  & 1 &  &  \\
DENVc1 &  &  & 1 &  &  & 1 &  &  &  &  & 1 &  &  &  \\
\rowcolor[HTML]{EFEFEF} 
DENVc2 &  &  & 1 &  &  & 1 &  &  &  &  & 1 &  &  &  \\
DENVc3 &  &  &  & 1 &  &  &  & 1 &  &  & 1 &  &  &  \\
\rowcolor[HTML]{EFEFEF} 
DENVc4 &  & 1 &  &  &  &  &  & 1 &  &  & 1 &  &  &  \\
DENVc5 &  & 1 &  &  &  &  &  & 1 &  &  & 1 &  &  &  \\
\rowcolor[HTML]{EFEFEF} 
DENVc6 &  &  &  & 1 &  &  &  & 1 &  &  &  &  &  &  \\
DENVc7 &  & 1 &  &  &  &  &  & 1 &  &  &  &  & 1 &  \\
\rowcolor[HTML]{EFEFEF} 
ZIKVc1 &  & 1 &  &  &  &  &  &  & 1 &  &  &  & 1 &  \\
ZIKVc2 &  & 1 &  &  &  &  &  & 1 &  &  & 1 &  &  &  \\
\rowcolor[HTML]{EFEFEF} 
ZIKVc3 &  & 1 &  &  &  &  &  & 1 &  &  & 1 &  &  &  \\
ZIKVc4 &  &  & 1 &  &  & 1 &  &  &  &  & 1 &  &  &  \\
\rowcolor[HTML]{EFEFEF} 
ZIKVc5 &  &  & 1 &  &  &  &  &  &  & 1 &  &  &  & 1 \\
ZIKVc6 &  &  & 1 &  &  &  &  & 1 &  &  & 1 &  &  &  \\
\rowcolor[HTML]{EFEFEF} 
ZIKVc7 &  & 1 &  &  &  & 1 &  &  &  &  & 1 &  &  &  \\
ZIKVc8 &  & 1 &  &  &  & 1 &  &  &  &  &  &  &  & 1 \\
\rowcolor[HTML]{EFEFEF} 
ZIKVc9 &  & 1 &  &  &  &  &  & 1 &  &  & 1 &  &  &  \\
ZIKVc10 &  &  & 1 &  &  &  &  & 1 &  &  &  &  &  & 1 \\
\bottomrule \\
\end{tabular}%
}
\label{table_dynamics_SEIDT}
\end{sidewaystable}
\FloatBarrier

\begin{table}[H]
\centering
\caption{Results of the Kolmogorov-Smirnov test: percentage of similar beta distributions in the infected state (I) for each scenario with a beta distribution mainly selected in I. Scenarios for which >50\% of p-value were > 0.05, corresponding to the percentage of distribution which are statistically similar, are shown in bold (12/19 for the SEIDT model and 9/15 for the SEID model)} 
\begin{tabular}{@{}llll@{}}
\toprule
Scenarios SEIDT model & Percentage p-value \textgreater 0.05 & Scenarios SEID model & Percentage p-value \textgreater 0.05 \\
\midrule
CHIKV\_LaR\_albo\_Rab\_7 & 33 & CHIKV\_IndOce\_albo\_Lyo\_3.94 & 50 \\
\textbf{CHIKV\_FrCarIs\_aeg\_Tha\_6} & \textbf{61} & CHIKV\_IndOce\_albo\_Lyo\_6.07 & 41 \\
\textbf{CHIKV\_Ind\_gen\_Tir\_8} & \textbf{61} & \textbf{DENV-1\_Tha2010a\_aeg\_Tha\_5.74} & \textbf{82} \\
\textbf{DENV-2\_Bang\_albo\_Rab\_7} & \textbf{57} & \textbf{DENV-1\_Tha2010b\_aeg\_Tha\_5.70} & \textbf{69} \\
DENV-2\_Bang\_albo\_Tun\_7 & 45 & \textbf{DENV-1\_Tha2013\_aeg\_Tha\_5.79} & \textbf{79} \\
\textbf{DENV-1\_Sing\_albo\_Rey\_5} & \textbf{56} & \textbf{DENV-1\_Lao2012\_aeg\_Tha\_5.84} & \textbf{87} \\
\textbf{DENV-3\_Cair\_1998\_aeg\_Cair\_4.7} & \textbf{62} & \textbf{DENV-1\_Nca2013\_aeg\_Tha\_5.77} & \textbf{62} \\
DENV-3\_Cair\_2008\_aeg\_Cair\_4.9 & 46 & DENV-1\_Gab2012\_aeg\_Tha\_5.82 & 42 \\
\textbf{DENV-1\_Sing\_albo\_Rom\_5} & \textbf{61} & \textbf{DENV-1\_Hai2012\_aeg\_Tha\_5.81} & \textbf{59} \\
\textbf{DENV-1\_Sing\_albo\_Mon\_5} & \textbf{55} & DENV-1\_Tha2012\_aeg\_Tha\_5.80 & 27 \\
ZIKV\_PueRic\_aeg\_PozRic\_7.2 & 44 & \textbf{ZIKV\_Asi\_albo\_Mar\_6.48} & \textbf{51} \\
\textbf{ZIKV\_FrP\_aeg\_Tah\_6.8} & \textbf{54} & \textbf{ZIKV\_Asi\_albo\_Mar\_6.87} & \textbf{56} \\
\textbf{ZIKV\_Ugan\_aeg\_Tow\_6.5} & \textbf{54} & \textbf{ZIKV\_Asi\_albo\_LaR\_5.90} & \textbf{64} \\
ZIKV\_Ncal\_aeg\_FrP\_7 & 45 & ZIKV\_Asi\_albo\_LaR\_6.87 & 47 \\
\textbf{ZIKV\_Ncal\_pol\_Wal\_7} & \textbf{57} & ZIKV\_Asi\_albo\_LaR\_8.37 & 50 \\
ZIKV\_Ncal\_albo\_Rab\_7.2 & 49 &  &  \\
\textbf{ZIKV\_Ncal\_aeg\_Sam\_7} & \textbf{51} &  &  \\
ZIKV\_Ncal\_aeg\_Ncal\_7 & 47 &  &  \\
\textbf{ZIKV\_Ncal\_albo\_Tun\_7} & \textbf{61} &  & \\
\bottomrule \\
\end{tabular}%
\label{table_ks_state_I_SEIDT}
\end{table}
\FloatBarrier

\begin{table}[H]
\centering
\caption{Results of the Kolmogorov-Smirnov test: percentage of similar beta distributions in the disseminated state (D) for each scenario with a beta distribution mainly selected in D. Scenarios for which >50\% of p-value were > 0.05, corresponding to the percentage of distribution which are statistically similar, are shown in bold (3/14).}
\begin{tabular}{@{}ll@{}}
\toprule
Scenario & Percentage p-value \textgreater 0.05 \\
\midrule 
DENV-2\_Bang\_albo\_Rab\_7 & 47 \\
DENV-2\_Bang\_albo\_Tun\_7 & 50 \\
DENV-1\_Sing\_albo\_Rey\_5 & 46 \\
\textbf{DENV-3\_Cair\_1998\_aeg\_Cair\_4.7} & \textbf{53} \\
DENV-3\_Cair\_2008\_aeg\_Cair\_4.9 & 45 \\
DENV-1\_Sing\_albo\_Mon\_5 & 50 \\
\textbf{ZIKV\_PueRic\_aeg\_PozRic\_7.2} & \textbf{54} \\
ZIKV\_FrP\_aeg\_Tah\_6.8 & 50 \\
\textbf{ZIKV\_Ugan\_aeg\_Tow\_6.5} & \textbf{52} \\
ZIKV\_Ncal\_aeg\_FrP\_7 & 49 \\
ZIKV\_Ncal\_pol\_Wal\_7 & 48 \\
ZIKV\_Ncal\_albo\_Rab\_7.2 & 46 \\
ZIKV\_Ncal\_aeg\_Sam\_7 & 48 \\
ZIKV\_Ncal\_aeg\_Ncal\_7 & 49 \\
\bottomrule \\
\end{tabular}%
\label{table_ks_state_D_SEIDT}
\end{table}
\FloatBarrier

\FloatBarrier




\end{document}